\documentclass[12pt]{article}
\usepackage[pdftex]{graphicx}
\usepackage{url}
\usepackage{braket}
\usepackage{latexsym}
\usepackage{mathtools}
\usepackage{url}
\usepackage{graphicx}
\usepackage{amsmath}
\usepackage{cite}

\usepackage[utf8]{inputenc}

\pdfoutput=1
\usepackage{tikz}
\usepackage{putex}
\usepackage{feyn}
\usepackage[vcentermath]{youngtab}
\usepackage{subfig}
\usepackage{lscape}

\usepackage{graphicx}
\usepackage{epstopdf}
\usepackage{enumerate}
\usepackage{cite}
\usepackage{tensor} 
\usepackage{slashed}
\usepackage{amsthm}
\usepackage{amsmath}
\usepackage{amssymb}
\usepackage{mathrsfs}
\usepackage{lgrind}
\usepackage{youngtab}

\usepackage{amsmath,amssymb,braket}

\usepackage{bbm}

\usepackage{hyperref}

\numberwithin{equation}{section}
\newtheorem*{con}{Conjecture}

\newcommand {\be} {\begin {equation}}
\newcommand {\ee} {\end {equation}}

\newcommand {\bes} {\begin {equation*}}
\newcommand {\ees} {\end {equation*}}

\def\CH{{\cal H}}

\newcommand{\beq}{\begin{equation}}
\newcommand{\eeq}{\end{equation}}

\def\be{ \begin{equation} }
\def\ee{ \end{equation} }

\def \be {\beta}

\def \beq { \begin{equation}}
\def \eeq {\end{equation}}

\DeclareMathOperator*{\Pf}{Pf}

\begin{document}

\preprint{PUPT-2581}

\institution{UCB}{Department of Physics, University of California, Berkeley, CA 94720}
\institution{PU}{Department of Physics, Princeton University, Princeton, NJ 08544}
\institution{PCTS}{Princeton Center for Theoretical Science, Princeton University, Princeton, NJ 08544}
\institution{HU}{Department of Physics, Harvard University, Cambridge, MA 02138}

\title{
Symmetry Breaking
in Coupled SYK or Tensor Models
}

\authors{Jaewon Kim,\worksat{\UCB} Igor R.~Klebanov,\worksat{\PU,\PCTS}
Grigory Tarnopolsky,\worksat{\HU} Wenli Zhao\worksat{\PU}
}

\abstract{
We study a large $N$ tensor model with $O(N)^3$ symmetry containing two flavors of Majorana fermions, $\psi_1^{abc}$ and $\psi_2^{abc}$. We also study its random counterpart consisting of two coupled Sachdev-Ye-Kitaev models, each one containing $N_{\rm SYK}$ Majorana fermions. In these models we 
assume tetrahedral quartic Hamiltonians which depend on a real coupling parameter $\alpha$.
We find a duality relation between two Hamiltonians with different values of $\alpha$, which allows us to restrict the model to the range of $-1\leq \alpha\leq 1/3$.
The scaling dimension of the fermion number operator $Q=i\psi_1^{abc} \psi_2^{abc}$ is complex and of the form $1/2 + i f(\alpha)$ in the range $-1\leq \alpha<0$,
indicating an instability of the conformal phase. Using Schwinger-Dyson equations to solve for the Green functions, we show that in the true low-temperature phase this operator acquires an expectation value. This demonstrates
the breaking of an anti-unitary particle-hole symmetry and other discrete symmetries.
We also calculate spectra of the coupled SYK models for 
values of $N_{\rm SYK}$ where exact diagonalizations are possible. For negative $\alpha$ we find a gap separating the two lowest energy states from the rest of the spectrum; this leads
to exponential decay of the zero-temperature correlation functions.
For $N_{\rm SYK}$ divisible by $4$, the two lowest states have a small splitting. They become degenerate in the large $N_{\rm SYK}$ limit, 
as expected from the spontaneous breaking of a $\mathbb{Z}_2$ symmetry. 
}

\date{}

\maketitle

\tableofcontents

\section{Introduction and Summary}

During the past several years there has been a flurry of activity on fermionic quantum mechanical models which are exactly solvable in the large $N$ limit because they are dominated by the so-called melonic Feynman diagrams. Work in this direction began with the Sachdev-Ye-Kitaev (SYK) models \cite{Sachdev:1992fk,Kitaev:2015,Sachdev:2015efa,Kitaev:2017awl}, which have random couplings.  
More recently, the tensor quantum mechanical models \cite{Witten:2016iux,Klebanov:2016xxf}, which have continuous symmetry groups and no randomness, were constructed following the
body of research on melonic large $N$ tensor models in $d=0$ 
\cite{Gurau:2009tw,Gurau:2011aq,Gurau:2011xq,Bonzom:2011zz,Tanasa:2011ur,Bonzom:2012hw,Carrozza:2015adg}
(for reviews, see  \cite{Gurau:2011xp,Tanasa:2015uhr,Delporte:2018iyf,Klebanov:2018fzb}). Both the random and non-random quantum mechanical models are
solvable via the melonic Schwinger-Dyson equations \cite{Polchinski:2016xgd,Maldacena:2016hyu,Gross:2016kjj,Jevicki:2016bwu,Kitaev:2017awl}, which indicate the existence of the nearly conformal phase which saturates the chaos bound. They shed new light on the dynamics of two-dimensional black holes \cite{Almheiri:2014cka,Jensen:2016pah,Maldacena:2016upp,Engelsoy:2016xyb}. 

These models may also have applications to a range of problems in condensed matter physics, including the strange metals 
\cite{Sachdev:2015efa,Gu:2016oyy,Kolovsky:2016irf,Bi:2017yvx,Haldar:2018qpn,Song:2017pfw,Jian:2017tzg,Patel:2018zpy}. 
With such applications in mind, it is interesting to study various dynamical phenomena in the SYK
and tensor models. For example, phase transitions in such models have been studied in \cite{Banerjee:2016ncu,Azeyanagi:2017drg, Fu:2018spl}.
In this paper we identify a simple setting where spontaneous symmetry breaking can occur: two SYK or tensor models coupled via a quartic interaction. 
We take this interaction to be purely melonic (i.e. tetrahedral in the tensor model case), so that the symmetry breaking can be deduced from
the large $N$ Schwinger-Dyson equations.

In the random case, we will study two coupled SYK models with the Hamiltonian
\begin{equation}
\label{SYK2fl}
H = \frac{1}{4!} J_{ijkl}\left (\chi_1^i \chi_1^j \chi_1^k \chi_1^l + \chi_2^i \chi_2^j \chi_2^k \chi_2^l  + 6 \alpha \chi_1^i \chi_1^j \chi_2^k \chi_2^l \right )\ ,
\end{equation}
where, as usual, all repeated indices are summed over.
The Majorana fermions are
$\chi_1^i$ and  $\chi_2^i$ with $i=1, \ldots, N_{\rm SYK}$, and $J_{ijkl}$ is a fully anti-symmetric real tensor with a Gaussian distribution.\footnote{This model seems similar to 
a coupled SYK model introduced in \cite{Gu:2016oyy}, but there each of the three terms in the Hamiltonian would have an independent random coupling. 
As a result, the Schwinger-Dyson equations are different
from those for theory  (\ref{SYK2fl}). The complex scaling dimension and symmetry breaking, which we describe in this paper, do not appear in the model of \cite{Gu:2016oyy}.}
 We will show that the real parameter $\alpha$ may be restricted to the range $-1\leq \alpha \leq 1/3$ by a duality symmetry. 
This quartic 
Hamiltonian, which couples $2N_{\rm SYK}$ Majorana fermions, is invariant under an anti-unitary particle-hole symmetry \cite{2009AIPC.1134...22K,Fidkowski:2009dba,Witten:2015aba,PhysRevB.95.115150,Fu:2016yrv,Cotler:2016fpe} generated by ${\cal P}$; see eq. (\ref{phsym}).
However, we will show that for $-1\leq \alpha <0$ 
this $\mathbb{Z}_2$ symmetry is spontaneously broken when $N_{\rm SYK}$ is divisible by 4 and taken to infinity.\footnote{When $N_{\rm SYK}$ is finite and not divisible by $4$, so that the total number of Majorana fermions is not divisible by $8$, the particle-hole symmetry is broken by a discrete anomaly \cite{2009AIPC.1134...22K,Fidkowski:2009dba,Witten:2015aba,PhysRevB.95.115150,
Fu:2016yrv,Cotler:2016fpe}.} 
 In this limit the fermion number operator $Q= i  \chi_1^j \chi_2^j$ acquires an expectation value.
 This leads to a gapped phase in two coupled SYK models similar to that found by Maldacena and Qi \cite{Maldacena:2018lmt} (for further results see \cite{Garcia-Garcia:2019poj});
however, instead of the quartic they assumed a quadratic coupling term $\mu Q$ which breaks the $\mathbb{Z}_2$ symmetry explicitly.
 This gapped phase was argued to be dual to a traversable wormhole in two-dimensional gravity \cite{Gao:2016bin,Maldacena:2017axo}, and our model (\ref{SYK2fl}) may have a similar interpretation for $-1\leq \alpha<0$.

As we show in section \ref{complexscaling}, a sign of instability of the conformal phase for $-1\leq \alpha <0$ is the presence of a complex scaling dimensions of the form $1/2 + i f(\alpha)$.  
Appearance of complex dimensions with real part equal to $d/2$ for some single-trace operators is a common phenomenon in large $N$ models  
\cite{Dymarsky:2005uh,Pomoni:2008de,Grabner:2017pgm,Prakash:2017hwq,Gorbenko:2018ncu}. 
Via the AdS/CFT correspondence \cite{Maldacena:1997re, Gubser:1998bc, Witten:1998qj}, such operators are related to scalar fields which violate the Breitenlohner-Freedman stability bound \cite{Breitenlohner:1982jf}. The fact that $\alpha=0$ is the lower edge of the conformal window is related to appearance of the marginal double-trace operator $Q^2$ there. 
For $0<\alpha\leq 1/3$ there are actually two fixed points connected by the flow of the coefficient of $Q^2$, but at $\alpha=0$ they merge and annihilate, as explained for
example in \cite{Kaplan:2009kr,Giombi:2015haa}.    

The complex scaling dimensions have been observed in bosonic tensor models \cite{Giombi:2017dtl, Giombi:2018qgp}, as well as in
a complex fermionic model introduced in \cite{Klebanov:2016xxf}  following the work in \cite{Gurau:2016lzk}.
This fermionic model is often called ``bipartite" because of the two types of interaction vertices (black and white) arranged in an alternating fashion, 
since the propagator must connect different vertices. 
The bipartite model was further studied in \cite{Klebanov:2018fzb} and shown to possess a complex scaling dimension of the operator $\bar \psi^{abc} \psi^{abc}$. 
Here we generalize this tensor model to one with a continuous parameter $\alpha$ in such a way that the bipartite model corresponds to $\alpha=-1$. This $O(N)^3$ symmetric model for Majorana fermions $\psi_1^{abc}$ and $\psi_2^{abc}$, with $a,b,c=1, \ldots, N$, has Hamiltonian
\begin{eqnarray}
\label{twoflavorHamilt}
& H = \frac{g}{4} \left ( \psi_1^{a_1b_1c_1}\psi_1^{a_1b_2c_2}\psi_1^{a_2b_1c_2}\psi_1^{a_2b_2c_1} +
\psi_2^{a_1b_1c_1}\psi_2^{a_1b_2c_2}\psi_2^{a_2b_1c_2}\psi_2^{a_2b_2c_1}\right ) \\
& +\frac{g\alpha }{2} \left ( \psi_1^{a_1b_1c_1}\psi_1^{a_1b_2c_2}\psi_2^{a_2b_1c_2}\psi_2^{a_2b_2c_1} +
\psi_1^{a_1b_1c_1}\psi_2^{a_1b_2c_2}\psi_1^{a_2b_1c_2}\psi_2^{a_2b_2c_1}+ \psi_1^{a_1b_1c_1}\psi_2^{a_1b_2c_2}\psi_2^{a_2b_1c_2}\psi_1^{a_2b_2c_1}\right )\ .\notag
\end{eqnarray}
For $\alpha=0$ this describes two decoupled copies of the basic Majorana $O(N)^3$ model with the tetrahedral interaction \cite{Klebanov:2016xxf}. The coupling term proportional to
$\alpha$ preserves its discrete symmetries and also has the tetrahedral structure, i.e. every two tensors have only one index contraction, so that the model (\ref{twoflavorHamilt}) is melonic. 
It is the tensor counterpart of the coupled SYK model (\ref{SYK2fl}), and in the large $N$ limit it is governed by the same Schwinger-Dyson equations for the two-point and four-point
functions.\footnote{In \cite{Bi:2017yvx,Jian:2017tzg} quartic interactions were added to SYK models, which have a ``double-trace" structure and contain an additional random tensor
$C_{ij}$. These interactions do not have a tensor counterpart because 
$\psi_1^{abc} \psi_1^{abc} $ is a c-number.}  

In section 2 we derive the Schwinger-Dyson equations and use them to study the scaling dimensions of various $O(N)^3$ invariant fermion bilinears. We also exhibit a duality symmetry which allows
us to restrict the model to the range $-1\leq \alpha \leq 1/3$. The nearly conformal phase of the theory is stable for $0\leq \alpha \leq 1/3$, but it is unstable for
$-1\leq \alpha <0$ as signaled by the complex scaling dimension of operator $i\psi_1^{abc}\psi_2^{abc}$.
The true behavior of the theory with negative $\alpha$ is the spontaneous breaking of the particle-hole $\mathbb{Z}_2$ symmetry, as we demonstrate
in section 3.
In section \ref{SDeqsandeffact} and \ref{SDeqsnumerics} we numerically study the large $N$ Schwinger-Dyson equations and exhibit the exponential decay of correlators at low temperature. 
We also ascertain the existence of second-order phase transitions by numerically computing the free energy. In section \ref{numericalspectrum} we study the numerical spectrum 
of the coupled SYK model (\ref{SYK2fl}) via exact diagonalizations
at finite $N_{\rm SYK}$. We observe that for $-1\leq \alpha<0$ there is a gap separating the two lowest energy states from the rest of the spectrum. 
For $N_{\rm SYK}$ divisible by $4$ there is also a small gap between the two lowest states, consistent with the fact that the ground state must be non-degenerate 
\cite{2009AIPC.1134...22K,Fidkowski:2009dba,Witten:2015aba,PhysRevB.95.115150,Fu:2016yrv,Cotler:2016fpe}, but
this gap decreases as $N_{\rm SYK}$ is increased. In the large $N_{\rm SYK}$ limit, the two lowest states become degenerate and give rise to the 
two inequivalent vacua, which are present due to the spontaneous breaking of the $\mathbb{Z}_2$ particle-hole symmetry.

This means that the low-temperature entropy is large for $0\leq \alpha\leq 1/3$ but vanishes
for $-1\leq \alpha<0$. It is tempting to suggest that the latter case is dual to a wormhole. This senstivity to the sign of the interaction coupling two CFTs is like in \cite{Gao:2016bin},
where the traversable wormhole appears only for one of the signs.\footnote{On the other hand, in the approach of \cite{Maldacena:2018lmt}, where the quadratic term $\mu Q$ 
was added to couple the two SYK
models, the gap (and therefore the wormhole) appeared for either sign of $\mu$.}

\section{Schwinger-Dyson Equations and Scaling Dimensions}

\begin{figure}[h!]
  \begin{center}  
    \includegraphics [width=0.95\textwidth, angle=0.]{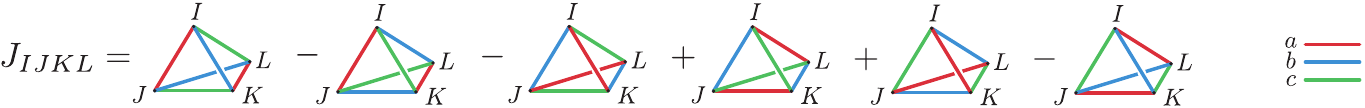}
  \end{center}
  \caption{Pictorial representation of the antisymmetric tensor $J_{IJKL}$. }
  \label{fig:jIJKL}
\end{figure}

In this section we study the two-flavor tensor model with Hamiltonian (\ref{twoflavorHamilt}).\footnote{This section is based in part on J.K.'s Princeton University senior 
thesis \cite{Jaewon:2018}.}  It can be compactly written in the form 
\begin{equation}
\label{twoflavorHamiltshort}
H= \frac{1}{4!} J_{IJKL}\left (\psi_{1}^{I} \psi_{1}^{J}\psi_{1}^{K}\psi_{1}^{L} + \psi_{2}^{I} \psi_{2}^{J}\psi_{2}^{K}\psi_{2}^{L} + 6 \alpha 
\psi_{1}^{I} \psi_{1}^{J}\psi_{2}^{K}\psi_{2}^{L} \right )\,,
\end{equation}
where the capital letters are a shorthand notation for three tensor indices: $I=a_{1}b_{1}c_{1}$, $J=a_{2}b_{2}c_{2}$, etc, and the non-random tetrahedral tensor coupling consists of six terms 
\begin{align}
J_{IJKL} =g \sum_{\sigma\in S_{3}}\textrm{sgn}(\sigma)\delta_{a_{1}a_{\sigma(2)}}\delta_{b_{1}b_{\sigma(3)}}\delta_{c_{1}c_{\sigma(4)}}\delta_{b_{\sigma(2)}b_{\sigma(4)}}\delta_{c_{\sigma(2)}c_{\sigma(3)}}\delta_{a_{\sigma(3)}a_{\sigma(4)}}\, .
\end{align}
The tensor $J_{IJKL}$ is antisymmetric under permutation of  indices $I,J,K,L$ and has a tetrahedron topology as shown in figure \ref{fig:jIJKL}. In the form (\ref{twoflavorHamiltshort}) the tensor model Hamiltonian is transparently similar  to the SYK  one (\ref{SYK2fl}).
In terms of the complex tensors
\begin{equation}
 \psi^{I} = {1\over \sqrt 2} (\psi_1^{I} + i \psi_2^{I})\ , \qquad \bar \psi^{I} = {1\over \sqrt 2} (\psi_1^{I} - i \psi_2^{I})\
\end{equation}
the Hamiltonian (\ref{twoflavorHamiltshort}) assumes the form
\begin{equation}
H=\frac{1}{4!} J_{IJKL} \bigg( \frac{1-3\alpha}{2} 
\left (\psi^I \psi^J \psi^K \psi^L + \bar \psi^I \bar \psi^J \bar \psi^K \bar \psi^L  \right)
+ 3 (1+ \alpha)  \bar \psi^I \bar \psi^J \psi^K \psi^L \bigg )
\ .
\end{equation}

The Hamiltonian (\ref{twoflavorHamiltshort}) is invariant under the $O(N)^3$ transformation
\begin{equation}
\psi_i^{abc} \rightarrow A^{a}_{a'}B^{b}_{b'}C^{c}_{c'}\psi_i^{a'b'c'}\ ,
\end{equation}
where $A$, $B$, and $C$ are orthogonal matrices. 
In addition, it has a particle-hole $\mathbb{Z}_2$ symmetry\footnote{The Hamiltonian also has discrete symmetries which do not involve $K$, which combine into the dihedral group $D_4$. 
This is discussed in detail for the coupled SYK counterpart in section \ref{symbreak} and in the Appendix.}
generated by \cite{2009AIPC.1134...22K,Fidkowski:2009dba,Witten:2015aba,PhysRevB.95.115150,Fu:2016yrv,Cotler:2016fpe}, 
\begin{equation}
\label{phsymtensor}
{\cal P}= K \prod_{I} (\psi^{I}+ \bar \psi^{I} )\ .
\end{equation}
where $K$ is the anti-unitary operator 
which acts by
\begin{equation}
K i K=- i\ , \qquad K\psi^{I}K= \psi^{I}\ , \qquad  K\bar{\psi}^{I}K=  \bar{\psi}^{I}\ .
\end{equation}
The fermion number operator 
\begin{equation}
Q=i \psi_1^{I} \psi_2^{I} =\frac{1}{2} [\bar \psi^{I}, \psi^{I}]
\end{equation}
does not in general commute with $H$, but it is conserved mod $4$. 
The particle-hole symmetry is not anomalous only if the total number of fermions $2 N^3$ is a multiple of $8$, i.e. when $N$ is even
\cite{2009AIPC.1134...22K,Fidkowski:2009dba,Witten:2015aba,PhysRevB.95.115150,Fu:2016yrv,Cotler:2016fpe}.
Even in this case, we will argue that in the large $N$ limit the symmetry is spontaneously broken for $-1\leq \alpha<0$ because
$Q$ acquires an expectation value.

 \subsection{Duality in the Two-Flavor Models}
\label{duality}

In this section we show that the two-flavor models with different values of $\alpha$ can be equivalent. We will demonstrate this explicitly 
in the tensor model case (\ref{twoflavorHamiltshort}), but the SYK case (\ref{SYK2fl}) works analogously. Let us perform the following transformation on the Majorana fermions:
\begin{equation}
\psi_1^{I} = \frac{1}{\sqrt{2}}(\tilde\psi_1^{I} + \tilde\psi_2^{I})\ , \qquad \psi_2^{I} = \frac{1}{\sqrt{2}}(\tilde\psi_1^{I} - \tilde\psi_2^{I})\ . 
\end{equation}
It preserves the anticommutation relations, and turns the Hamiltonian (\ref{twoflavorHamiltshort}) into \footnote{Using antisymmetry of the tensor $J_{IJKL}$ one can operate with Majorana 
fermions as commuting variables but keeping order of $I,J,K,L$ indices fixed. }
\begin{equation}
\label{Hamiltchoice}
H = \frac{1}{4!}J_{IJKL} \frac{1+3\alpha}{2}\left (\tilde{\psi}_{1}^{I} \tilde{\psi}_{1}^{J}\tilde{\psi}_{1}^{K}\tilde{\psi}_{1}^{L} + \tilde{\psi}_{2}^{I} \tilde{\psi}_{2}^{J}\tilde{\psi}_{2}^{K}\tilde{\psi}_{2}^{L} + \frac{6 (1-\alpha)}{1+3\alpha} 
\tilde{\psi}_{1}^{I} \tilde{\psi}_{1}^{J}\tilde{\psi}_{2}^{K}\tilde{\psi}_{2}^{L} \right )\,.
\end{equation}
Thus the energy levels are symmetric under the duality transformation 
\begin{equation}
J \to \frac{1+3\alpha}{2}J, \qquad \alpha \to \frac{1-\alpha}{1+3\alpha}\,. \label{duality1}
\end{equation}

Defining 
\begin{equation}
\tilde{\alpha}=\frac{1}{2}(1+3\alpha)\ , \qquad \tilde J = J \sqrt{ |\tilde{\alpha}| }\ ,
\end{equation}
we find that the duality transformation
\begin{equation}
\label{dualrel}
\tilde{\alpha}\to 1/\tilde{\alpha}\ , \qquad \tilde J \to \tilde J\ 
\end{equation}
acts on the rescaled Hamiltonian $\tilde H= H/\sqrt{|\tilde \alpha|}$:\footnote{For the original Hamiltonian (\ref{Hamiltchoice} this transformation rescales the energy levels. 
Therefore, our results for dimensionful quantities, like energy levels and Green functions, will not respect the duality under (\ref{dualrel}).}  
\begin{equation}
\label{Hamiltchoice}
\tilde H = \frac{1}{4!}\tilde J_{IJKL} 
\left (\tilde{\psi}_{1}^{I} \tilde{\psi}_{1}^{J}\tilde{\psi}_{1}^{K}\tilde{\psi}_{1}^{L} + \tilde{\psi}_{2}^{I} \tilde{\psi}_{2}^{J}\tilde{\psi}_{2}^{K}\tilde{\psi}_{2}^{L} + 
\left (-2 + \frac{4} {\tilde \alpha} \right ) 
\tilde{\psi}_{1}^{I} \tilde{\psi}_{1}^{J}\tilde{\psi}_{2}^{K}\tilde{\psi}_{2}^{L} \right )\,.
\end{equation}

This means that the fundamental domain is $-1 \leq \tilde{\alpha} \leq 1$. Thus, we may restrict
$\alpha$ to the domain
\begin{equation}
-1 \leq \alpha \leq \frac{1}{3}\ .
\end{equation}
The values of $\alpha$ outside of this domain are related to it by the duality. For $\alpha=-1$ the transformation (\ref{duality1}) maps the theory into itself, but with 
$H\rightarrow -H$.

In fact, the case $\alpha = -1$ corresponds to the complex bipartite model \cite{Gurau:2016lzk, Klebanov:2018fzb}:
\begin{equation}
H_{\alpha=-1} = 2\frac{1}{4!}J_{IJKL}\left(\psi^{I}\psi^{J}\psi^{K}\psi^{L}+\bar{\psi}^{I}\bar{\psi}^{J}\bar{\psi}^{K}\bar{\psi}^{L}\right)\,,
\label{bipartiteH}
\end{equation}

where we introduced the complex tensor $\psi^I=\frac{1}{\sqrt{2}} (\psi_1^I+i\psi_2^I)$.

The theory with $\alpha = 1/3$ is mapped into itself by  (\ref{duality1}).
In this case the Hamiltonian is
\begin{equation}
H_{\alpha=\frac{1}{3}} = 4\frac{1}{4!}J_{IJKL} \bar{\psi}^{I}\bar{\psi}^{J} \psi^{K}\psi^{L}\ ,
\end{equation}
which has $O(N)^3\times U(1)$ symmetry.
In the three-index notation
\begin{equation}
H_{\alpha=\frac{1}{3}} =
\frac{g }{3} \left ( \bar \psi^{a_1b_1c_1}\bar \psi^{a_1b_2c_2} \psi^{a_2b_1c_2}\psi^{a_2b_2c_1} -
\bar \psi^{a_1b_1c_1}\bar \psi^{a_2 b_1 c_2} \psi^{a_ 1 b_2 c_2}\psi^{a_2 b_2c_1}+ \bar \psi^{a_1b_1c_1}\bar \psi^{a_2 b_2 c_ 1} \psi^{a_1 b_2 c_2}\psi^{a_2b_1 c_2 }\right )\ .
\end{equation}
This is different from the $SU(N)^2\times O(N)\times U(1)$ symmetric complex tensor model \cite{Klebanov:2016xxf}; the latter involves taking only the first term in this Hamiltonian.

\subsection{Feynman rules and two-point functions}
\begin{figure}[h!]
  \begin{center}  
    \includegraphics [width=0.8\textwidth, angle=0.]{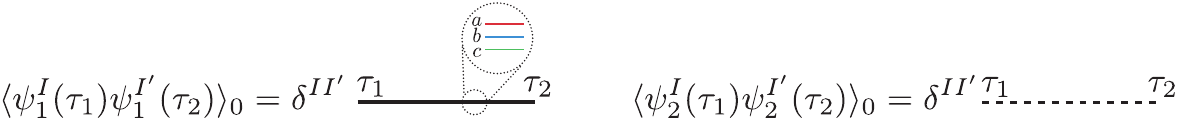}
  \end{center}
  \caption{Bare propagators for the Majorana tensor fields. Each thick black solid or dashed line caries three tensor indices $a,b,c$. }
  \label{fig:props}
\end{figure}

\begin{figure}[h!]
  \begin{center}  
    \includegraphics [width=0.75\textwidth, angle=0.]{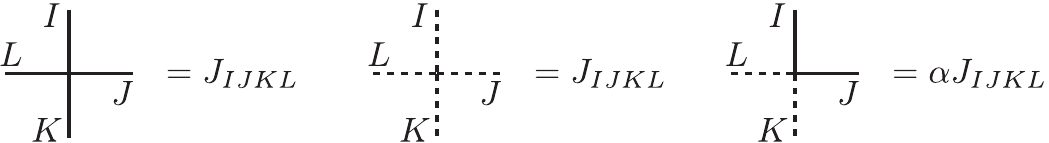}
  \end{center}
  \caption{Interaction vertices. }
  \label{fig:vert}
\end{figure}

\begin{figure}[h!]
  \begin{center}  
    \includegraphics [width=0.45\textwidth, angle=0.]{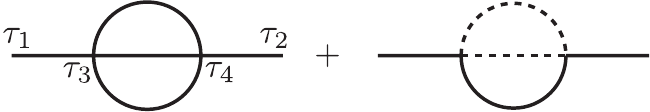}
  \end{center}
  \caption{The leading melonic correction to the full two-point function $\textbf{G}_{11}$. }
  \label{fig:firstcor}
\end{figure}

\begin{figure}[h!]
  \begin{center}  
    \includegraphics [width=0.45\textwidth, angle=0.]{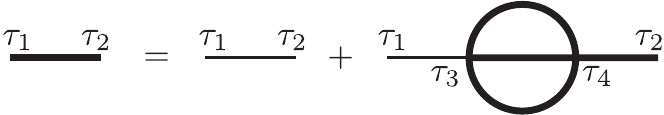}
  \end{center}
  \caption{Schwinger-Dyson equation for the full two-point function $\textbf{G}(\tau_{12})$. }
  \label{fig:SDeq}
\end{figure}

At first we list the Feynman rules which follow from the Hamiltonian (\ref{twoflavorHamilt}).  In figures (\ref{fig:props}) and (\ref{fig:vert}) 
we define propagators and interaction vertices for the given two-flavour tensor model.
\noindent 
Since the interaction  terms have a tetrahedral tensor structure the melonic Feynman diagrams 
dominate in the large $N$ limit.  Let us define bare two-point functions
\begin{align}
G_{11}(\tau_{12})= \frac{1}{N^{3}}\langle T \psi^{I}_{1}(\tau_{1}) \psi^{I}_{1}(\tau_{2})\rangle_{0}, \quad 
G_{22}(\tau_{12})= \frac{1}{N^{3}}\langle T \psi^{I}_{2}(\tau_{1}) \psi^{I}_{2}(\tau_{2})\rangle_{0}\,,
\end{align}
where the sum over indices $I$ is assumed.  
The leading melonic correction to the full two-point function $\textbf{G}_{11}$ is represented in figure \ref{fig:firstcor}.  
Using that 
\begin{align}
J_{IJKL}J_{IJKL} = g^{2}(6N^{6}-18 N^{4}+12 N^{3})
\end{align}
we find 
\begin{align}
\textbf{G}_{11}(\tau_{12}) = G_{11}(\tau_{12})+g^{2}N^{3}\int d\tau_{3} d\tau_{4}G_{11}(\tau_{13})\big(G_{11} (\tau_{34})^{3}+3\alpha^{2}
G_{11} (\tau_{34})G_{22} (\tau_{34})^{2}\big)G_{11} (\tau_{42})+\dots\,.
\end{align}
A similar expression can be derived for $\textbf{G}_{22}$. Since there is a symmetry $\psi_{1}\leftrightarrow \psi_{2}$ we can assume that 
$\textbf{G}_{11}=\textbf{G}_{22}=\textbf{G}$ and obtain a Schwinger-Dyson equation for the full two-point function (see figure \ref{fig:SDeq})
\begin{align}
\textbf{G}(\tau_{12})= G(\tau_{12}) + (1+3\alpha^2)g^2N^3\int d\tau_{3} d\tau_{4}  G(\tau_{13})\textbf{G}(\tau_{34})^3 \textbf{G}(\tau_{42})\,, \label{SDeq1}
\end{align}
where $G(\tau)=\frac{1}{2}\textrm{sgn}(\tau)$ is the bare propagator. 

In writing this Schwinger-Dyson equation we implicitly made an important assumption that  the two-point functions
\begin{align}
\textbf{G}_{12}(\tau_{12})= \frac{1}{N^{3}}\langle T \psi^{I}_{1}(\tau_{1}) \psi^{I}_{2}(\tau_{2})\rangle, \quad 
\textbf{G}_{21}(\tau_{12})= \frac{1}{N^{3}}\langle T \psi^{I}_{2}(\tau_{1}) \psi^{I}_{1}(\tau_{2})\rangle\,,
\end{align}
are zero $\textbf{G}_{12}(\tau)=\textbf{G}_{21}(\tau)=0$. This follows from the $\mathbb{Z}_{2}$ symmetry $\psi_{2}\to -\psi_{2}$. As we will see
below, the  $\mathbb{Z}_{2}$ symmetry can be spontaneously broken for some range of parameter $\alpha$
 and dimensionless coupling $\beta J$, where $\beta =1/T$ is the inverse temperature and $J^{2}=g^{2}N^{3}$ is effective coupling constant.

Let us first assume that $\mathbb{Z}_{2}$ symmetry is not broken and analyze the SD equation (\ref{SDeq1}).
At large coupling constant $\beta J$ and intermediate distances $1/J\ll \tau \ll \beta$ the solution to  this equation  is given by

\begin{equation}
\textbf{G}(\tau) = \left(\frac{1}{4\pi(1+3\alpha^2)}\right)^{\frac{1}{4}}\frac{\textrm{sgn}(\tau)}{|J \tau|^{1/2}}\,.
\end{equation}

\subsection{Scaling dimensions of bilinear operators}
We can use the large $N$ Schwinger-Dyson equations for the three-point functions to deduce the scaling dimensions of four families of bilinear operators:
\begin{equation}
\label{bilinears}
\begin{split}
& O^{2n+1}_{1} = \psi_1^{}\partial_{\tau}^{2n+1}\psi_1^{} + \psi_2^{} \partial_{\tau}^{2n+1}\psi_2^{}\ , \qquad
O^{2n+1}_{2} = \psi_1^{} \partial_{\tau}^{2n+1}\psi_1^{} - \psi_2^{} \partial_{\tau}^{2n+1}\psi_2^{}\ , \\
&O_3^{2n+1} = \psi_1 \partial_{\tau}^{2n+1} \psi_2 + \psi_2 \partial_{\tau}^{2n+1} \psi_1 \ , \qquad 
O_4^{2n} = \psi_1^{} \partial_{\tau}^{2n} \psi_2^{} - \psi_2^{} \partial_{\tau}^{2n} \psi_1^{}\,,
\end{split}
\end{equation}
where $n=0,1, 2, \ldots,$ and the sum over tensor indices is assumed.\footnote{In the coupled SYK model (\ref{SYK2fl}) the same expressions for bilinear operators are applicable after
replacement of $\psi_A^I$ by $\chi_A^i$, with $A=1,2$.} 
Each of these operators is invariant under the $O(N)^3$ symmetry, but they are distinguished by their transformations to discrete symmetry.

\begin{figure}[h!]
  \begin{center}  
    \includegraphics [width=0.6\textwidth, angle=0.]{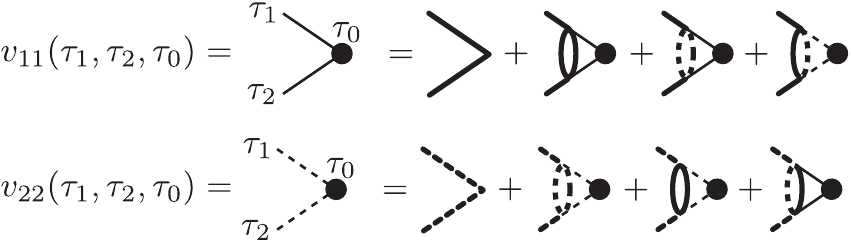}
  \end{center}
  \caption{The Bethe-Salpeter equations for the three-point functions $v_{11}$ and $v_{22}$. }
  \label{fig:scaldim1}
\end{figure}

We take some operator $O(\tau)$ and consider two three-point functions of the form
\begin{align}
v_{11}(\tau_{1},\tau_{2},\tau_{0}) = \langle \psi^{I}_{1}(\tau_{1})\psi^{I}_{1}(\tau_{2})O(\tau_{0})\rangle, \quad 
v_{22}(\tau_{1},\tau_{2},\tau_{0}) = \langle \psi^{I}_{2}(\tau_{1})\psi^{I}_{2}(\tau_{2})O(\tau_{0})\rangle\,, \label{v1122}
\end{align}
where we assume summation over the index $I$. In the large $N$ limit  the functions (\ref{v1122}) obey the melonic Bethe-Salpeter equations. They are schematically represented in figure \ref{fig:scaldim1}. 
In the conformal limit one can ignore the first diagram on the right  and obtain 
 \begin{align}
  \left(\begin{array}{c}
    v_{11} \\ 
    v_{22} \\ 
  \end{array}\right) =  \left( \begin{array}{cc}
    K_{11,11} & K_{11,22} \\ 
    K_{22,11} & K_{22,22} \\ 
  \end{array}\right)* \left(\begin{array}{c}
    v_{11} \\ 
    v_{22} \\ 
  \end{array}\right) \,, \label{BSeq1}
\end{align}
where assuming that $\textbf{G}_{11}=\textbf{G}_{22}=\textbf{G}$ we find 
\begin{align}
K_{11,11}=K_{22,22} = \frac{1+\alpha^{2}}{1+3\alpha^{2}}K_{c}, \quad  K_{11,22}=K_{22,11} =\frac{2\alpha^{2}}{1+3\alpha^{2}} K_{c}\,,
\end{align}
and $K_{c}$ is the kernel of the SYK model, defined in conformal limit as 
\begin{align}
K_{c}(\tau_{1},\tau_{2};\tau_{3},\tau_{4}) = -\frac{3}{4\pi} \frac{\textrm{sgn}(\tau_{13})\textrm{sgn}(\tau_{24})}{|\tau_{13}|^{2\Delta}|\tau_{24}|^{2\Delta}|\tau_{34}|^{2-4\Delta}}, \quad \Delta =\frac{1}{4}\,.
\end{align}
An arbitrary conformal three-point function of the form (\ref{v1122}) with an operator of scaling dimension $h$ has the form 
\begin{align}
v_{h}(\tau_{1},\tau_{2},\tau_{0}) = \frac{c\,\textrm{sgn}(\tau_{12})}{|\tau_{01}|^{h}|\tau_{02}|^{h}|\tau_{12}|^{2\Delta-h}}\,,
\end{align}
and obviously must be antisymmetric under $\tau_{1}\leftrightarrow \tau_{2}$. 
This three-point function is an eigenvector of the kernel $K_{c}$ with the eigenvalue $g(h)$:\footnote{To take the integrals one should use star-triangle identities twice \cite{Kitaev:2017awl}.}
\begin{align}
g(h)\int d\tau_{3}d\tau_{4}K_{c}(\tau_{1},\tau_{2};\tau_{3},\tau_{4})   v_{h}(\tau_{3},\tau_{4},\tau_{0}) =  v_{h}(\tau_{1},\tau_{2},\tau_{0}) \,.
\end{align}
To solve (\ref{BSeq1}) one has to find eigenvalues of  the matrix and equate them to unity. This gives an equation for possible scaling dimensions.  It easy to see that  this matrix acquires diagonal form in the basis of vectors $v_{11}+v_{22}$ and $v_{11}-v_{22}$ and we find two equations for the scaling dimensions 
\begin{align}
g_{A}(h)=1, \quad \frac{1-\alpha^{2}}{1+3\alpha^{2}}g_{A}(h) =1 , \quad g_{A}(h)= -\frac{3}{2}\frac{\tan (\frac{\pi}{2}(h-\frac{1}{2}))}{h-1/2} \,.
\label{opdimeq1}
\end{align}
The scaling dimensions of the operator  $O^{2n+1}_{1} = \psi_1^{}\partial_{\tau}^{2n+1}\psi_1^{} + \psi_2^{} \partial_{\tau}^{2n+1}\psi_2^{}$
satisfy $g_{A}(h)=1$ and are independent of $\alpha$.  They are given by the well-known series  $h= 2.00, 3.77, 5.68, 7.63, 9.60, \dots$ which approaches $2n+\frac{3}{2}$. 
These are the same scaling dimensions as in the basic $O(N)^3$ tensor model \cite{Klebanov:2016xxf} and the SYK model.
On the other hand, the scaling dimensions of operators  $O^{2n+1}_{2} = \psi_1^{} \partial_{\tau}^{2n+1}\psi_1^{} - \psi_2^{} \partial_{\tau}^{2n+1}\psi_2^{}$ are given by $ \frac{1-\alpha^{2}}{1+3\alpha^{2}}g_{A}(h) =1$ and depend on $\alpha$. 
As a check we note that for $\alpha = 0$ the spectra of $O_{2}$ and $O_{1}$ are the same; this is as expected since the two flavors are decoupled.

\begin{figure}[h!]
  \begin{center}  
    \includegraphics [width=0.69\textwidth, angle=0.]{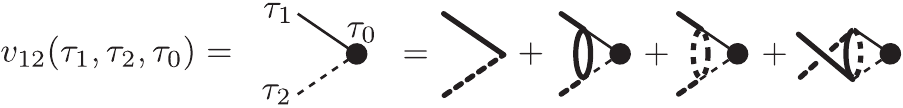}
  \end{center}
  \caption{The Bethe-Salpeter equations for the three-point functions $v_{12}$. }
  \label{fig:scaldim2}
\end{figure}

Now consider the last possible  three-point function
\begin{align}
v_{12}(\tau_{1},\tau_{2},\tau_{0}) = \langle \psi^{I}_{1}(\tau_{1})\psi^{I}_{2}(\tau_{2})O(\tau_{0})\rangle \,.\label{v1221}
\end{align}
The melonic  Bethe-Salpeter equation for this three-point function is represented in figure \ref{fig:scaldim2} 
and in the conformal limit, neglecting the first diagram on the right we get 
\begin{align}
v_{12}(\tau_{1},\tau_{2},\tau_{0}) =  \int d\tau_{3}d\tau_{4} \frac{2}{1+3\alpha^{2}}(\alpha K_{c}(\tau_{1},\tau_{2};\tau_{3},\tau_{4}) -\alpha^{2}K_{c}(\tau_{1},\tau_{2};\tau_{4},\tau_{3}))v_{12}(\tau_{3},\tau_{4},\tau_{0})\,.
\end{align}
In this case there are two general possibilities for conformal three-point function, namely  anti-symmetric and symmetric cases
\begin{align}
v^{A}_{h}(\tau_{1},\tau_{2},\tau_{0}) = \frac{c\,\textrm{sgn}(\tau_{12})}{|\tau_{01}|^{h}|\tau_{02}|^{h}|\tau_{12}|^{2\Delta-h}}\,, \quad v^{S}_{h}(\tau_{1},\tau_{2},\tau_{0}) = \frac{c\,\textrm{sgn}(\tau_{01})\textrm{sgn}(\tau_{02})}{|\tau_{01}|^{h}|\tau_{02}|^{h}|\tau_{12}|^{2\Delta-h}}\,.
\end{align}
Therefore, we find equations which determine spectra of antisymmetric and symmetric operators
\begin{align}
\frac{2(\alpha+\alpha^{2})}{1+3\alpha^{2}}g_{A}(h)=1, \quad  \frac{6(\alpha-\alpha^{2})}{1+3\alpha^{2}}g_{S}(h)=1, \quad g_{S}(h) =-\frac{1}{2}\frac{\tan (\frac{\pi}{2}(h+\frac{1}{2}))}{h-1/2}\,.
\label{opdimeq2}
\end{align}
The scaling dimensions of operators $O_3^{2n+1}$ satisfy the first equation above and $O_4^{2n}$ the second.
We can check this result by comparing with the results for the complex bipartite fermion model (\ref{bipartiteH}). It was found \cite{Klebanov:2018fzb} that 
the scaling dimensons of $O_4^{2n}$ are determined by
\begin{equation}
g_{\textrm{sym}}(h) = \frac{3}{2}\frac{\tan(\frac{\pi}{2}(h+\frac{1}{2}))}{h-1/2} =1\ , 
\end{equation}
and indeed for $\alpha=-1$ we get $\frac{6(\alpha-\alpha^{2})}{1+3\alpha^{2}}g_S(h) = g_{\textrm{sym}}(h)$.  

To summarize, we 
have found that scaling dimensions of the operators (\ref{bilinears}) 
can be obtained by solving equations $g_{i}(h)=1$, where 
\begin{equation}
\label{gforoper}
(g_{1}(h), g_{2}(h), g_{3}(h),g_{4}(h)) = \left(g_{A}(h),\frac{1-\alpha^{2}}{1+3\alpha^{2}}g_{A}(h), \frac{2\alpha(1+\alpha)}{1+3\alpha^{2}}g_{A}(h), \frac{6\alpha(1-\alpha)}{1+3\alpha^{2}}g_{S}(h)\right)\,.
\end{equation}
The duality relation (\ref{duality1}) is reflected in the behavior of functions $g_{i}(h)$, which define scaling dimensions of the operators $O_{i}$. Using  (\ref{gforoper}) and (\ref{duality1}) one finds
\begin{align}
\left(g_{1}(h),g_{2}(h),g_{3}(h),g_{4}(h)\right) \to \left(g_{1}(h),g_{3}(h),g_{2}(h),g_{4}(h)\right) \,.
\end{align}
Indeed, under  $\psi \to \tilde\psi$  the operators $O_{i}$ transform as $
\left(O_{1},O_{2},O_{3},O_{4}\right) \to \left(O_{1},O_{3},O_{2},O_{4}\right) $.

\subsection{Complex scaling dimensions}
\label{complexscaling}

In this section, we examine if there exist any complex solutions of the equations $g_{i}(h) = 1$ defined in (\ref{gforoper}). If such a complex root exists, then a conformal primary has a complex scaling dimension, which leads to a destabilization of the model. Indeed, a complex scaling dimension of the form $\frac{1}{2}\pm i f$ corresponds to a scalar fields in \(AdS_2\) whose \(m^2\) is below the Breitenlohner-Freedman bound \(m_{BF}^2 = -\frac{1}{4}\). Since $\Delta = \frac{1}{2} \pm \sqrt{\frac{1}{4}+m^2}$ \cite{Maldacena:1997re, Gubser:1998bc, Witten:1998qj}, 
\begin{equation}
m^2 = -\frac{1}{4} - f^2< m_{BF}^2\ .
\end{equation}
In such a case one may expect ``tachyon condensation" in AdS space. In the dual CFT the operator dual to the tachyon acquires an expectation value, leading to symmetry breaking. We will obtain some support to this picture.  

First of all we notice that the functions $g_{A}(h)$ and $g_{S}(h)$ are real only if $h$ is real or $h=\frac{1}{2}+if$ for real $f$. Next it is easy to check that 
\begin{equation}
-\frac{3\pi}{4} \leq g_{A}(1/2+i f) < 0 , \quad -\infty \leq g_{S}(1/2+i f) < 0 \,.
\end{equation}
Using the fact that $-\frac{1}{3}\leq \frac{1-\alpha^{2}}{1+3\alpha^{2}} \leq 1$ (and the same  for $\frac{2\alpha(1+\alpha)}{1+3\alpha^{2}}$ due to duality) we conclude that equations $g_{i}(1/2+i f)=1$ for $i=1,2,3$ do not have solutions, thus scaling dimensions of the operators $O_{1}$, $O_{2}$ and $O_{3}$
are always real.

\begin{figure}[h!]
  \begin{center}
    \includegraphics [width=0.4\textwidth, angle=0.]{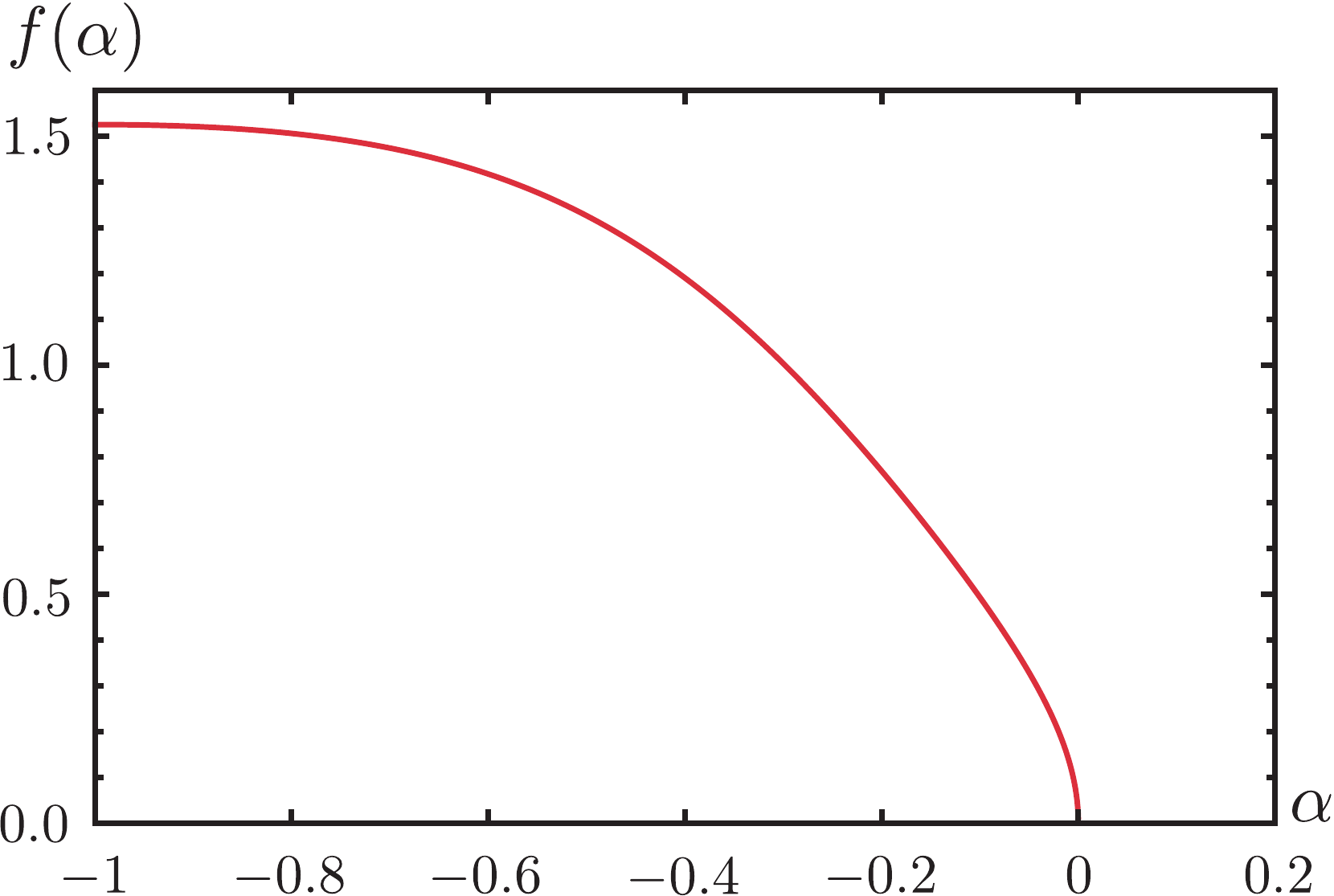}
  \end{center}
  \caption{Imaginary part of the scaling dimension of the fermion number operator $Q$. At $\alpha=-1$ it reaches its maximum value $\approx 1.5251$.}
  \label{fgraph}
\end{figure}

On the other hand, since $\frac{6\alpha(1-\alpha)}{1+3\alpha^{2}}<0$ for negative $\alpha$, the equation $g_{4}(h)=1$ has solutions $h=1/2\pm if(\alpha)$, where $f(\alpha)$ can be found from the equation 
\begin{equation}
f \tanh (\pi  f/2)  = -\frac{3\alpha(1-\alpha)}{1+3\alpha^2}\,.
\end{equation}
The plot of $f(\alpha)$ is shown in figure \ref{fgraph}. For slightly negative $\alpha$ we find 
\begin{equation}
f (\alpha) = \sqrt {\frac{-6 \alpha}{\pi}} \left (1+ O(\alpha) \right )
\ ,
\end{equation}
while $f(-1) \approx 1.5251$ in agreement with the result for the bipartite model found in \cite{Klebanov:2018fzb}.
Thus, for $-1 \leq \alpha<0$ there is an operator with the complex scaling dimension $h=1/2+ i f(\alpha)$ or its complex conjugate: the fermion number operator $Q= iO_4^0$.
This makes the conformal large $N$ limit unstable.

For $0 \leq \alpha\leq 1/3$ there are no complex solutions of $g_4(h)=1$. The two lowest positive real solutions, $h_\pm (\alpha)$, satisfy $h_+ + h_- =1$. These two roots are the scaling dimensions of operator $Q$ in two different large $N$ CFTs \cite{Klebanov:1999tb}, as we explain below.
We find
\begin{equation}
h_+ (\alpha)= 1/2+ \sqrt {\frac{6 \alpha}{\pi}}+ \ldots
\end{equation}
when $\alpha$ is small and positive, and $h_+ (1/3)=1$. 
The fact that $\alpha=0$ is the lower edge of the conformal window is related to the behavior of the scaling dimension of the ``double-trace" operator $Q^2$. 
In the large $N$ limit, $\Delta_{Q^2}= 2 \Delta_Q$. In one of the CFTs, $\Delta_Q=h_+$, so that $\Delta_{Q^2}> 1$. Since the operator $Q^2$ is irrelevant, this CFT is stable.
There is RG
flow leading to it, which originates from another large $N$ fixed point where $Q^2$ is relevant \cite{Witten:2001ua,Gubser:2002vv}.
At this UV fixed point, $\Delta_Q= h_-$. 
When $\alpha=0$ the two fixed points merge and annihilate, 
as in various other theories, for example \cite{Dymarsky:2005uh,Pomoni:2008de,Kaplan:2009kr,Giombi:2015haa}. 
For $-1\leq \alpha<0$ there are two different theories containing complex dimension $\Delta_Q=1/2+ i f(\alpha)$ or its complex conjugate. They may be formally regarded as ``complex CFTs" \cite{Gorbenko:2018ncu}, but we will see in the next section 
that their true physics includes symmetry breaking, which leads to a gap in the energy spectrum.


\section{Symmetry Breaking }
\label{symbreak}

In section \ref{complexscaling}, we showed that for the coupled tensor model (\ref{twoflavorHamilt}) in the range $-1\leq \alpha<0$ 
the fermion number operator $Q=i \psi_1^{I}\psi_2^{I}$ has a complex scaling dimension, signaling an instability of the conformal phase of the model. 
In this section we show that this operator acquires a vacuum expectation value (VEV) in the true low-temperature phase of the large $N$ model. 
Based on this, it is tempting to make the following conjecture. 

\begin{con}
If the assumption of conformal invariance in a large $N$ theory leads to a single-trace operator with a complex scaling dimension of the form $d/2+ i f$, 
then in the true low-temperature phase this operator 
acquires a VEV.
\end{con}

In our case, the $O(N)^3$ symmetry implies that 
\begin{equation}
\label{bilinVEV}
 \langle i \psi_1^I\psi_2^J\rangle= \delta^{IJ} A , 
\end{equation}
where we used the short-handed notation $I=abc$, and $A$ is of order $1$ in the large $N$ limit. 

This leads to an exponential decay of correlation functions and signifies a gap in the energy spectrum. Furthermore, the VEV (\ref{bilinVEV}) implies that various discrete symmetries, including the particle-hole symmetry (\ref{phsymtensor}), the interchange symmetry between $\psi_1, \psi_2$, and the reflection symmetry $\psi_2\to -\psi_2$, are spontaneously broken. Therefore, one should expect a second-order phase transition between the broken and unbroken symmetry phases. In addition, the spontaneously broken symmetry also implies a ground state degeneracy in the large $N$ energy spectrum. \footnote{Due to a technicality we only expect a two-fold degeneracy although multiple $\mathbb{Z}_2$ have been broken. We will comment on this issue below.} 

In this section we extensively analyze the phenomenon of symmetry breaking, sometimes using the SYK counterpart (\ref{SYK2fl}) of the $O(N)^3$ 
tensor model (\ref{twoflavorHamilt}). 
The two models have many similarities at large $N$: they share the same Schwinger-Dyson equations, and the spectra of bilinear operators.
The SYK formulation, however, is advantageous for the purpose of exact numerical diagonalizations: we can study cases where the integer $N_{\rm SYK}$ is not the cube of an integer.

Let us first demonstrate the connection between the tensor model and the SYK counterpart.
For the one-flavor $O(N)^3$ tensor model the analogous SYK model has the random tensor $J_{ijkl}$ which is fully antisymmetric. The mixed term
$A_{ijkl} \chi_1^i \chi_1^j \chi_2^k \chi_2^l$ has only the symmetries
\begin{equation}
\label{reducedsym}
 A_{ijkl}=- A_{jikl}=- A_{ijlk} = A_{klij}\ ,
\end{equation}
which are the same as for the Riemann tensor.
However, the full interaction term following from (\ref{twoflavorHamilt}) is
\begin{equation}
A_{ijkl} ( \chi_1^i \chi_1^j \chi_2^k \chi_2^l +  \chi_1^i \chi_2^j \chi_1^k \chi_2^l  + \chi_1^i \chi_2^j \chi_2^k \chi_1^l )= 
(A_{ijkl}+ A_{iljk}+  A_{iklj}) \chi_1^i \chi_1^j \chi_2^k \chi_2^l\ .
\end{equation}
Since $A_{ijkl}+ A_{iljk}+  A_{iklj}$ is fully antisymmetric due to (\ref{reducedsym}), 
the mixed term has a fully antisymmetric random coupling. 
We will assume that it is proportional to the coupling in the diagonal term of (\ref{twoflavorHamilt}), and
are thus led to the random model (\ref{SYK2fl}). This model is the special $M=2$ case of a periodic SYK chain model
\begin{equation}
\label{SYK2chain}
H_{\rm chain} = \frac{1}{4!} J_{ijkl}\sum_{x=1}^M \left (\chi_x^i \chi_x^j \chi_x^k \chi_x^l + 3 \alpha \chi_x^i \chi_x^j \chi_{x+1}^k \chi_{x+1}^l \right )\ ,
\end{equation}
where the integer $x$ labels the lattice site, and $\chi^i_{M+1}\equiv \chi^i_1$. This can be obtained from the model of \cite{Gu:2016oyy} by identifying the separate
random couplings up to a factor of $\alpha$.

Introducing the complex combination
$\psi^j = \frac{1}{\sqrt{2}}(\chi_1^j + i\chi_2^j)$, we may write the Hamiltonian (\ref{SYK2fl}) as 
\begin{equation}
H=\frac{1}{4!} J_{ijkl} \bigg( \frac{1-3\alpha}{2} 
\left (\psi^i \psi^j \psi^k \psi^l + \bar \psi^i \bar \psi^j \bar \psi^k \bar \psi^l  \right)
+ 3 (1+ \alpha)  \bar \psi^i \bar \psi^j \psi^k \psi^l \bigg )
\ .
\end{equation}
As usual, we will assume that each variable $J_{ijkl}$ has a gaussian distribution with variance 
$6 J^2 N_{\rm SYK}^{-3}$. We will typically state energies in units of $J$, or equivalently set $J=1$.

The duality symmetry described in section \ref{duality} applies to the coupled SYK model (\ref{SYK2fl}), and again allows us to restrict $\alpha$ to the range from $-1$ to $1/3$.
There are two interesting limiting cases.
For $\alpha=-1$ the transformation (\ref{duality1}) maps $H\rightarrow -H$. This means that, for any random choice of $J_{ijkl}$ the energy spectrum is exactly symmetric under
$E\rightarrow -E$. This can be seen in the histograms of the spectrum shown in fig. (\ref{espectrum15}); in particular, there are many states whose energy is exactly zero. 
For $\alpha=-1$ the model is a random counterpart of the complex bipartite model: 
\begin{equation}
\label{bipartiteHamilt}
H_{\alpha=-1}=\frac{2}{4!} J_{ijkl}  \left (\psi^i \psi^j \psi^k \psi^l + \bar \psi^i \bar \psi^j \bar \psi^k \bar \psi^l  \right)\ .
\end{equation}

The fermion number operator 
\begin{equation}
\label{fermnum}
Q=i \chi_1^{j} \chi_2^{j} =\frac{1}{2} [\bar \psi^{j}, \psi^{j}]
\end{equation}
does not in general commute with $H$, but it is conserved mod $4$, just like in the Maldacena-Qi model \cite{Garcia-Garcia:2019poj}. 
For $\alpha=1/3$, however, we find the Hamiltonian
\begin{equation}
\label{newcompSYK}
H_{\alpha=\frac{1}{3}}= \frac{4}{4!} J_{ijkl} \bar \psi^i \bar \psi^j \psi^k \psi^l \ , 
\qquad [Q, H_{\alpha=\frac{1}{3}}]=0\ .
\end{equation}
Thus, we have enhanced $U(1)$ symmetry $\psi^j \rightarrow e^{i\gamma}  \psi^j$.\footnote{This model is similar to the complex SYK model \cite{Sachdev:2015efa}, 
but in (\ref{newcompSYK}) the coupling 
$J_{ijkl}$ is taken to be fully antisymmetric.} 
We note that, for $\alpha=1/3$ the scaling dimension of operator $Q=O_4^0$ is $h=1$ consistent with charge conservation. Also, here
$g_2(h)=g_3(h)$, so that the scaling dimensions of $O^{2n+1}_{2}$ and $O^{2n+1}_{3}$ are equal. This is because 
\begin{equation}
O^{2n+1}_{2} + i O^{2n+1}_{3} = 2 \psi^{j}\partial_t^{2n+1}\psi^{j} \ .
\end{equation}
Furthermore, the transformation (\ref{duality1}) maps $H_{\alpha=\frac{1}{3}}$ into itself, so the theory is selfdual.

For general $\alpha$, the model (\ref{SYK2fl}) has multiple discrete symmetries, which are discussed in more detail in the appendix. These discrete symmetries can be spontaneously broken due to a VEV of $Q$ if $Q$ is not invariant under them.  In the model (\ref{SYK2fl}), there are two symmetries that are not broken by  a VEV of $Q$: 
the anti-unitary time-reversal symmetry $K$, and a $\mathbb{Z}_4$ symmetry generated by $\frac{\pi}{2}$ rotation $R$ in $\chi_1, \chi_2.$
\begin{equation}
\label{piovertwo}
R\chi_1 R^{\dagger}=\chi_2\ , \qquad R\chi_2 R^{\dagger}=-\chi_1\ . 
\end{equation}
They both preserve $Q. $ The model (\ref{SYK2fl}) also has multiple reflection symmetries that are spontaneously broken by the VEV of $Q$, which we list in the appendix. In fact all unitary discrete symmetries of the model (\ref{SYK2fl}) form the Dihedral group of order 8, $D_4$.
In our case, any two broken symmetries that can be related by an unbroken symmetry do not produce extra ground state degeneracy, and therefore it is enough to focus on one of them. 

Let us focus on the particle-hole symmetry \cite{2009AIPC.1134...22K,Fidkowski:2009dba,Witten:2015aba,PhysRevB.95.115150,Fu:2016yrv,Cotler:2016fpe} generated by
\begin{equation}
\label{phsym}
{\cal P}= K \prod_{i=1}^{N_{\rm SYK}} (\psi^i+ \bar \psi^i )\ , \qquad {\cal P}^2 =(-1)^{N_{\rm SYK} (N_{\rm SYK}-1)/2}\ .
\end{equation}
It acts on the fermion number as
\begin{equation}
{\cal P} Q {\cal P} = -{\cal P}^2 Q\ .
\end{equation}

For $N_{\rm SYK}$ not divisible by $4$, there is a two-fold degeneracy of the ground state in section \ref{numericalspectrum}, 
due to an anomaly in the particle-hole symmetry \cite{2009AIPC.1134...22K,Fidkowski:2009dba,Witten:2015aba,PhysRevB.95.115150,Fu:2016yrv,Cotler:2016fpe}. 
For $N_{\rm SYK}$ divisible by $4$ this symmetry is not anomalous, and we find a non-degenerate ground state,
which is followed by a nearby state when $-1\leq \alpha < 0$. The two lowest states become degenerate in the large $N_{\rm SYK}$ limit, and they are separated by a gap
from the remaining states. This leads to a spontaneous symmetry breaking
through the formation of an expectation value of $Q$. We will demonstrate this effect by solving the large $N_{\rm SYK}$ Schwinger-Dyson equations for the Green functions, and
with diagonalizations at finite $N_{\rm SYK}$.

\subsection{Schwinger-Dyson equations and the effective action}
\label{SDeqsandeffact}

In this section we derive the large $N_{\rm SYK}$ effective action of $G\Sigma$ type, and the Schwinger-Dyson equations, for the coupled SYK model (\ref{SYK2fl}). 
Following \cite{Maldacena:2018lmt}, we introduce bi-local variables
\begin{equation}
G_{ab}(\tau,\tau')=\frac{1}{N_{\text{SYK}}}\langle T \chi_a^i(\tau)\chi_b^i(\tau')\rangle \ , 
\end{equation} 
and the corresponding Lagrange multipliers $\Sigma_{ab}(\tau, \tau')$, where $a,b=1,2$.
The effective action  is given by 
\begin{equation}
\begin{split}
-\frac{\beta S_{\rm eff}}{N_{\text{SYK}}} =& \log \Pf(\partial_\tau \delta_{ab} - {\Sigma_{ab}}) - \frac{1}{2} \int d\tau d\tau'\bigg(\sum_{a,b} \Sigma_{ab}(\tau,\tau') G_{ab}(\tau,\tau') -\frac{J^2}{4} \Big(\sum_{a,b} G_{ab}^4(\tau,\tau')\\&+6\alpha(G_{12}^2(\tau,\tau')+G_{21}^2(\tau,\tau'))(G_{11}^2(\tau,\tau')+G_{22}^2(\tau,\tau'))
+6\alpha^2\big(G_{11}^2(\tau,\tau')G_{22}^2(\tau,\tau')\notag\\
&+G_{12}^2(\tau,\tau') G_{21}^2(\tau,\tau')+4G_{11}(\tau,\tau')G_{22}(\tau,\tau')G_{12}(\tau,\tau')G_{21}(\tau,\tau')\big)\Big)\bigg)\,.
\end{split}
\end{equation}
By translation invariance
\begin{equation}
G_{ab}(\tau,\tau')= G_{ab}(\tau-\tau')\ , \qquad \Sigma_{ab}(\tau,\tau')= \Sigma_{ab}(\tau-\tau')\ .
\end{equation}
We also have the general properties
\begin{equation}
G_{11}(\tau)=- G_{11}(-\tau)\ , \qquad G_{22}(\tau)=- G_{22}(-\tau)\ , \qquad G_{12}(\tau)=-G_{21}(-\tau)\ .
\end{equation}
The Schwinger Dyson (SD) equations for the two point functions assume the form\footnote{These equations are also valid in the two-flavor tensor model (\ref{twoflavorHamilt}), where
$G_{ab}(\tau)=\frac{1}{N^3}\langle T \psi_a^I(\tau)\psi_b^I(0)\rangle$.
}
\begin{align}
\label{fullSD}
&\partial_\tau G_{11}(\tau) - \int d\tau' \big (\Sigma_{11}(\tau - \tau')  G_{11}(\tau') + \Sigma_{12}(\tau-\tau') G_{21}(\tau')\big ) = \delta(\tau)\ ,\nonumber \\
&\partial_\tau G_{12}(\tau)  - \int d\tau' \big (\Sigma_{11}(\tau-\tau') G_{12}(\tau') + \Sigma_{12}(\tau-\tau') G_{22}(\tau') \big ) = 0\ , \nonumber \\
& J^{-2}\Sigma_{11} = G_{11}^3  + 3\alpha G_{11}  (G_{12}^2  +G_{21}^2 ) + 3\alpha^2  G_{11} G_{22}^2   + 
6\alpha^2 G_{22}  G_{12} G_{21} \ , \nonumber \\
& J^{-2} \Sigma_{12} = G_{12}^3 + 3\alpha G_{12}  (G_{11}^2+ G_{22}^2) + 3\alpha^2 G_{12}G_{21}^2 + 6\alpha^2  G_{11} G_{22} G_{21} \ ,
\end{align}
and similarly for $1\leftrightarrow 2$. These equations and the effective action are invariant under $1\leftrightarrow 2$ and $G_{12}\to -G_{12}, G_{21}\to -G_{21}.$

\subsection{ Solutions of Schwinger-Dyson equations and symmetry breaking}
\label{SDeqsnumerics}

For $0 \leq \alpha \leq 1/3$ there are no operators with complex scaling dimensions, so it is consistent to assume that the discrete symmetries are 
unbroken and set $G_{12}=0$, and $G_{11}= G_{22}$, to obtain a nearly conformal solution in the low energy limit. 
However, the appearance of a complex scaling dimension for $-1\leq \alpha<0$ shows that such a conformal phase is unstable. We will show that, in this range of
$\alpha$ the true phase of the theory exhibits spontaneous symmetry breaking. 

In order to exhibit it, we have to allow the possibility that $G_{12}(\tau)\neq 0.$ The underlying  $\mathbb{Z}_2$ symmetry of the Hamiltonian (\ref{SYK2fl}) implies that such solutions must  come in pairs related by 
$G_{12} (\tau)\rightarrow - G_{12} (\tau)$ (in our numerical work we will typically exhibit only one of these two solutions).  
They correspond to working around the two inequivalent vacua, which we will call 
$|0_+\rangle$ and $|0_-\rangle$.
They are distinguished by the sign of the expectation value of operator $Q=i \chi_1^i \chi_2^i$:
\begin{equation}
\label{ineqvac}
\langle 0_+ | Q |0_+\rangle = A N_{\textrm{SYK}}\ ,\qquad
\langle 0_- | Q |0_-\rangle = -A N_{\textrm{SYK}}\ , \qquad  \langle 0_- | Q |0_+\rangle = 0 \ .
\end{equation}
 
The unbroken symmetry $R$ in (\ref{piovertwo}) implies 
\begin{equation}
G_{12}(-\tau)=-G_{21}(\tau)=G_{12}(\tau)\ , \qquad
G_{22}(\tau)= G_{11}(\tau)\ ,
\end{equation}
 and similarly for $\Sigma_{ab}$. Using these constraints,  
 we obtain for the effective action
 \begin{align}
-\frac{\beta S_{\textrm{eff}}}{N_{\textrm{SYK}}} =& \log \textrm{Pf}(\delta_{ab}\partial_{\tau}-\Sigma_{ab}) -\beta\int_{0}^{\beta}d\tau \Big(\Sigma_{11}G_{11}+\Sigma_{12}G_{12}\notag\\
&- \frac{J^{2}}{4}\big((1+3\alpha^{2})(G_{11}^{4}+G_{12}^{4})+12 \alpha (1-\alpha)G_{11}^{2}G_{12}^{2}\big)\Big)\, .
\end{align}
The Schwinger Dyson equations become
\begin{align}
\label{kineticSD}
&\partial_\tau G_{11}(\tau) - \int d\tau' \big (\Sigma_{11}(\tau - \tau')  G_{11}(\tau') - \Sigma_{12}(\tau-\tau') G_{12}(\tau')\big ) = \delta(\tau)\ ,\nonumber \\
&\partial_\tau G_{12}(\tau)  - \int d\tau' \big (\Sigma_{11}(\tau-\tau') G_{12}(\tau') + \Sigma_{12}(\tau-\tau') G_{11}(\tau') \big ) = 0\ , 
\end{align}
and
\begin{align}
\label{otherSD}
& J^{-2}\Sigma_{11}(\tau)  = (1+3\alpha^2) G_{11}^3(\tau) + 6\alpha (1-\alpha) G_{11}(\tau) G_{12}^2(\tau) \ , \nonumber \\
& J^{-2}  \Sigma_{12}(\tau) = (1+3\alpha^2) G_{12}^3(\tau) +6\alpha (1-\alpha ) G_{11}^2(\tau) G_{12}(\tau) \ . 
\end{align}
(\ref{kineticSD}) may be written in momentum space as 
\begin{align}
& G_{11}(\omega_{n})=\frac{-i\omega_{n} -\Sigma_{11}(\omega_{n})}{(-i\omega_{n} -\Sigma_{11})^{2}+\Sigma_{12}^{2}}, \quad  
G_{12}(\omega_{n})=  \frac{\Sigma_{12}(\omega_{n})}{(-i\omega_{n} -\Sigma_{11})^{2}+\Sigma_{12}^{2}}\ .
\end{align}

These equations, together with (\ref{otherSD}), can be solved numerically using the method of weighted iterations used in \cite{Maldacena:2016hyu}.\footnote{In this case we find it more convenient to use a slow decay rate on the weight $x$. }  To trigger the spontaneous symmetry breaking, we start our iteration   
process with a tiny non-zero $G_{12}(\tau)$ which is purely imaginary. If we are in the unbroken phase, after the iterations $G_{12}$ becomes zero; whereas 
if we are in the broken phase we find a non-zero purely imaginary solution for $G_{12}$.

\begin{figure}[h!]
    \centering
        \includegraphics[width=0.328\textwidth, angle=0.]{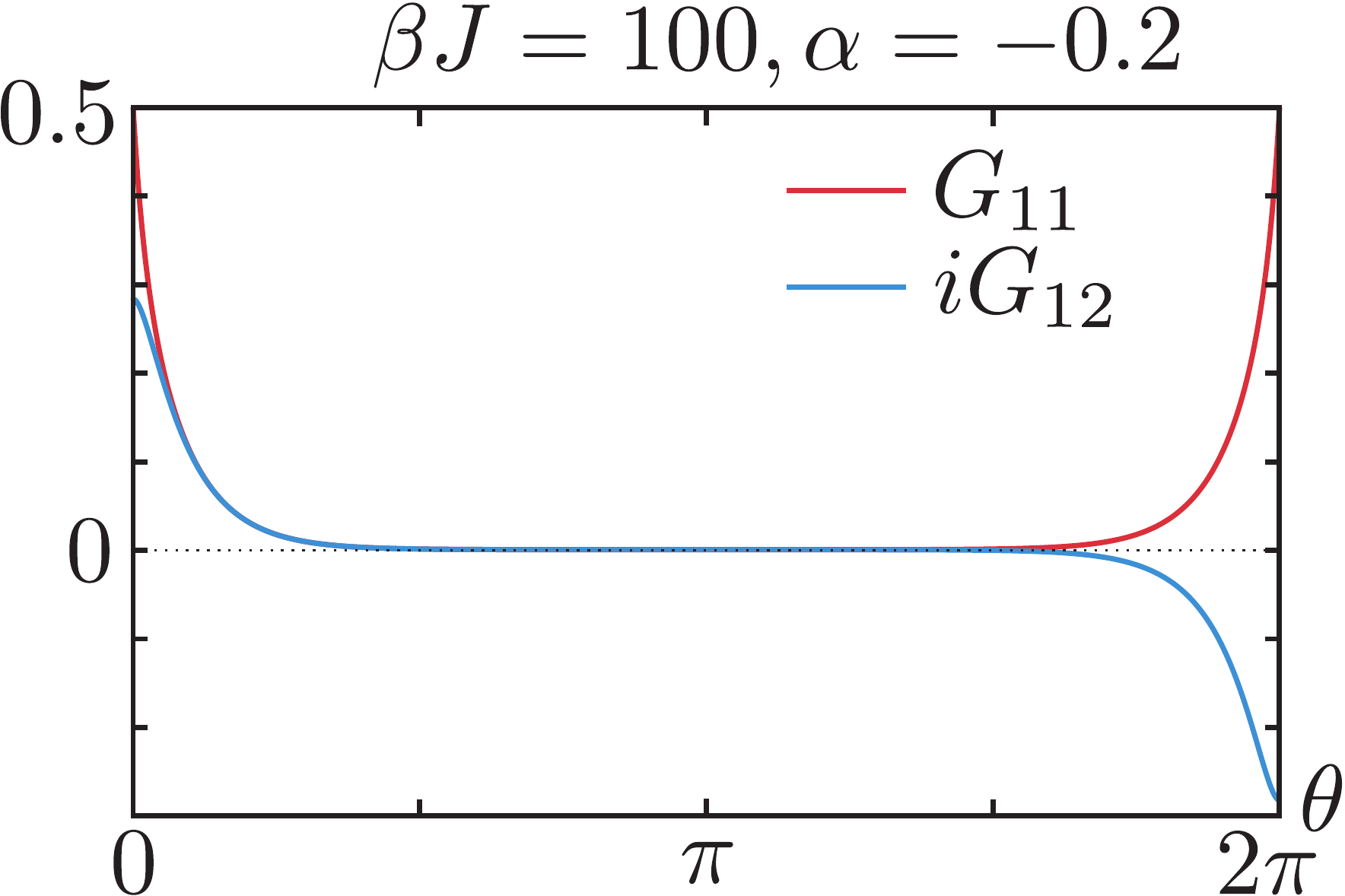}
    \includegraphics[width=0.328\textwidth, angle=0.]{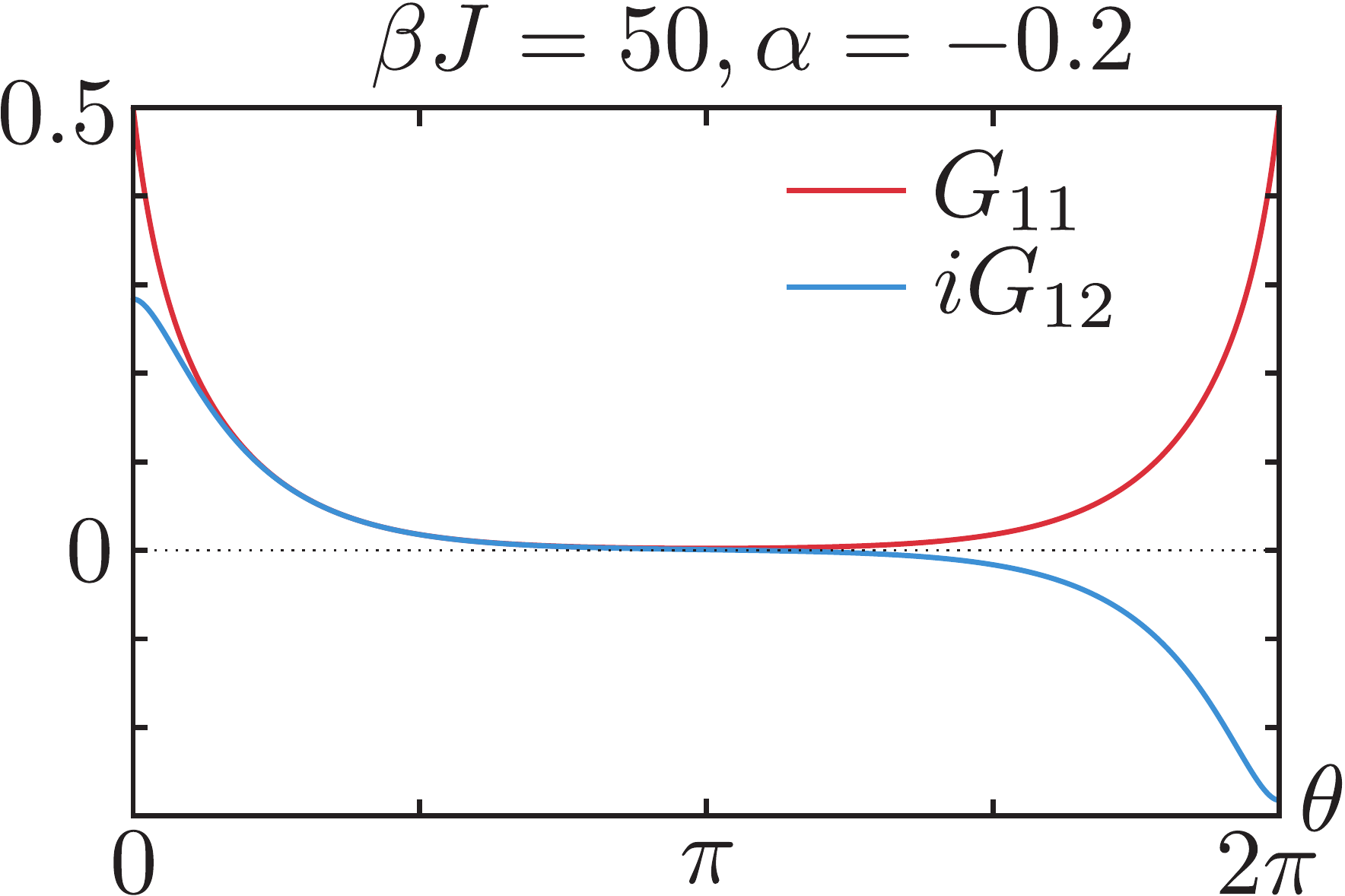}
    \includegraphics[width=0.328\textwidth, angle=0.]{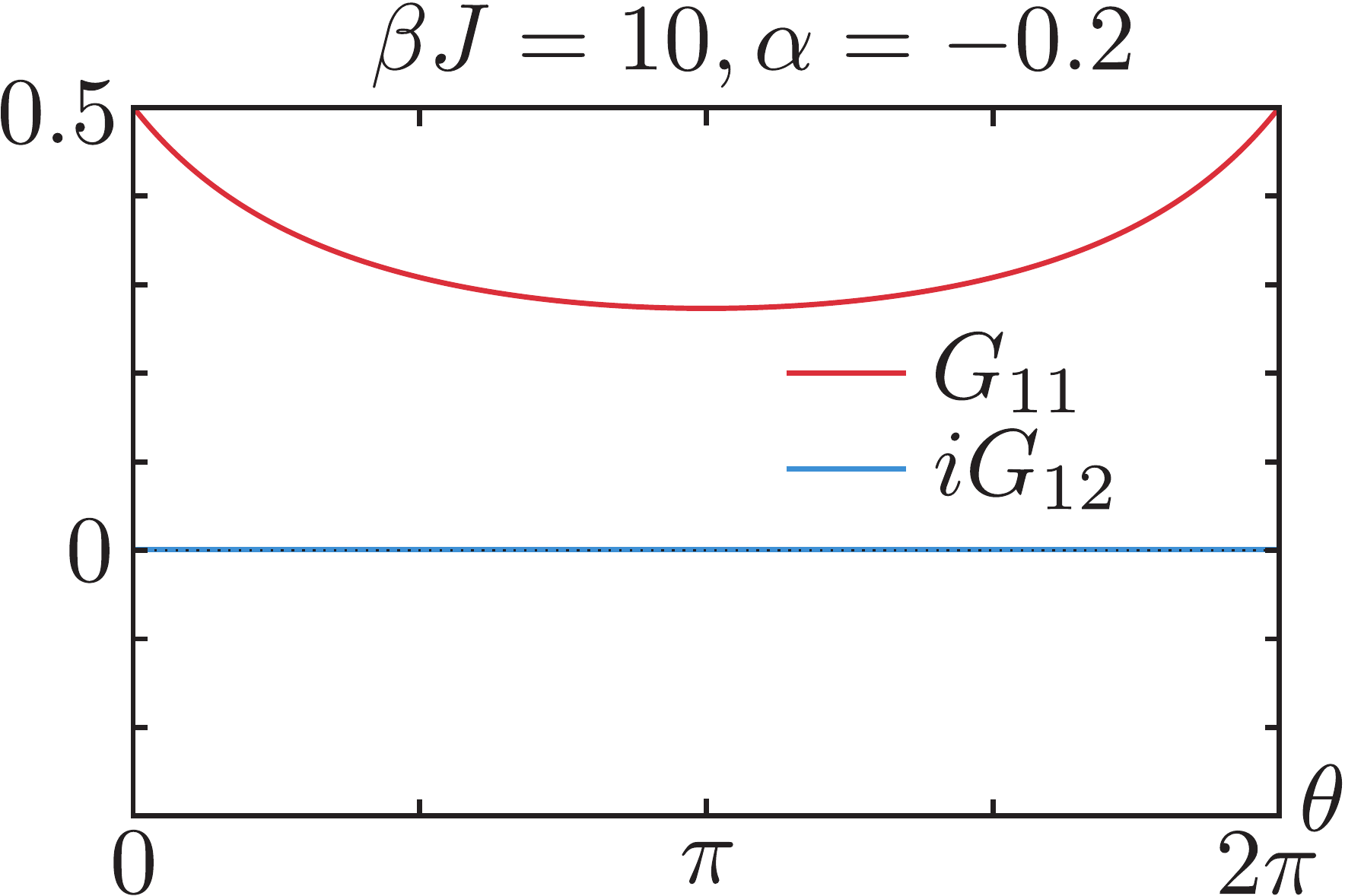}\\
         \includegraphics[width=0.328\textwidth, angle=0.]{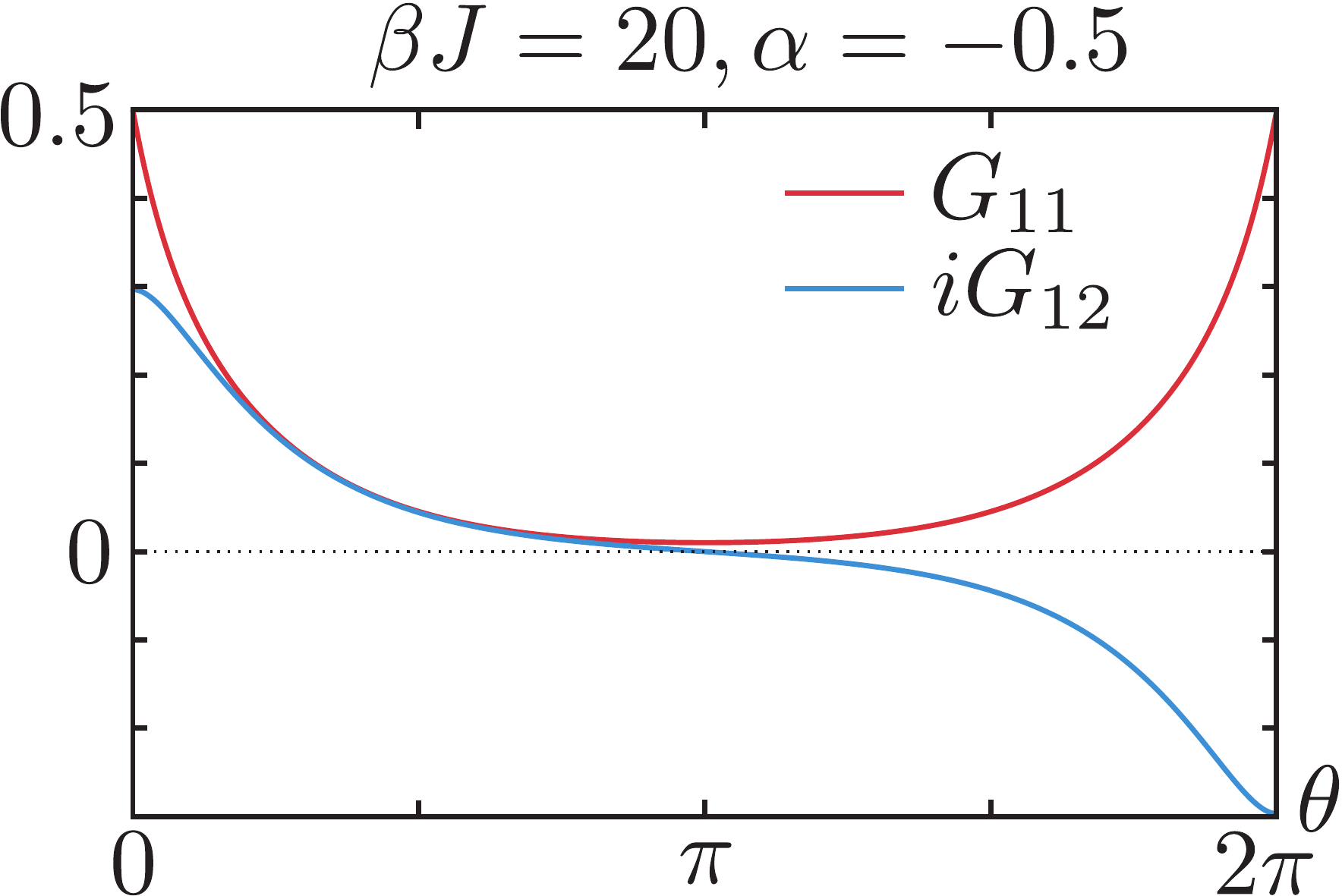}
        \includegraphics[width=0.328\textwidth, angle=0.]{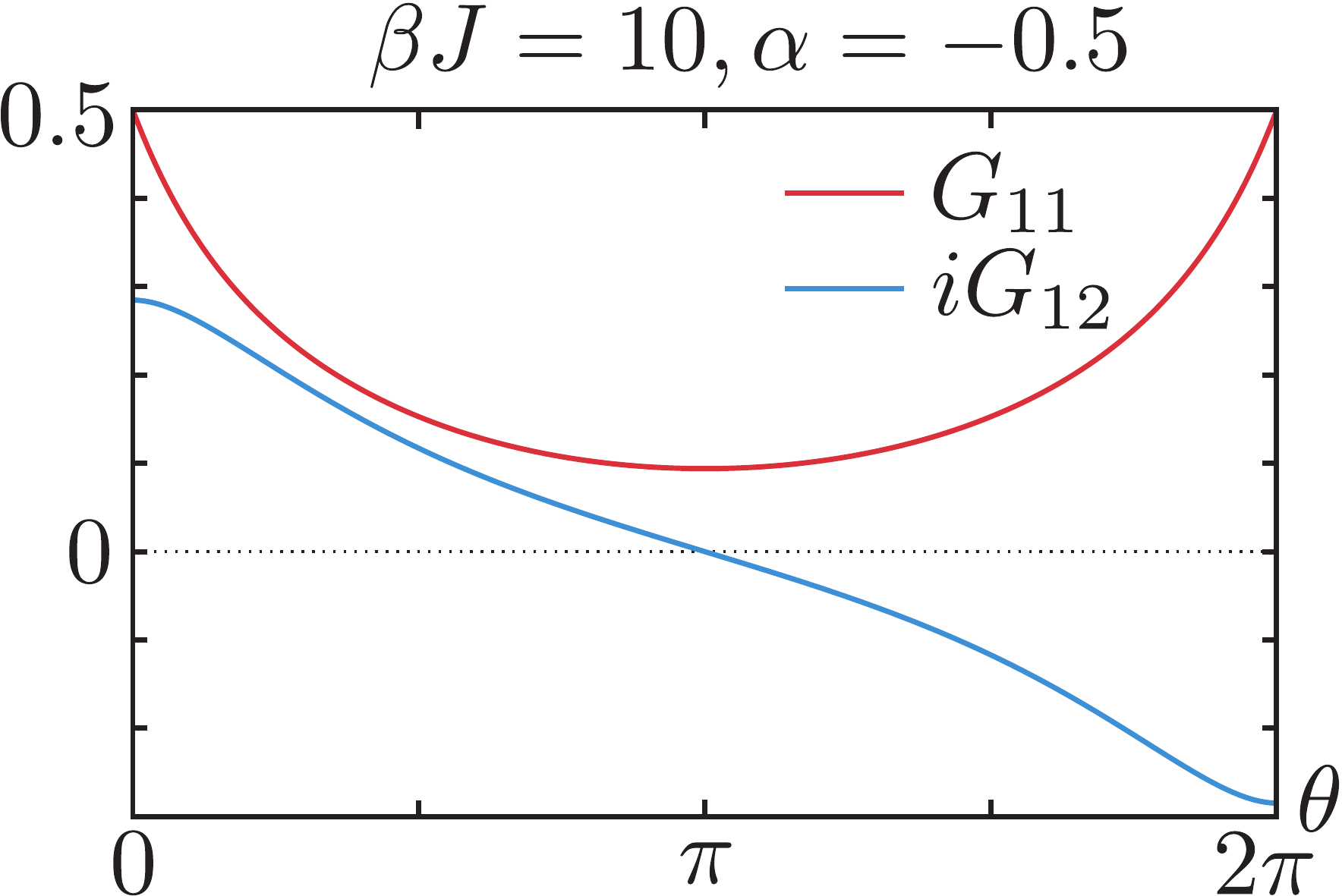}
        \includegraphics[width=0.328\textwidth, angle=0.]{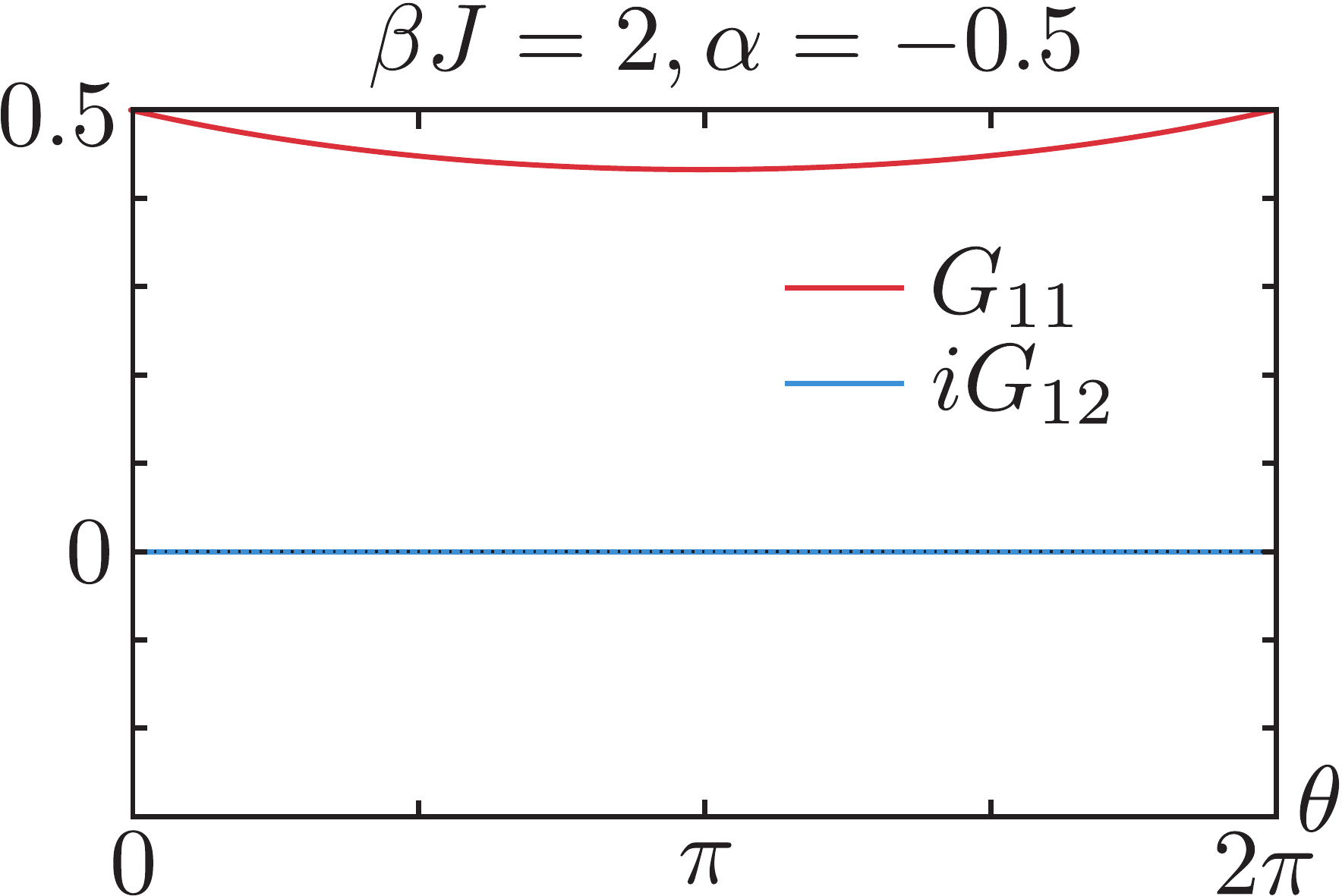}\\
    \includegraphics[width=0.328\textwidth, angle=0.]{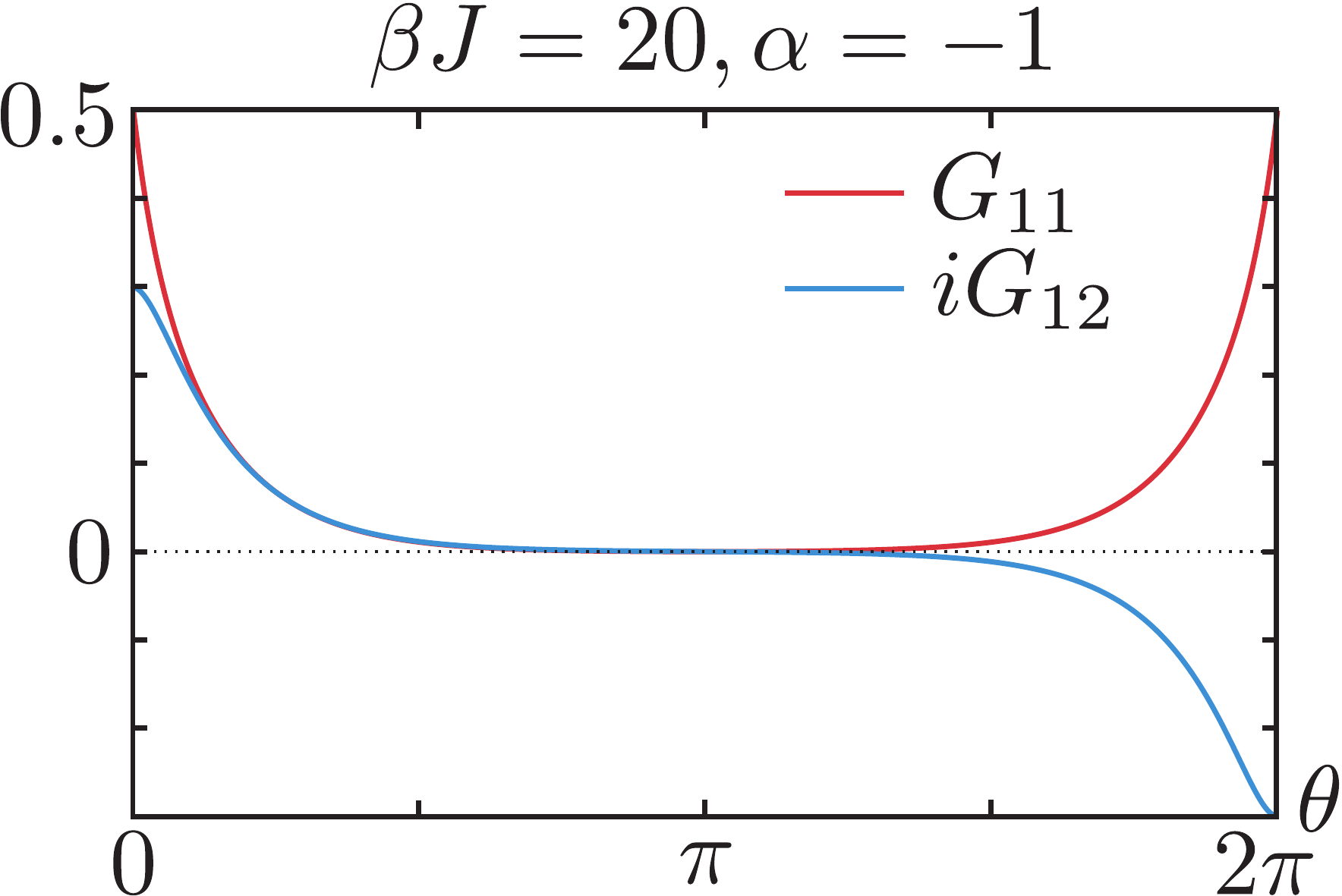}
    \includegraphics[width=0.328\textwidth, angle=0.]{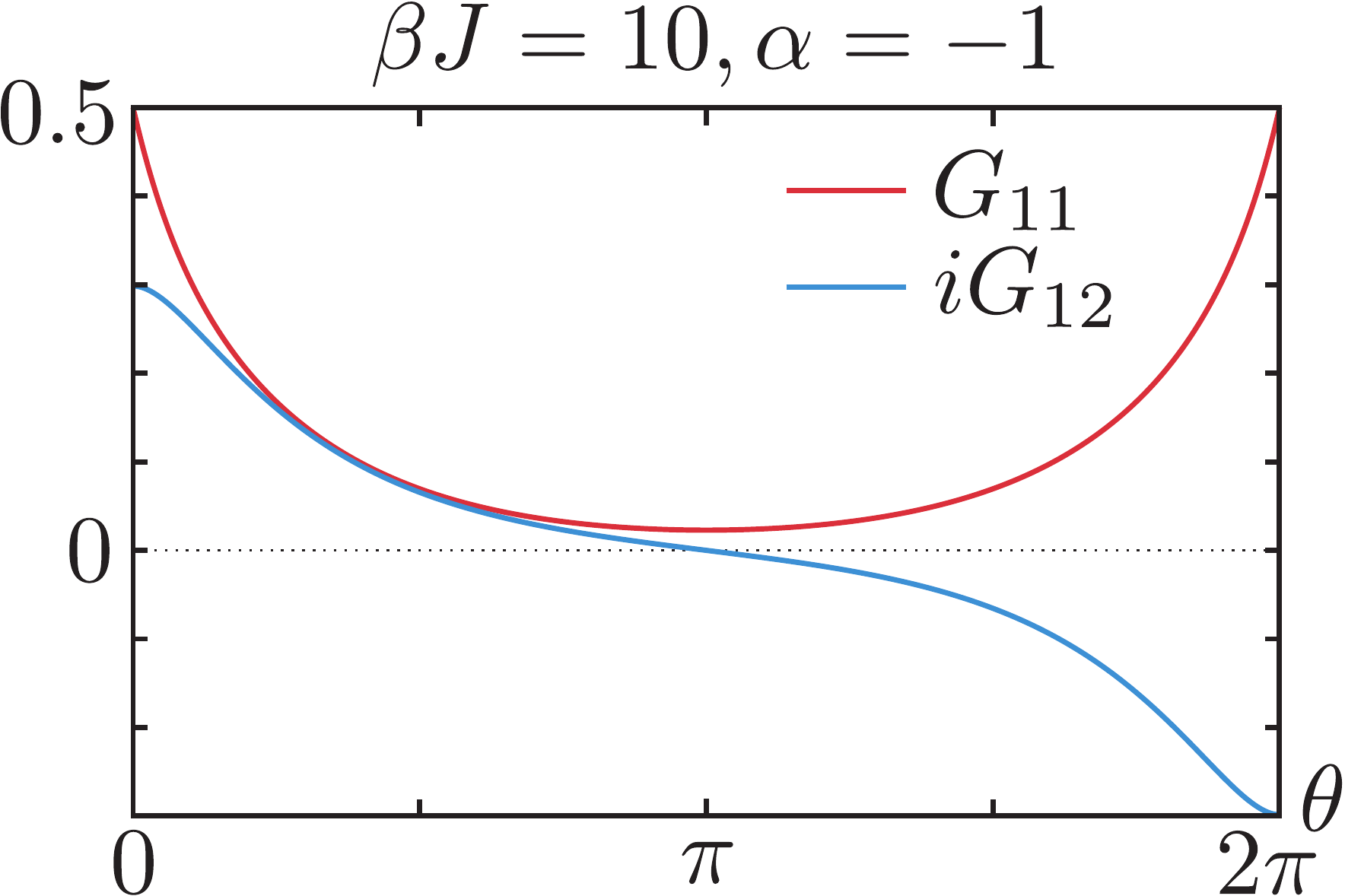}
    \includegraphics[width=0.328\textwidth, angle=0.]{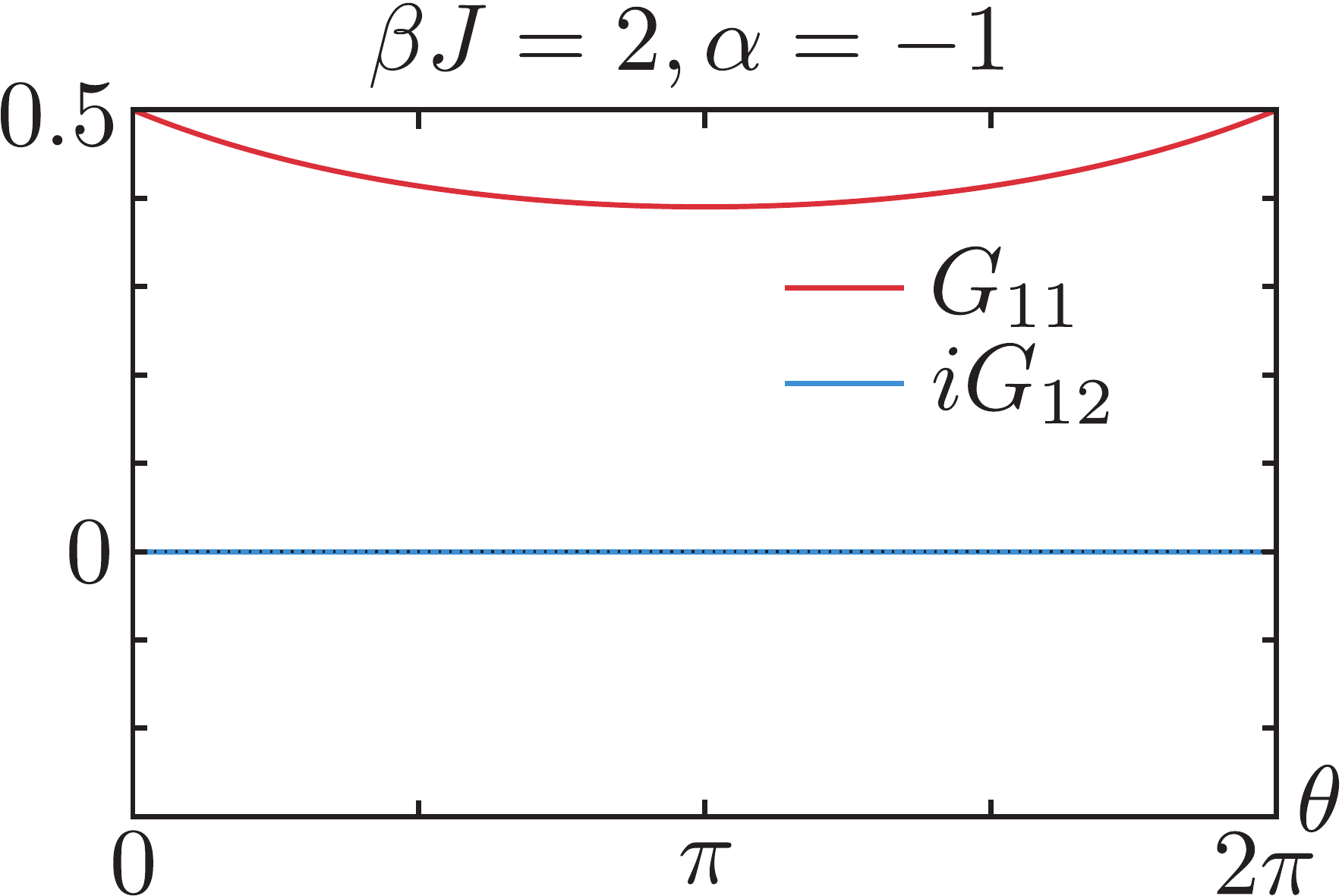}
    \caption{Numerical solutions  for $\alpha =-1, -0.5, -0.2$ and various values of $\beta J$. }
    \label{G11G12lowT}
\end{figure}

\begin{figure}[h!]
    \centering
         \includegraphics[width=0.5\textwidth, angle=0.]{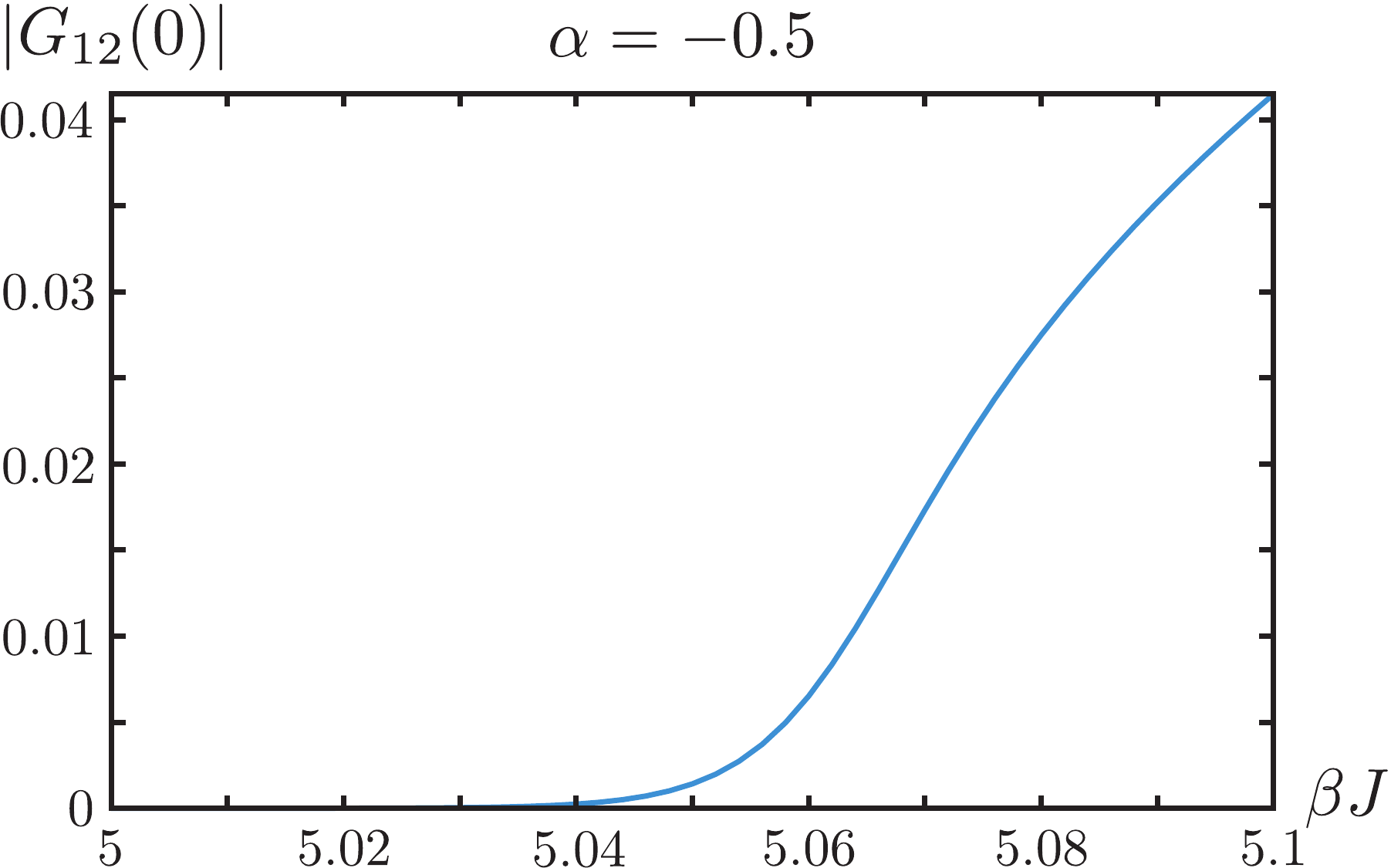}
          \caption{The expectation value of $Q/N_{\textrm{SYK}}$, i.e. $|G_{12}(0)|$, as a function of $\beta J$ for $\alpha = - 0.5$. The region near $(\beta J)_{\rm crit}$ is shown.}
    \label{G12critical}
\end{figure}

\begin{figure}[h!]
    \centering
         \includegraphics[width=0.5\textwidth, angle=0.]{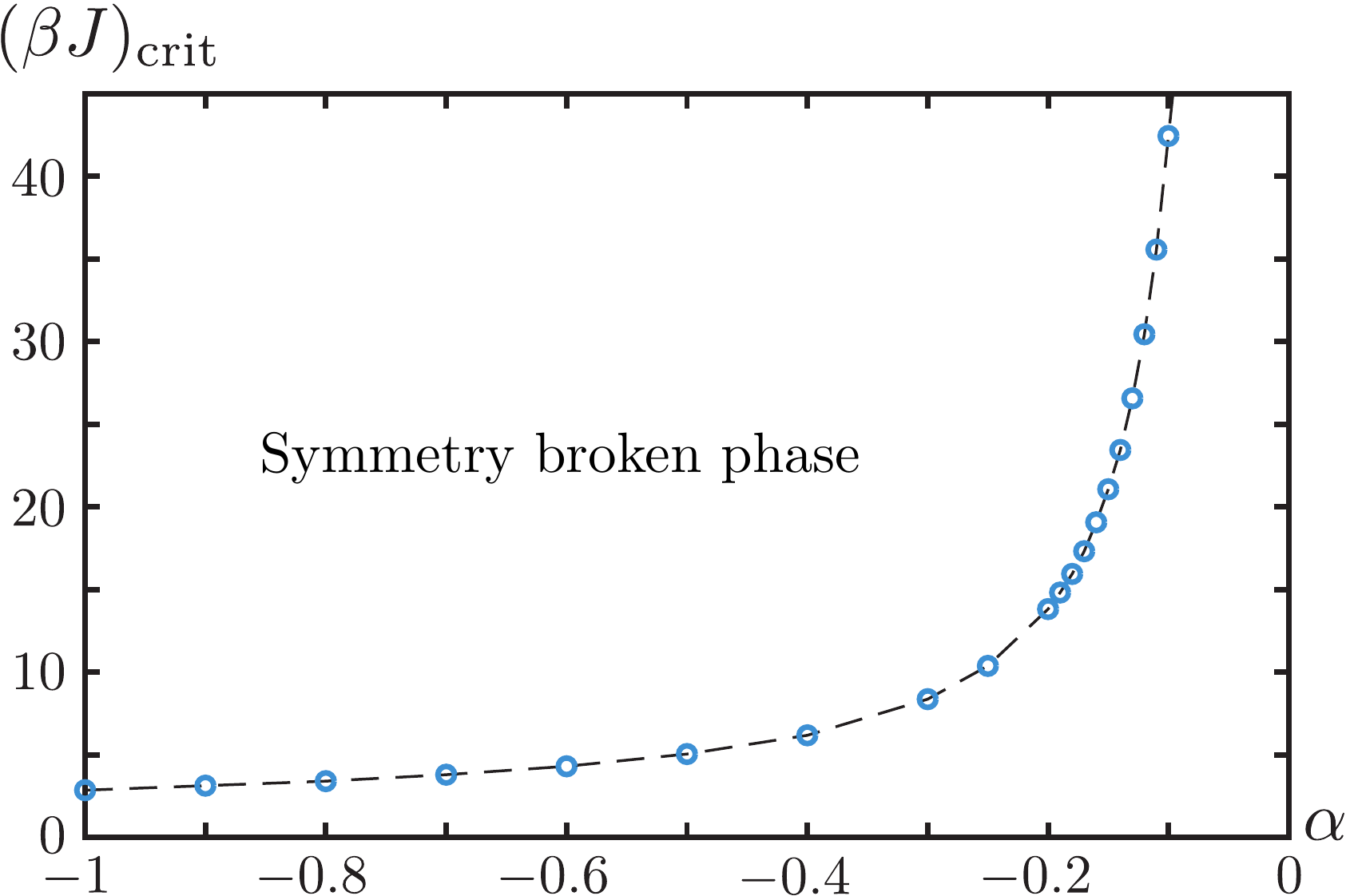}
          \caption{Critical value of $\beta J$ as a function of $\alpha$. }
   \label{alphabJcri}
\end{figure}

The plots of $G_{11}$ and $i G_{12}$ for different values of $\alpha$ and $\beta J$ are shown in fig. \ref{G11G12lowT}. 
For each value of $\alpha$ between $-1$ and $0$
there are two phases.
 In the low temperature phase (large $\beta J$),  there are three distinct solutions: two solutions with non-vanishing $i G_{12}$ related by
$G_{12}(\tau)\rightarrow - G_{12}(\tau)$ 
and the one where $G_{12}(\tau)=0$. The solutions with non-vanishing $i G_{12}$ are the ones with the lower free energy.
As $\beta J$ decreases, $|G_{12}(\tau)|$ decreases everywhere for the non-trivial solution (see figure \ref{G11G12lowT}, \ref{G12critical}), and at the critical value becomes exactly zero. 
For $\beta J < (\beta J)_{\rm crit}$ the only possible solution is $G_{12}(\tau)=0$.
Thus, the $\mathbb{Z}_2$ symmetry is restored, and this is a second-order phase transition. 
The plot of $ (\beta J)_{\rm crit}$ vs. $\alpha$ is shown in figure \ref{alphabJcri};  it blows up as
$\alpha$ approaches zero from below.\footnote{We note that this function does not have a vanishing derivative at the self-dual value of $\alpha=-1$.
Had we plotted the critical value of $\beta \tilde J= \beta J \sqrt{|\tilde \alpha|}$, this derivative would vanish but the plot would not be monotonic.}

Using the solutions of the Schwinger-Dyson equations we can numerically compute the large $N$ free energy 
\begin{align}
-\frac{\beta F}{N_{\textrm{SYK}}} =& \log 2 +\frac{1}{2} \sum_{n=-\infty}^{+\infty}\log\bigg( \Big(1+\frac{\Sigma_{11}(\omega_{n}) }{i\omega_{n}}\Big)^{2}-\frac{\Sigma_{12}^{2}(\omega_{n})}{\omega_{n}^{2}}\bigg)\notag\\
&+\frac{3}{4} \sum_{n=-\infty}^{+\infty}\big(\Sigma_{11}(\omega_{n})G_{11}(\omega_{n})-\Sigma_{12}(\omega_{n})G_{12}(\omega_{n})\big)\,, \label{freeSD}
\end{align}
where the sum $\sum_n \log(-i \omega_{n})$ is replaced by $\log(2)$. The  energy can be computed with the formula 
\begin{align}
\label{gseform}
\frac{E}{N_{\textrm{SYK}}} = \frac{1}{2\beta} \sum_{n=-\infty}^{+\infty}\big(\Sigma_{11}(\omega_{n})G_{11}(\omega_{n})-\Sigma_{12}(\omega_{n})G_{12}(\omega_{n})\big)\,
\end{align}
and at low temperatures it should converge to the energy of the ground state $E_{0}$ divided by $N_{\textrm{SYK}}$.

Now one can compare the  free energy in the symmetry broken phase, $F_{G_{12} \neq 0}$ with that of the symmetry unbroken phase, $F_{G_{12} = 0}$.
In particular, the free energy of the latter is simply twice that of a single SYK with a rescaling $J\to \sqrt{1+3\alpha^2}J.$ It follows that in the ``conformal window" $0\leq \alpha \leq 1/3$ the low-temperature limit of the entropy is 
\begin{equation}
S_0 = 2 c_0 N\ , \qquad c_0 =\frac{1}{8}\log 2 + \frac{K} {2 \pi} \approx 0.2324 \ ,
\end{equation}
which is twice that of the single SYK model. The fact that this is independent of $\alpha$ means that the $g$-theorem\cite{Affleck:1991tk} is obeyed to leading order in $N$, even though the theory is not exactly conformal due to the peculiarities of the $h=2$ mode. 
As a further check, one can consider a large $q$ expansion \cite{Maldacena:2016hyu, Tarnopolsky:2018env}, 
\begin{equation}
\beta F_{G_{12}=0}^{(q)}=-\log 2-\frac{1}{q^2} 2\pi \nu\big(\tan\frac{\pi \nu}{2}-\frac{\pi \nu}{4}\big)-\frac{1}{q^3}2\pi \nu \Big(\pi \nu-2\tan\pi \nu\big(1-\frac{\pi^2\nu^2}{12}\big)\Big)+\dots,
\end{equation}
where $\beta J \sqrt{(1+3\alpha^2)2^{1-q} q}=\frac{\pi \nu}{\cos\frac{\pi \nu}{2}}.$ The free energy of the symmetry unbroken phase $F_{G_{12}=0}$ is seen to agree well numerically with $F^{(4)}_{G_{12}=0}$. 
  
\begin{figure}[h!]
    \centering
    \includegraphics[width=0.46\textwidth, angle=0.]{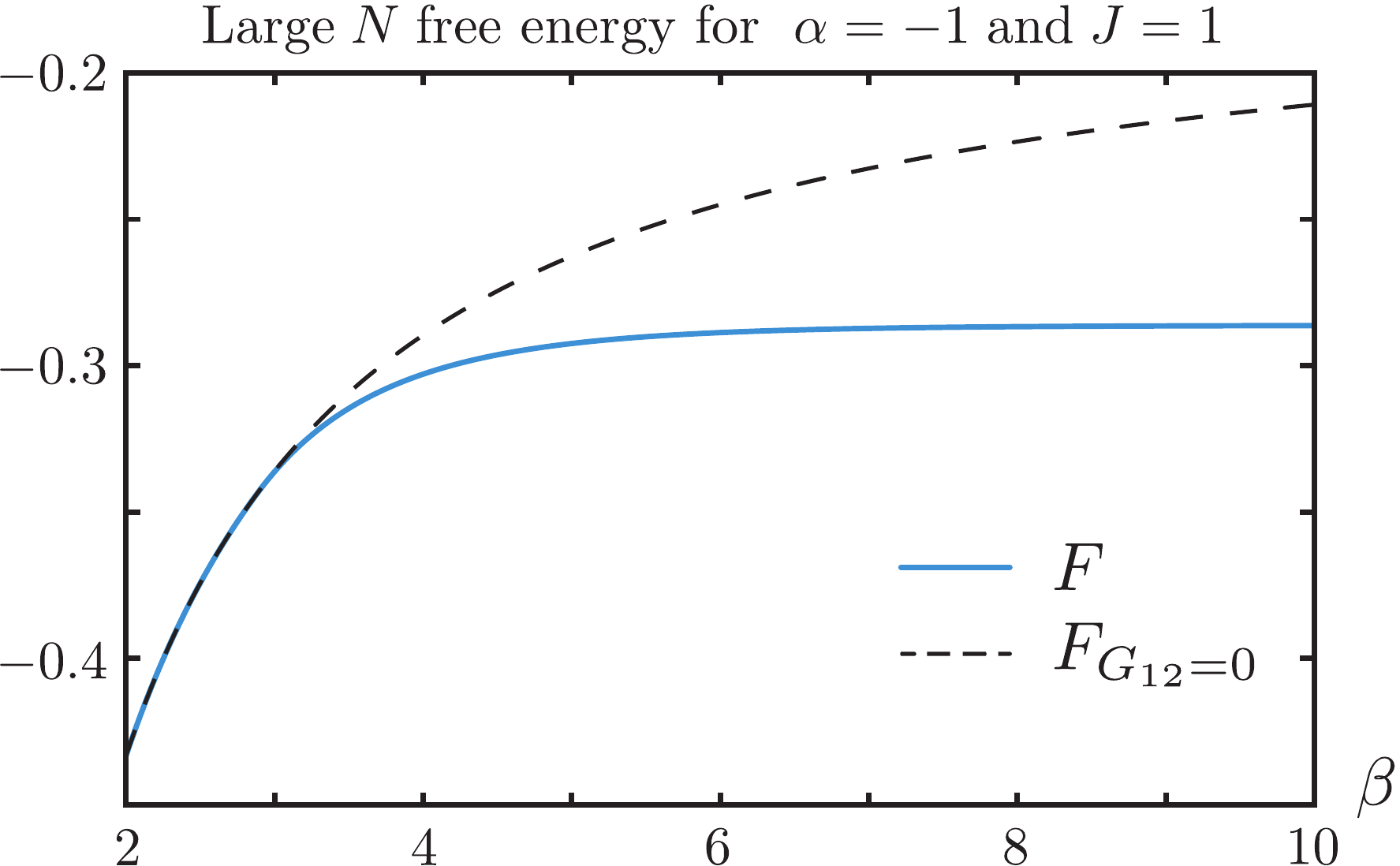}
    \includegraphics[width=0.51\textwidth, angle=0.]{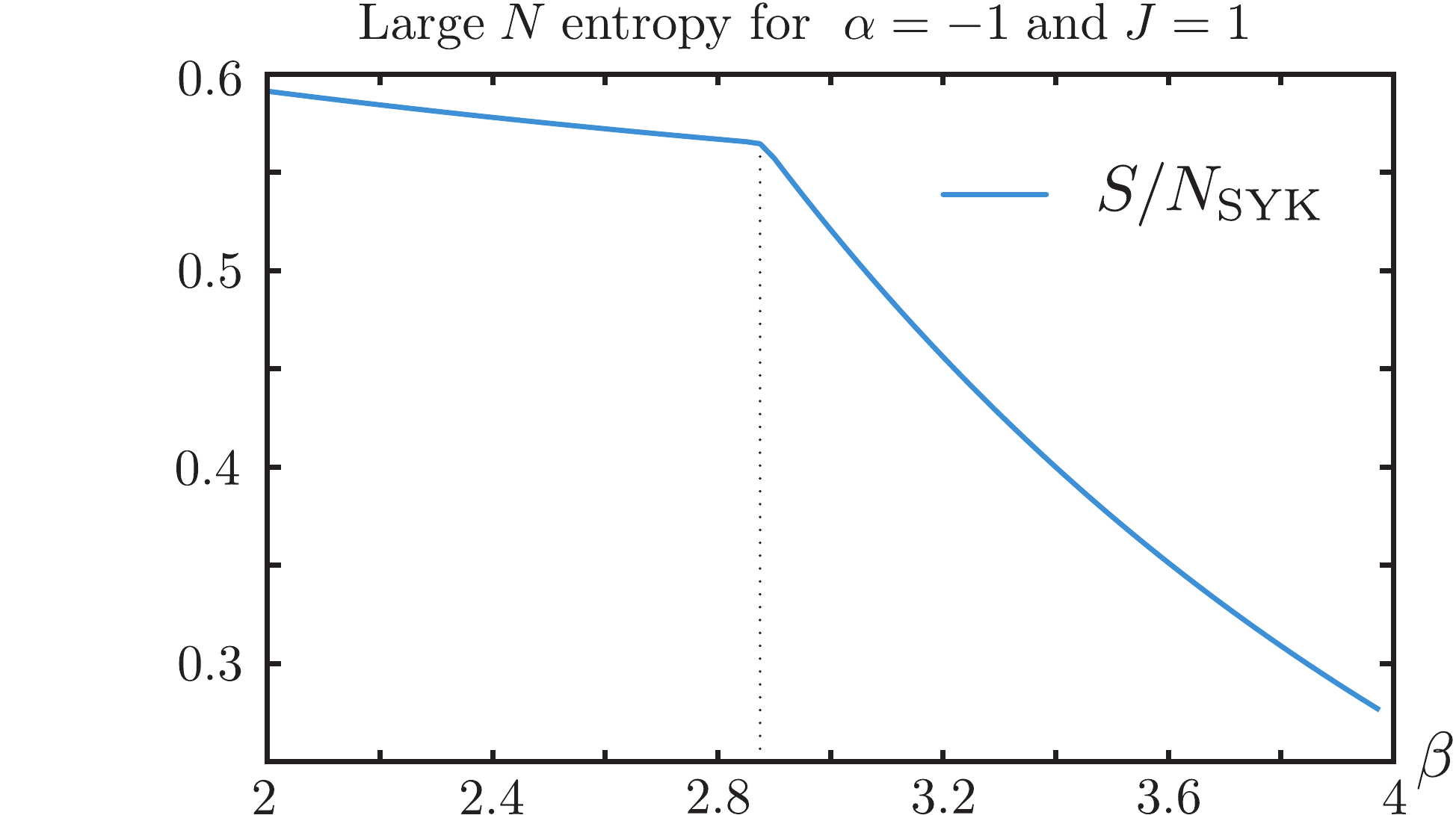}
    \caption{Large $N$ free energies of the true numerical solution and the solution with $G_{12}=0$ for $\alpha=-1, J=1.$ The graph on the right shows the entropy; we can clearly see a second order phase transition, 
as there is a discontinuity in its derivative near critical temperature.   }
    \label{freeE}
\end{figure}

In figure \ref{freeE} we plot for $\alpha=-1$ the free energy of the symmetry broken phase (\ref{freeSD}) as a function of $\beta J$ and compare it with that of the unbroken phase, obtained by setting $G_{12}=0$ in the SD equations (\ref{kineticSD}) and (\ref{otherSD}). We also show the entropy as a function of $\beta J$.
The plot shows a clear second order phase transition at 
$ (\beta J)_{\rm crit}\approx 2.87$, and the derivative of the entropy is discontinuous. We will systematically study the critical exponents in future work.

\begin{figure}[h!]
    \centering
    \includegraphics[width=0.5\textwidth, angle=0.]{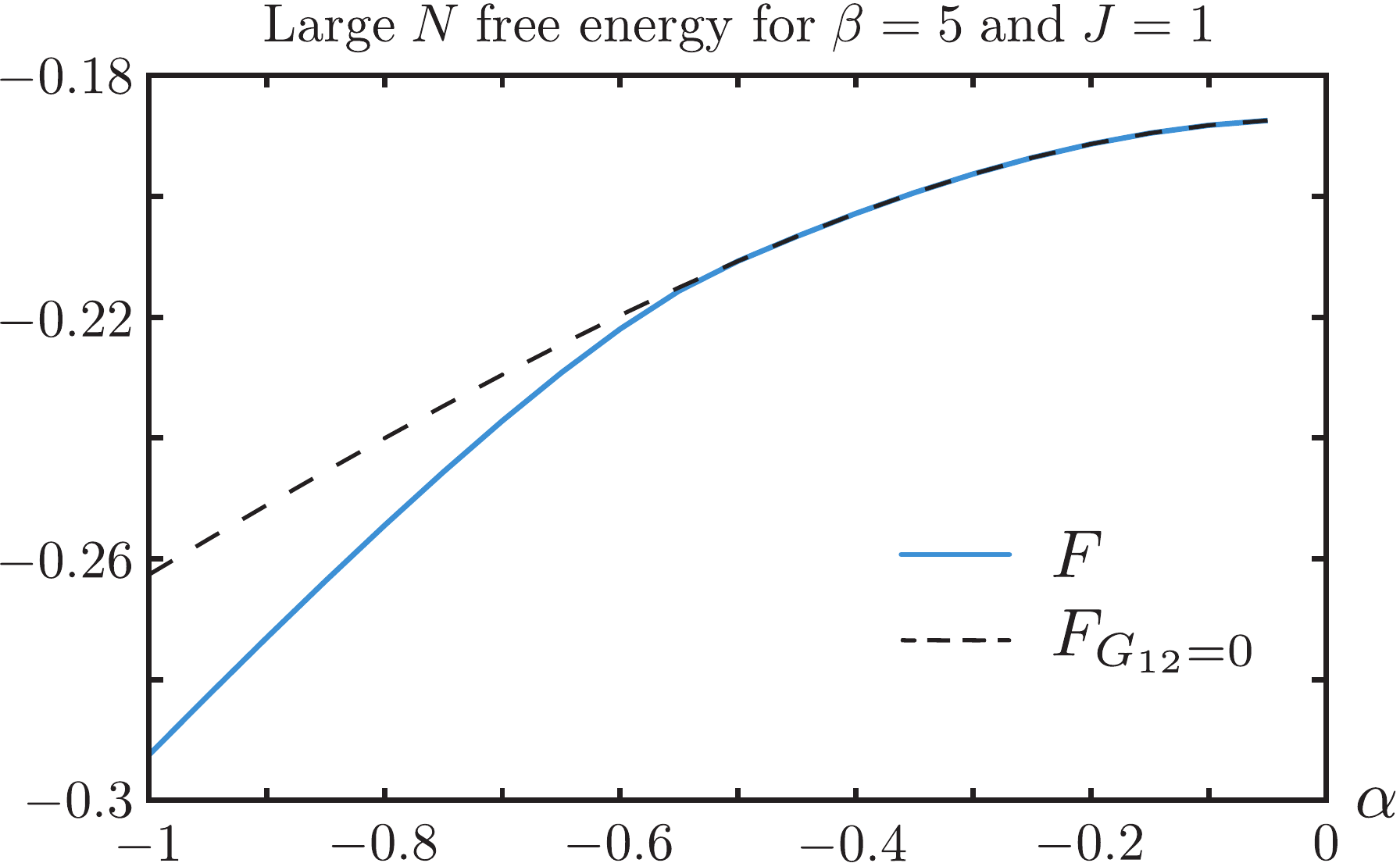}
    \caption{Large $N$ free energies at fixed $\beta$ and $J.$ We take $\beta=5,J=1,$ and decrease $\alpha.$ We observe also a second order phase transition. }
    \label{freeEagain}
\end{figure}
We notice that at sufficiently large $\beta J$, there is a range of $\tau$ where both $i G_{12}(\tau)$ and $G_{11}(\tau)$ decay exponentially and share the same decay rate. To explain this fact, let us study the $T=0$ case and insert the complete set of states
\begin{equation}
G_{11} (\tau) = \langle 0_+| e^{-H \tau }\chi_1^1 (0) e^{H \tau} |n \rangle \langle  n | \chi_1^1 (0) | 0_+\rangle \ .
\end{equation}  
For large $\tau$ the sum is dominated by the lowest excited state, and we find 
\begin{equation}
G_{11} (\tau) \rightarrow e^{-(E_1- E_0) \tau}  \langle 0_+| \chi_1^1 (0)  |1 \rangle \langle  1 | \chi_1^1 (0)|0_+ \rangle\ . 
\end{equation}  
Similarly, we find that the large $\tau$ behavior of $G_{12}$ is
\begin{equation}
G_{12} (\tau) \rightarrow e^{-(E_1- E_0) \tau}  \langle 0_+| \chi_1^1 (0)  |1 \rangle \langle  1 | \chi_2^1 (0)| 0_+ \rangle\ . 
\end{equation} 
Thus the universal decay rate among correlators signifies a mass gap in the spectrum.

\begin{figure}[h!]
    \centering
    \includegraphics[width=0.7\textwidth, angle=0.]{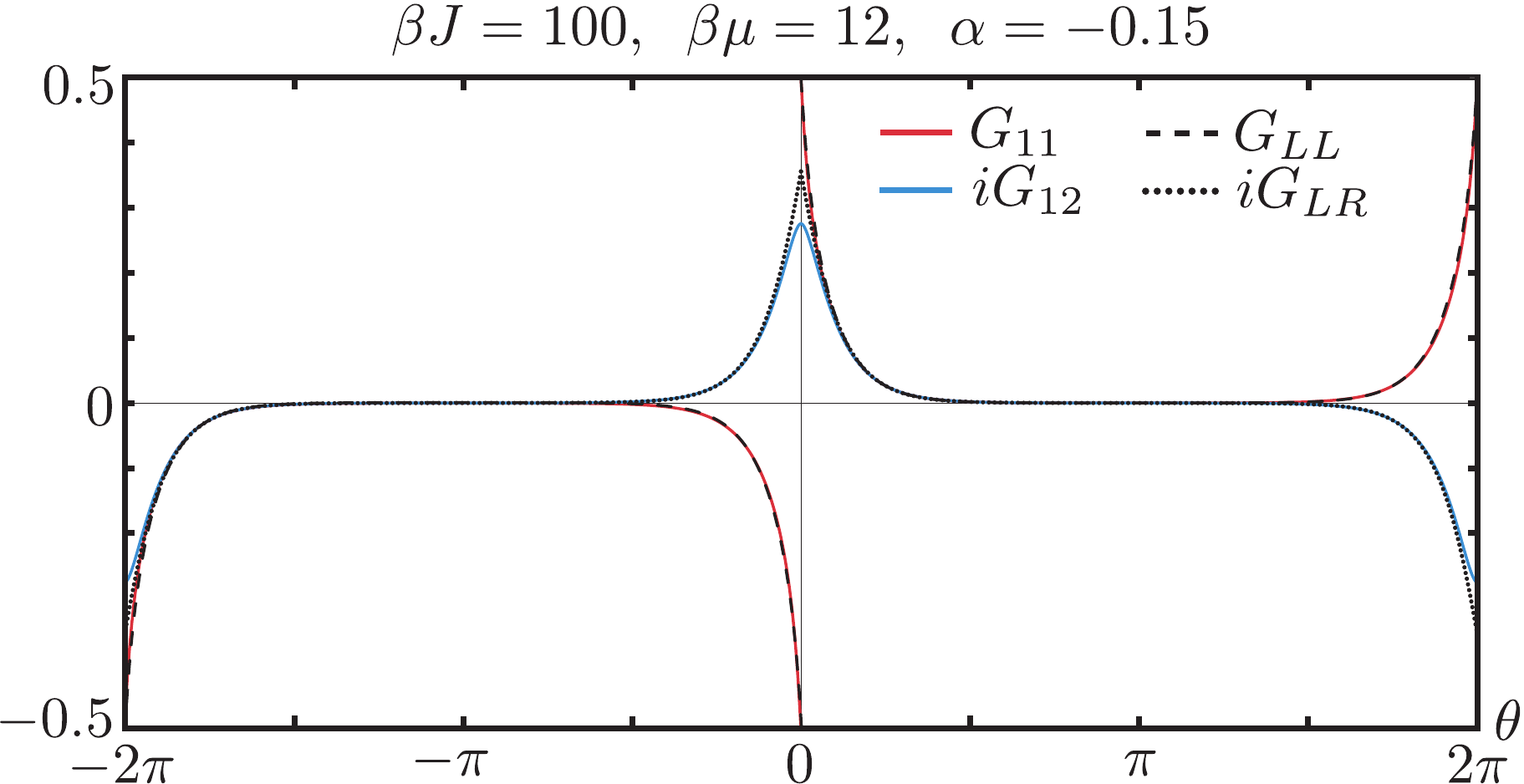}
    \caption{Plot of solutions $G_{LL}$ and $iG_{LR}$ for the model in \cite{Maldacena:2018lmt}  superimposed with $G_{11}$ and $iG_{12}$.
Parameters chosen so that the
solutions are close for most of the range of $\theta$.}
    \label{CompareG}
\end{figure}

In the work of Maldacena and Qi \cite{Maldacena:2018lmt} the functions $G_{11}$ and $G_{12}$ were also found to be exponentially decreasing for sufficiently large $\beta J$.
In fig. \ref{CompareG} we exhibit superimposed plots of the low temperature solutions to our system of equations and those from \cite{Maldacena:2018lmt}, with parameters chosen so that the
solutions are close to one another for most of the range. We observe a difference in the behavior of $iG_{12}(\tau)$ and $iG_{LR}(\tau)$ at small $\tau$: in our case the function is smooth with a vanishing derivative at $\tau=0$, 
while in \cite{Maldacena:2018lmt}
its derivative is discontinuous at $\tau=0$; this is due to the fact that their Hamiltonian includes a quadratic term. 

\begin{figure}[h!]
    \centering
    \includegraphics[width=0.45\textwidth, angle=0.]{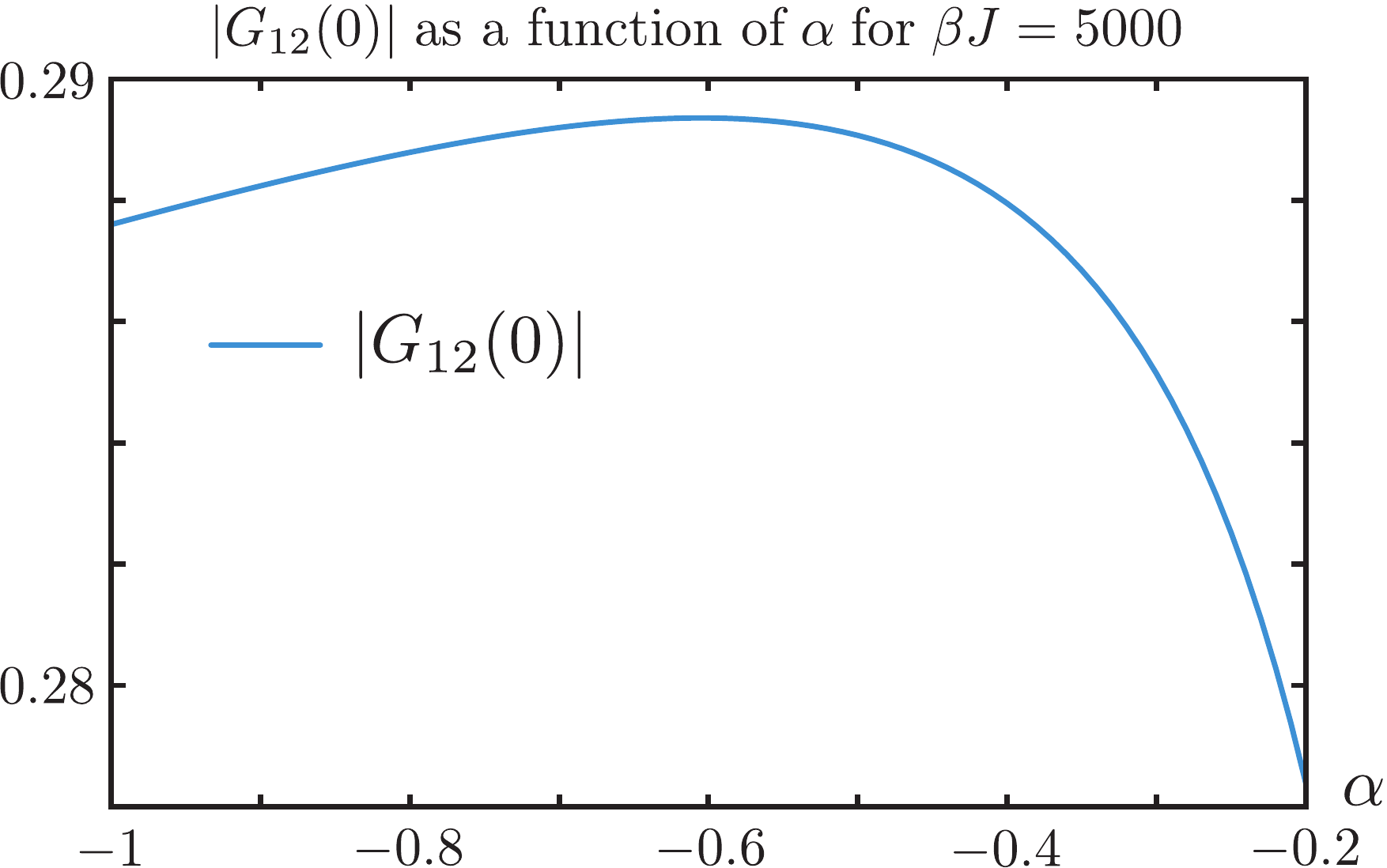}
    \caption{The expectation value of $Q/N_{\textrm{SYK}}$, i.e. $|G_{12}(0)|$, 
as a function of $\alpha$ for $\beta J = 5000$. }
    \label{condvsalpha}
\end{figure}

\begin{figure}[h!]
    \centering
             \includegraphics[width=0.42\textwidth, angle=0.]{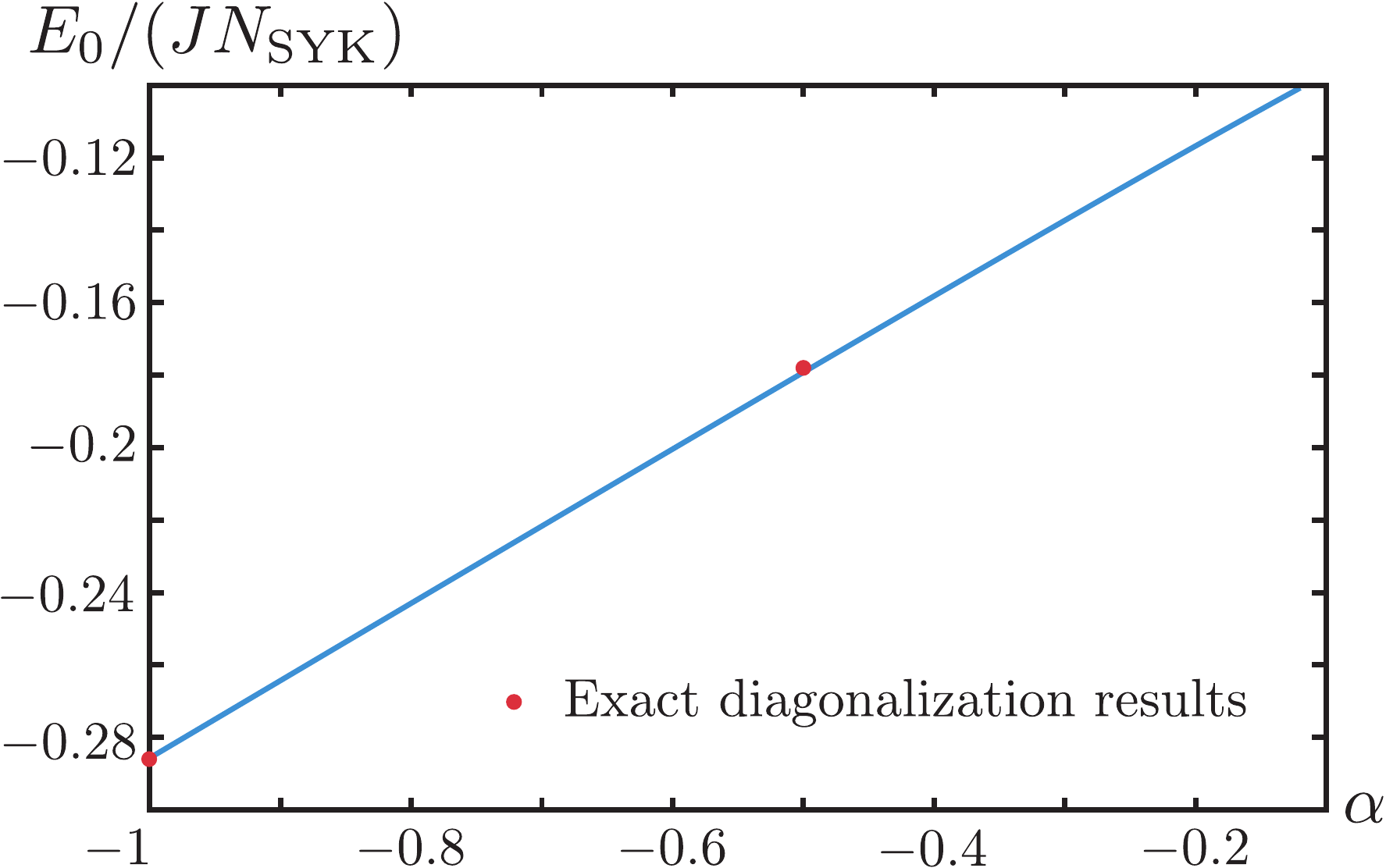}\hspace{3em}
         \includegraphics[width=0.4\textwidth, angle=0.]{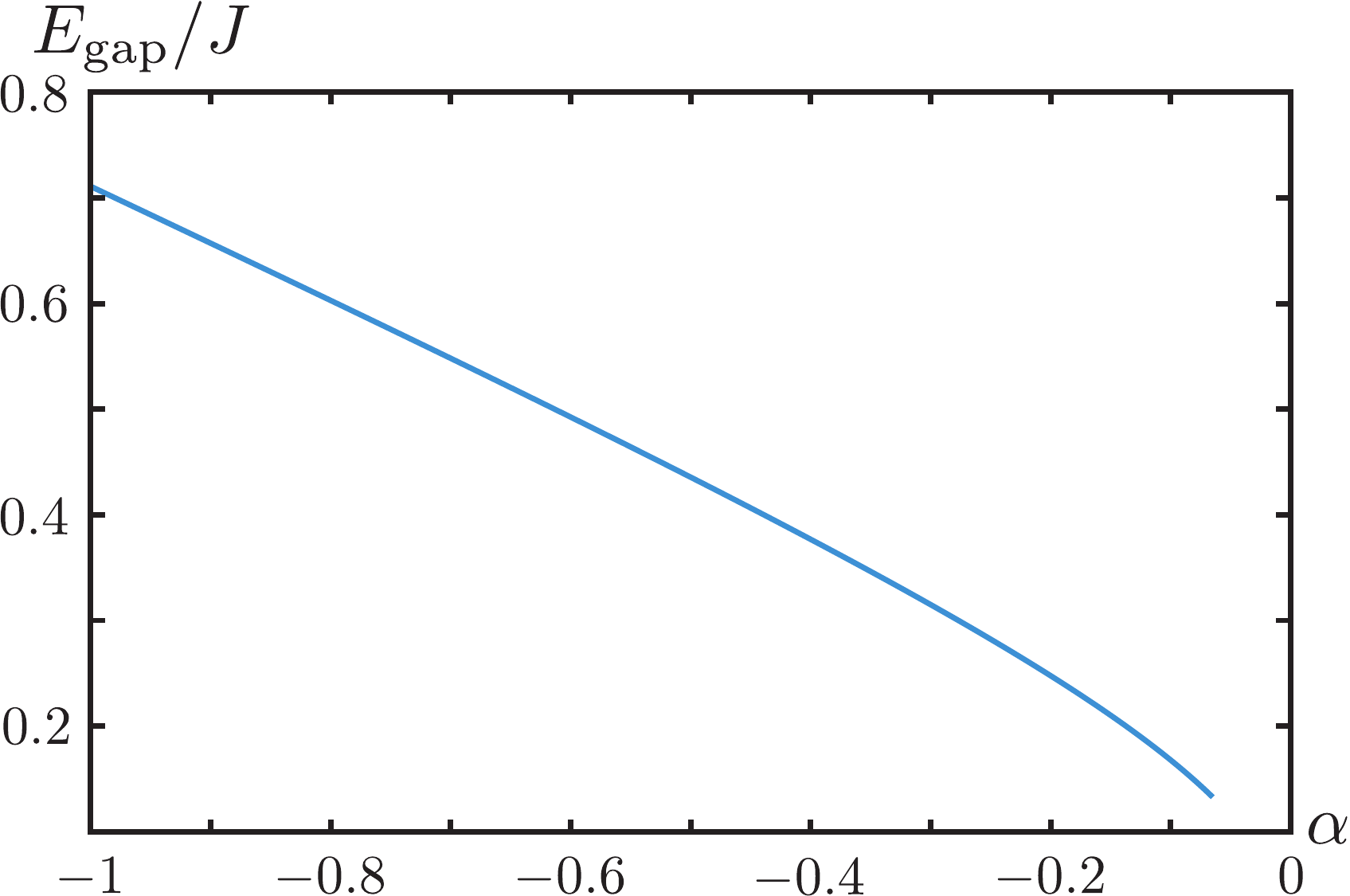}
          \caption{Right: 
the value of $E_0/(J N_{\rm SYK})$ as a function of $\alpha$. Both graphs are approximately linear in $\alpha$ for $\alpha$ not too small. 
Results from exact diagonalizations, (\ref{gslinear}), are shown with dots. Left: the large $N_{\rm SYK}$ energy gap in the spectrum, computed from the exponential decay of the Green functions.  }
    \label{egapvsalpha}
\end{figure}

We can also study what happens at low temperatures (large $\beta J$) as a function of $\alpha$. 
In figure \ref{condvsalpha} we plot $i G_{12}(0)$, which is the expectation value of the order parameter 
$Q/N_{\textrm{SYK}}$, for a large $\beta J$. This quantity becomes small as $\alpha$ is increased towards zero. 
In figure \ref{egapvsalpha} we plot the large $N_{\rm SYK}$ limit of the energy gap $E_{\rm gap}$ divided by $J$, calculated from the exponential decay of the Green functions.
We also plot the
ground state energy $E_0$ divided by $JN_{\rm SYK}$ calculated using (\ref{gseform}).
Results from exact diagonalizations extrapolated to large $N_{\rm SYK}$,  (\ref{gslinear}), are shown with dots and demonstrate very good agreement.
The exact diagonalizations for finite $N_{\rm SYK}$ are discussed in the next section.

\subsection{Exact diagonalization for finite $N_{\rm SYK}$}
\label{numericalspectrum}

In this section we present numerical results for the spectra of two coupled SYK models with Hamiltonian (\ref{SYK2fl}). 
We first check that the results from exact diagonalizations agree well with expectations: the spectrum for $\alpha=0$ and $N_{\rm SYK} = 30$, and the ground state energy of $\alpha = -1$ for various $N_{\rm SYK}$ concur well with analytical arguments, and with the results from \ref{SDeqsnumerics}. Then we present our results on the energy gap and broken symmetry.

\begin{figure}[h!]
    \centering
         \includegraphics[width=0.44\textwidth, angle=0.]{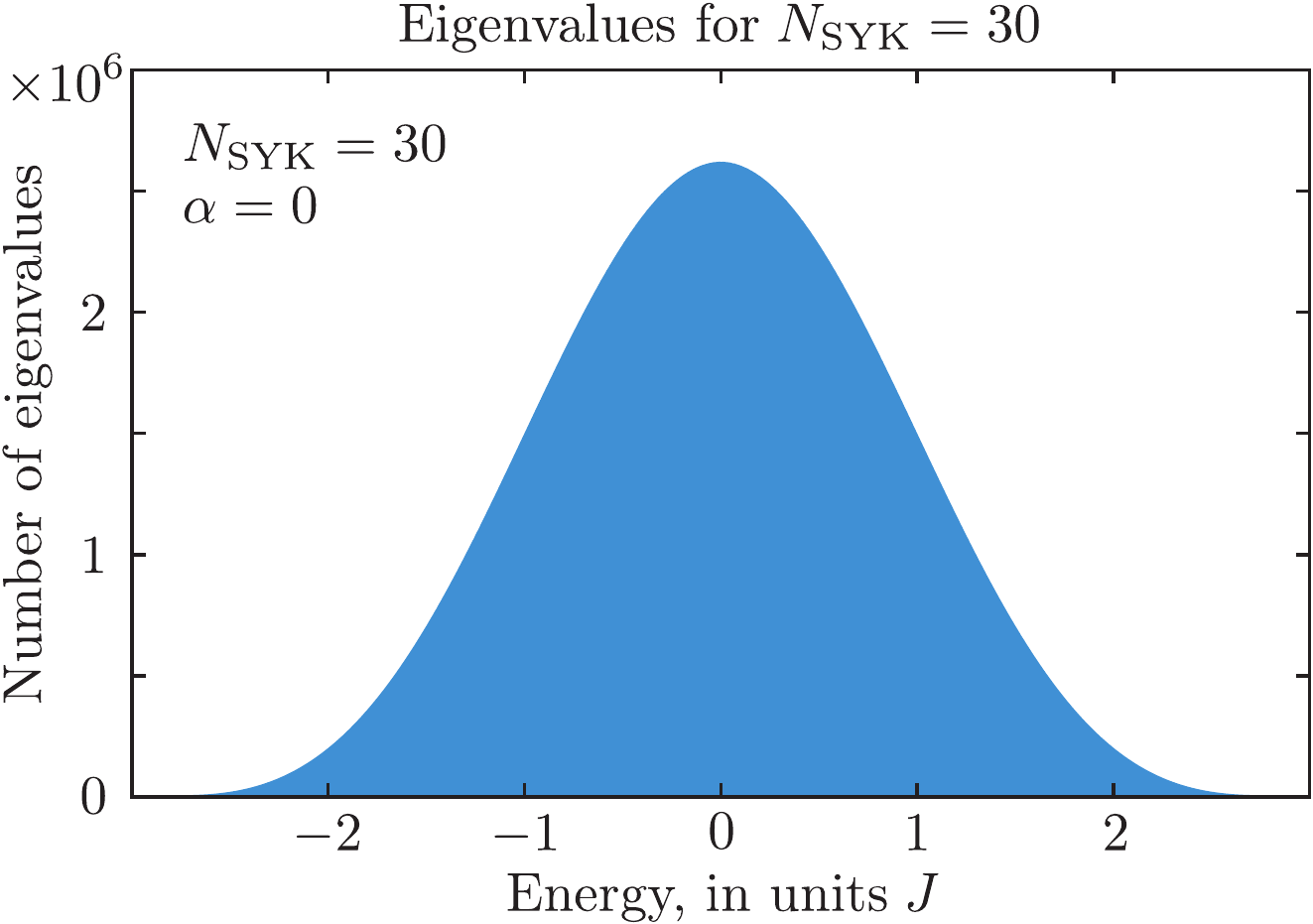}\hspace{3em}
         \includegraphics[width=0.45\textwidth, angle=0.]{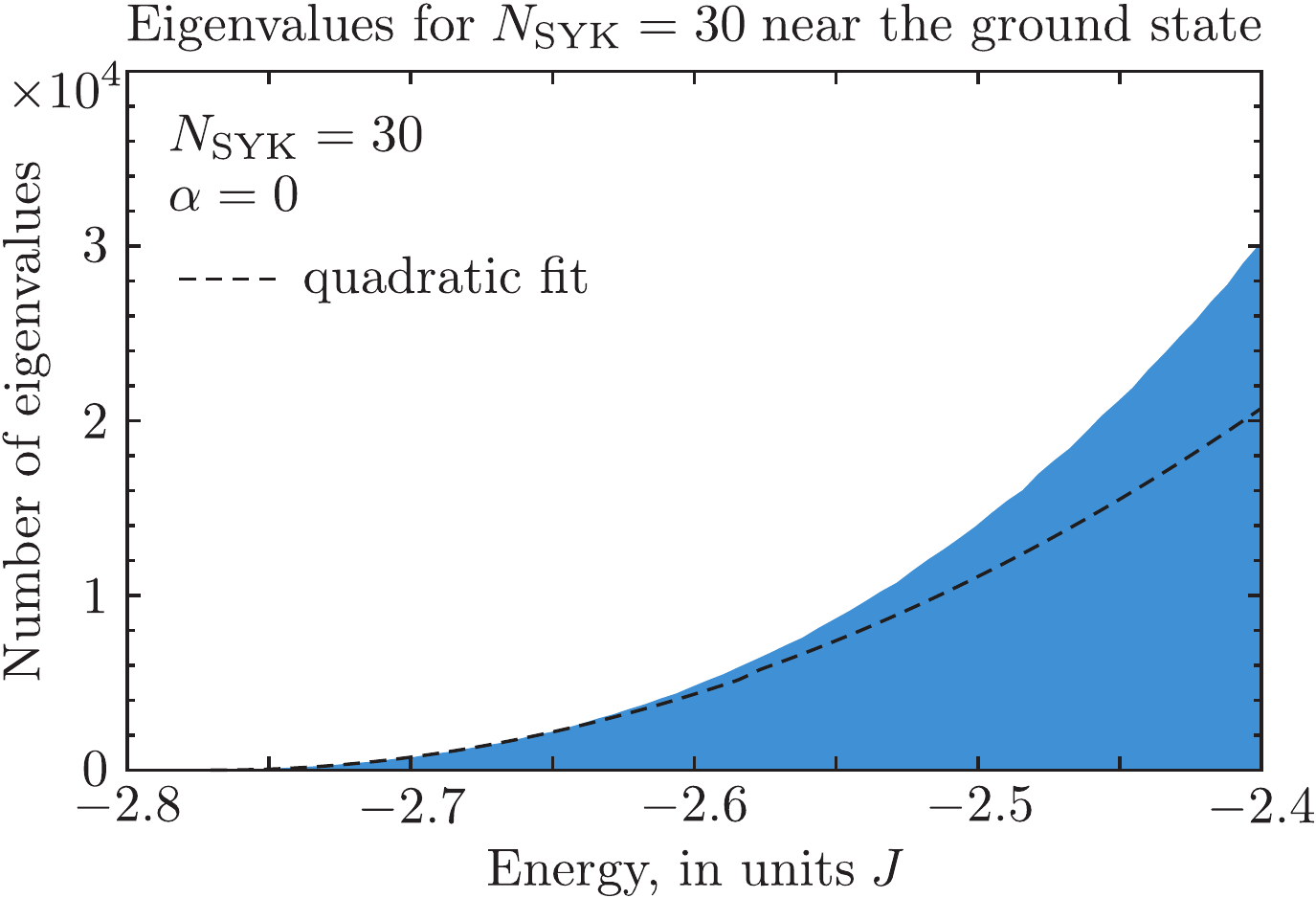}
          \caption{Left: The energy spectrum for $\alpha=0$, i.e. for two decoupled SYK models, for a single sampling of $N_{\rm SYK} = 30$.  Right: 
The same spectrum magnified near the lower edge. }
    \label{decoupled}
\end{figure}

The biggest number we are able to access via exact diagonalization of the coupled SYK models is $N_{\rm SYK}=16$. In this case the discrete symmetry (\ref{phsym}) is not anomalous, and the ground state is non-degenerate.
However, for $-1\leq \alpha<0$ we observe a nearby excited state followed by a gap. We will interpret this as indication of approach to spontaneous symmetry breaking, which takes
place in the large $N_{\rm SYK}$ limit.
We will also present spectra for $N_{\rm SYK}=15$, where the discrete symmetry (\ref{reflectioneven}) is anomalous, so that the states are doubly degenerate. 
There is again a gap in the spectrum present for $-1\leq \alpha <0$.
Furthermore, we will present numerical results on the VEV of operator $i\chi_1^i \chi_2^i$ for $N_{\rm SYK} = 14$, which demonstrates that it is non-vanishing for $-1\leq \alpha< 0$.

\begin{figure}[h!]
  \begin{center}
  \includegraphics [width=0.4\textwidth, angle=0.]{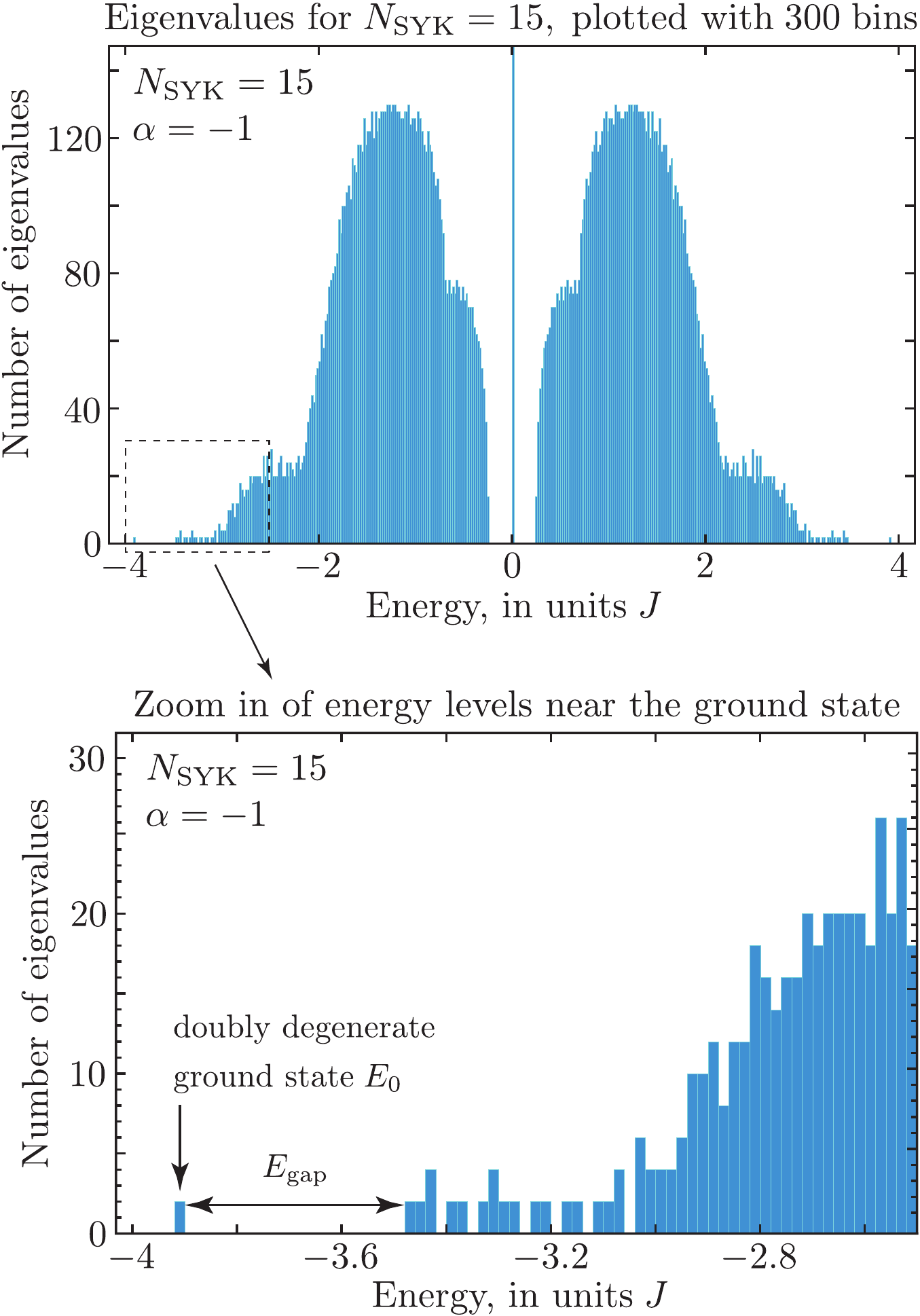}\hspace{3em}
    \includegraphics [width=0.4\textwidth, angle=0.]{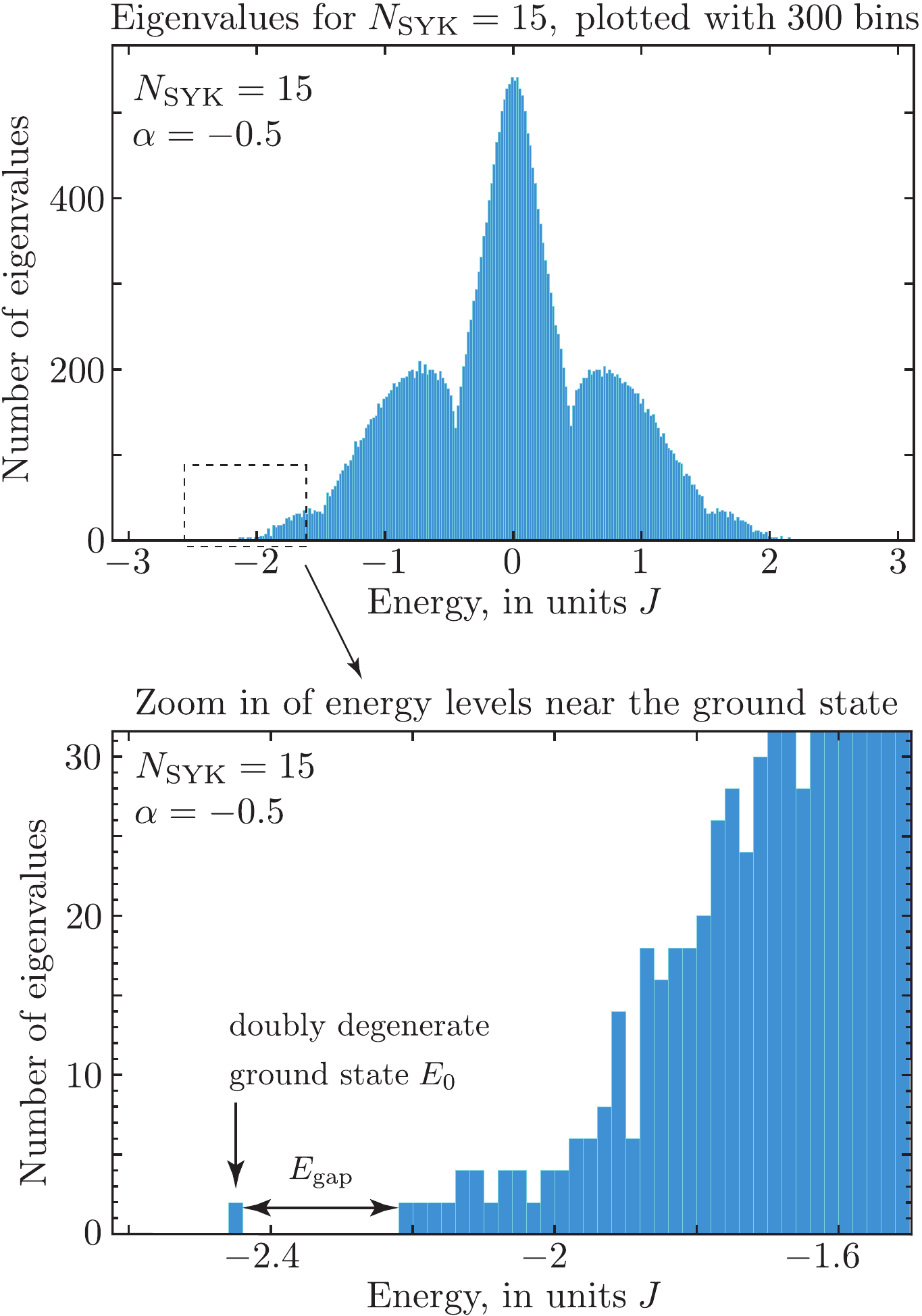}\\
      \includegraphics [width=0.394\textwidth, angle=0.]{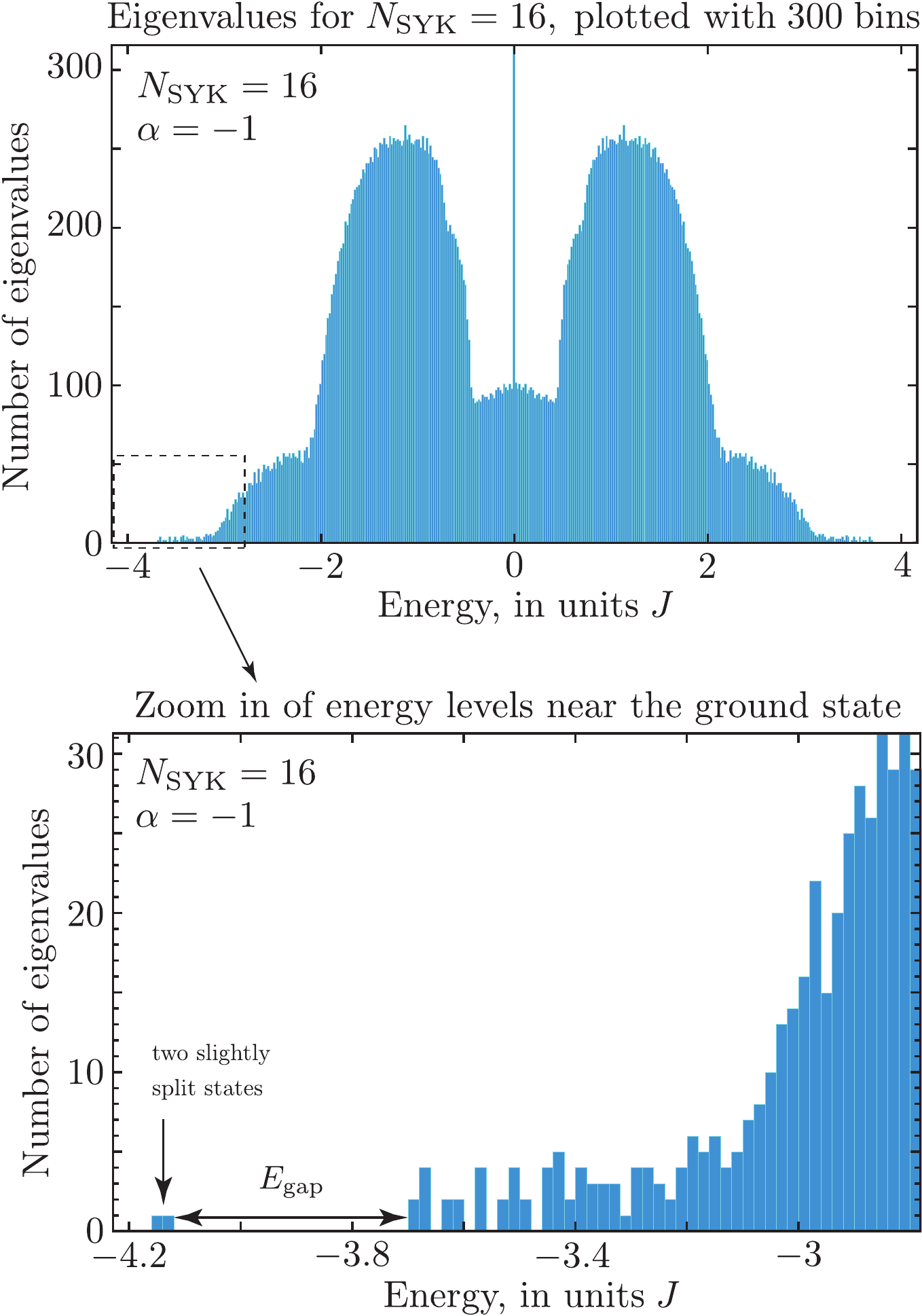}\hspace{3em}
    \includegraphics [width=0.4\textwidth, angle=0.]{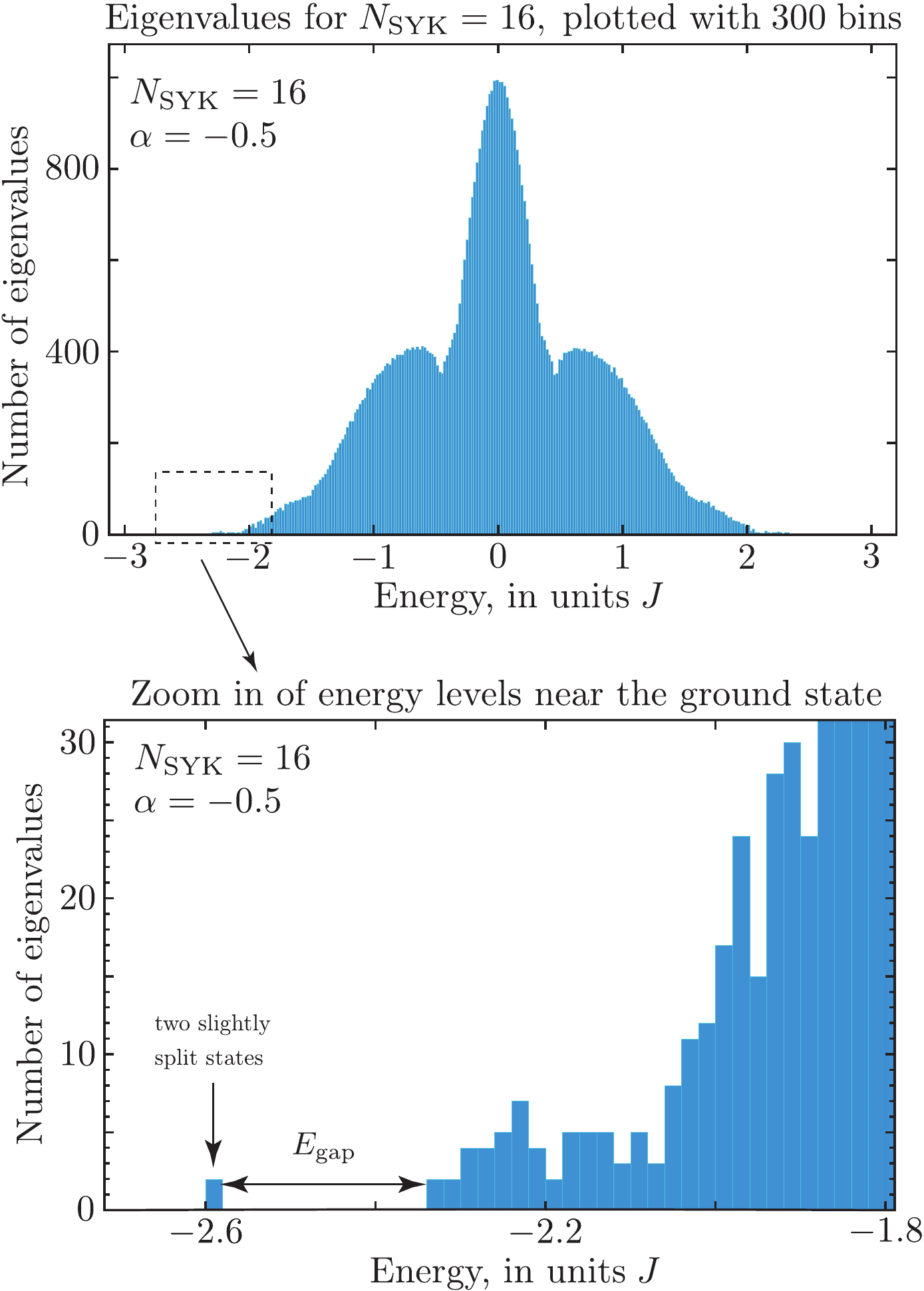}
  \end{center}
  \caption{The spectrum for a single realization with $N_{\rm SYK} = 15,\; 16$ and $\alpha=-1,\;-0.5$. For $\alpha=-1$, the spectrum exhibits a gap near $E=0$  when $N_{\text{SYK}}$ is odd and a large number of states with $E=0$.  }
  \label{espectrum15}
\end{figure}

\begin{figure}[h!]
    \centering
         \includegraphics[width=0.45\textwidth, angle=0.]{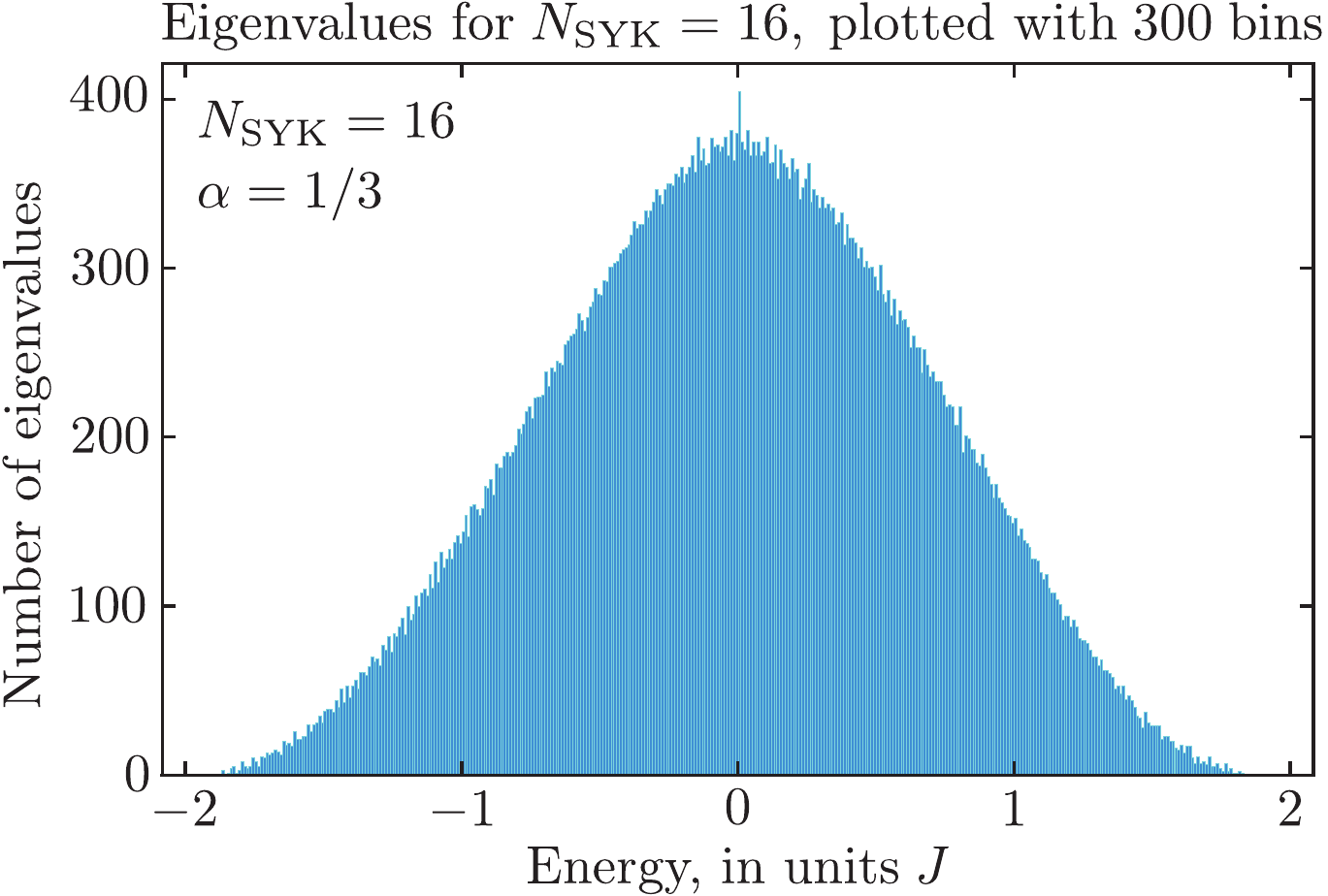}
          \caption{The spectrum for a single realization for  $N_{\rm SYK} = 16$ and $\alpha=1/3$.  }
    \label{espectrumthird}
\end{figure}

First, let us consider $\alpha=0$, where we find the spectrum of two SYK model with the same random couplings. The density of states for this model is simply given by the convolution of that of the single SYK model:\footnote{We thank D. Stanford for a useful discussion about this.}
\begin{equation}
\rho_{\rm double} (E) = \int de \rho(e) \rho(E-e)
\end{equation}
This in particular helps us determine the behavior of $\rho_{\rm double} (E)$ near the ground state.
Shifting the energy so that the ground state is at zero, we know that $\rho(E)\rightarrow A\sqrt{E}$ for small $E$. 
Therefore, for small $E$
\begin{equation}
\rho_{\rm double} (E) \rightarrow A^2 \int_0^E de \sqrt{e (E-e)} = {\pi A^2 E^2\over 8}\ .
\end{equation} 
The numerical density of states, shown in figure \ref{decoupled} for $N_{\rm SYK}=30$, is in good agreement with the $E^2$ dependence near the ground state.

Let us proceed to the spectra for non-vanishing values of $\alpha$. In figs. \ref{espectrum15}, \ref{espectrumthird} we plot
the spectra of energy divided by $J$ for $\alpha=-1, -1/2, 1/3$ and different values of $N_{\rm SYK}$. These energy distributions have interesting and unusual shapes. For the special values 
$\alpha=-1$ and $1/3$ we observe large numbers of states with $E=0$; this creates the zero-energy peaks seen in the graphs.
For $\alpha=-1$ and
odd $N_{\rm SYK}$ we find that the $E=0$ peak is separated by gaps from the remaining states, but for even $N_{\rm SYK}$ it is not.

\begin{figure}[h!]
  \begin{center}
 \includegraphics [width=0.4\textwidth, angle=0.]{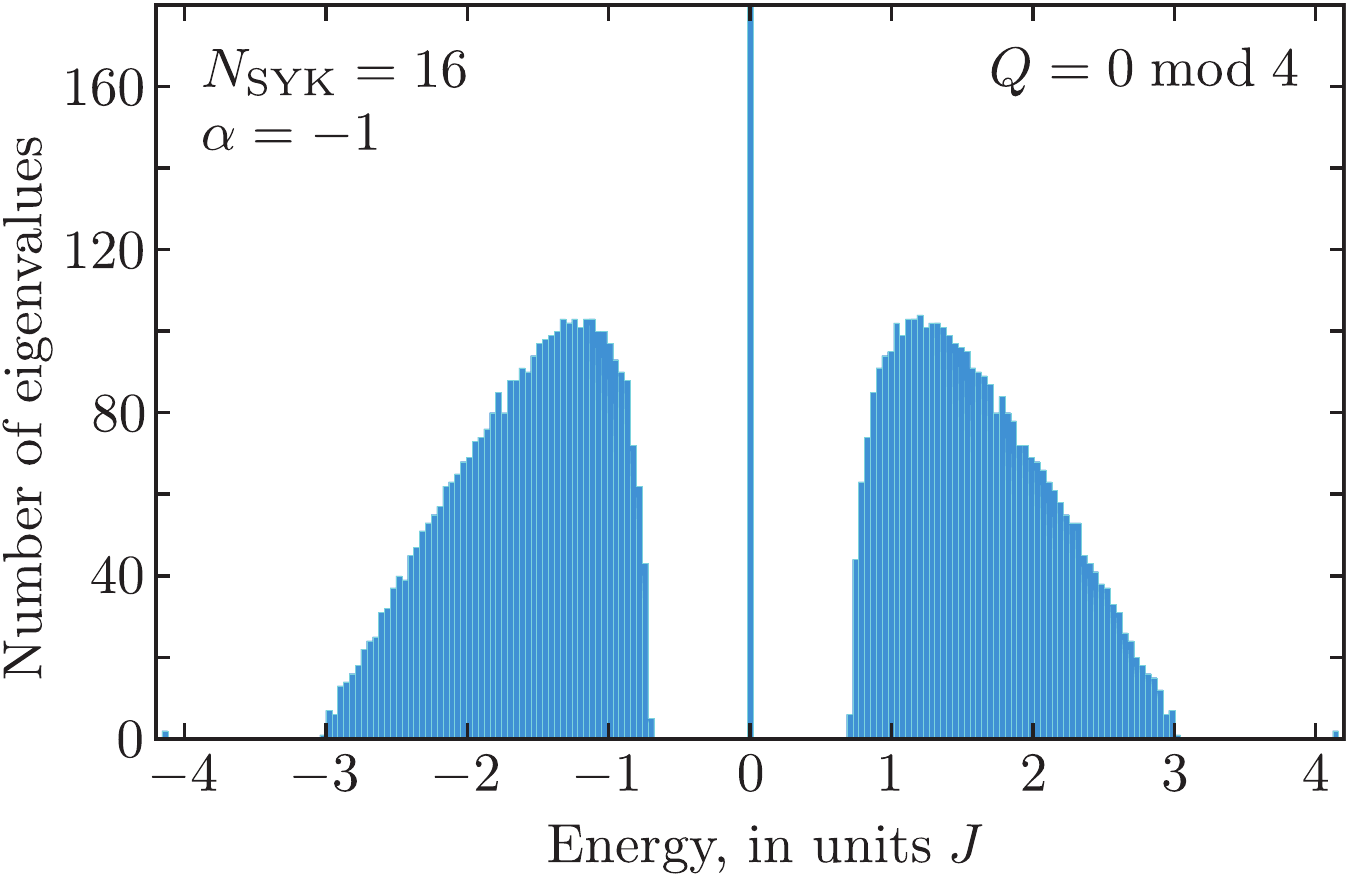}\hspace{3em}
 \includegraphics [width=0.4\textwidth, angle=0.]{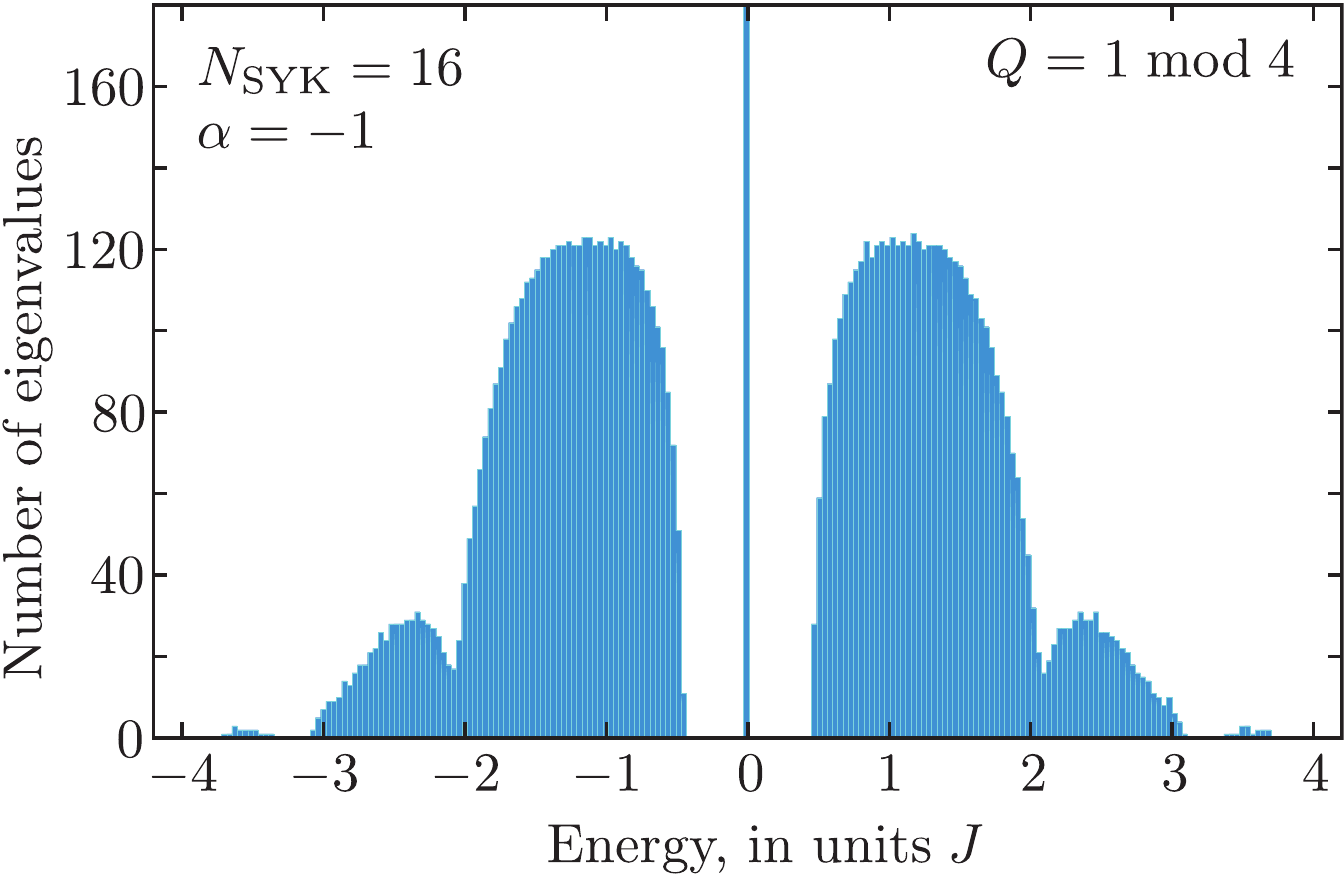}\\
 \vspace{1em}
 \includegraphics [width=0.4\textwidth, angle=0.]{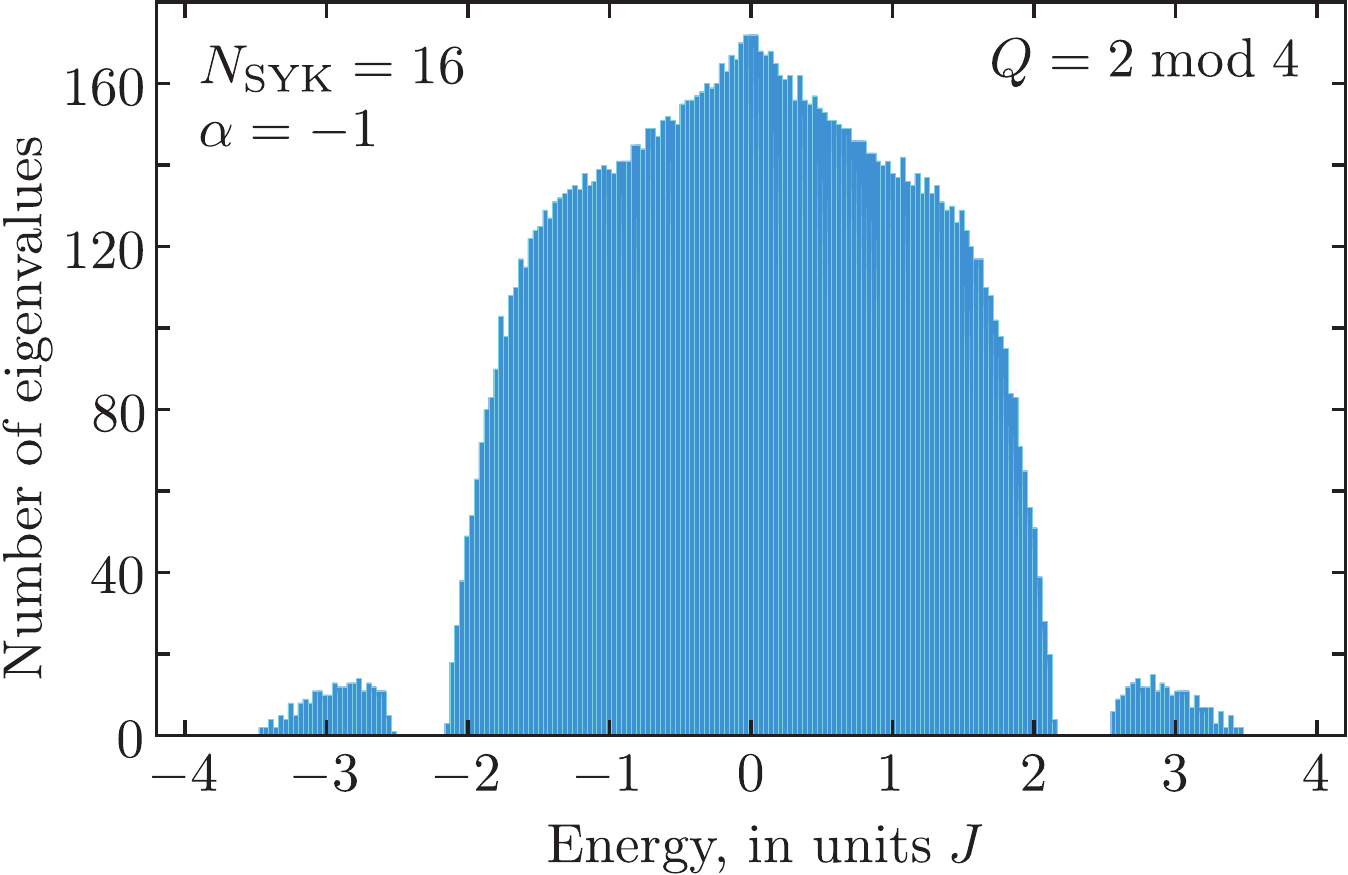}\hspace{3em}
 \includegraphics [width=0.4\textwidth, angle=0.]{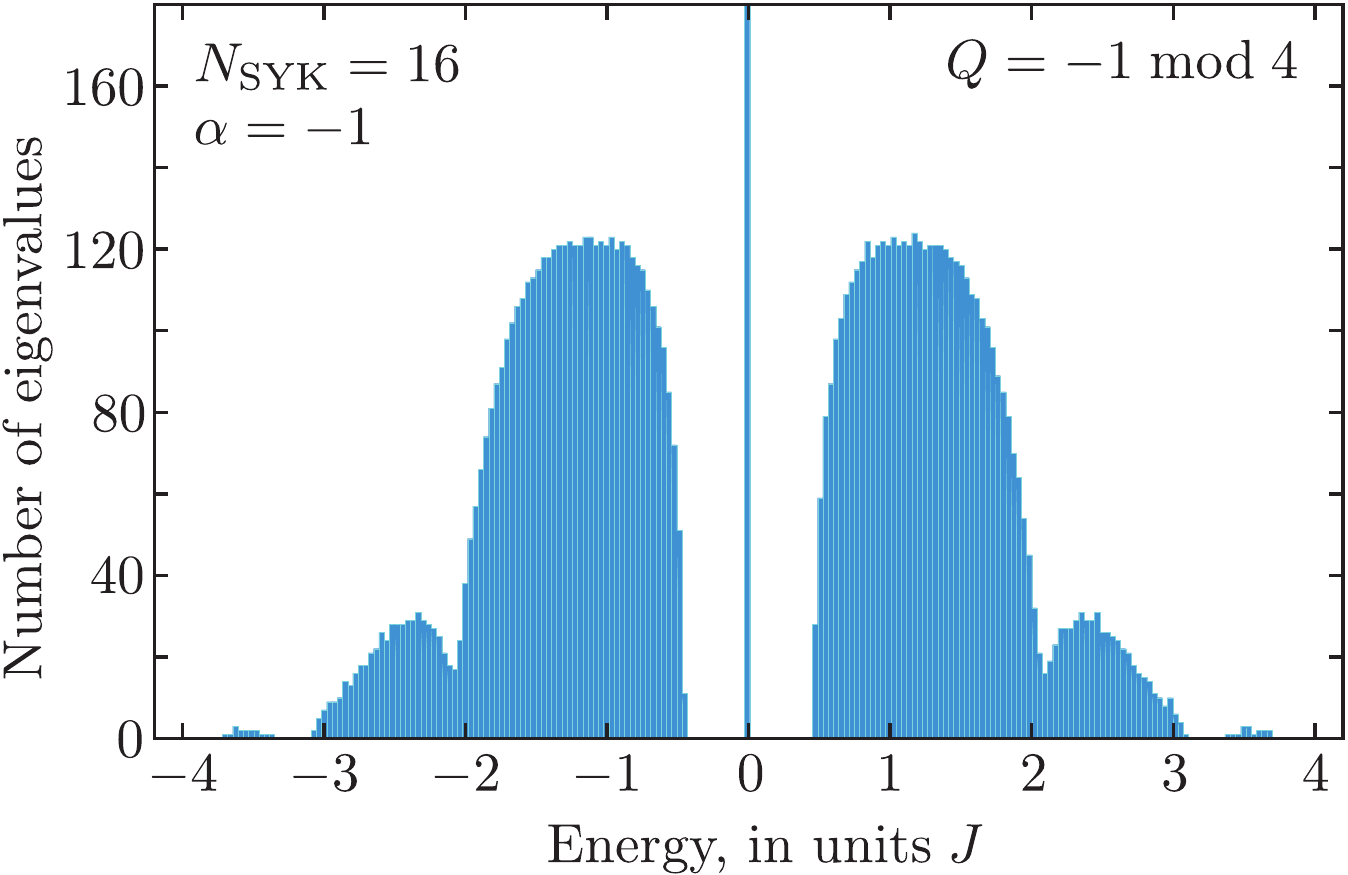}
  \end{center}
  \caption{The spectrum for a single realization of the coupled SYK model with $N_{\rm SYK} = 16$ and $\alpha=-1$ separated into four $Z_4$ symmetry sectors. In each of the sectors the spectrum is symmetric under $E\rightarrow -E$. 
The $Z_4$ invariant sector shows two nearly degenerate lowest states separated by a gap from the rest of the states.  }
  \label{separatespectra}
\end{figure}

In order to clarify the peculiar shapes of the energy distributions in fig. \ref{espectrum15}, it is useful to separate them into distinct $Z_4$ symmetry sectors\footnote{We are very grateful to J. Verbaarschot for raising a question about separation of the spectra into sectors.} labeled by the eigenvalue of 
$e^{\pi i Q/2}$, as shown in fig. \ref{separatespectra} for $N_{\rm SYK}=16$. The sectors where $e^{i\pi  Q/2}=\pm i$, i.e. $Q=\pm 1$ mod $4$, have identical energy spectra which are shown on the right. They contain the symmetric bumps, which produce the ``rabbit ears" pattern in the overall distribution. For $\alpha=-1$ these sectors also contain large numbers of states with $E=0$ (they are discussed in Appendix B). 
On the left in fig. \ref{separatespectra} we show the states with $e^{i\pi Q/2}=\pm 1$. For $e^{i\pi Q/2}=- 1$ the distribution is smooth and does not contain a sharp peak at $E=0$.  
The $Z_4$ invariant sector $e^{i \pi  Q/2}=1$ contains the two nearly degenerate lowest states separated by a very clear gap from the remaining states. For $\alpha=-1$ this sector also contains a large number of $E=0$ states.\footnote{If we gauge the $Z_4$ symmetry, then only the sector with $e^{i\pi Q/2}=1$ will remain in the spectrum.}

\begin{figure}[h!]
  \begin{center}
 \includegraphics [width=0.4\textwidth, angle=0.]{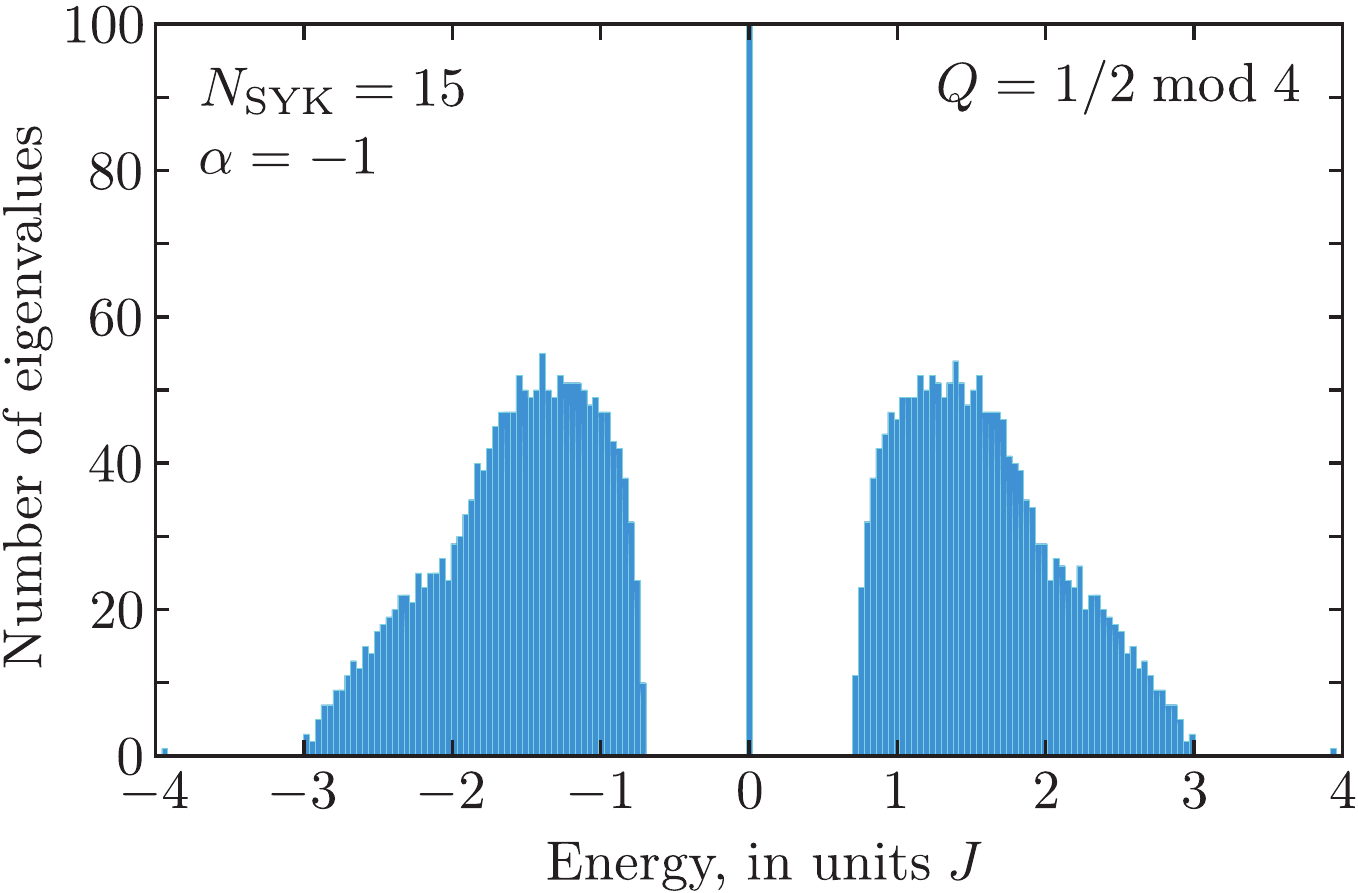}\hspace{3em}
 \includegraphics [width=0.4\textwidth, angle=0.]{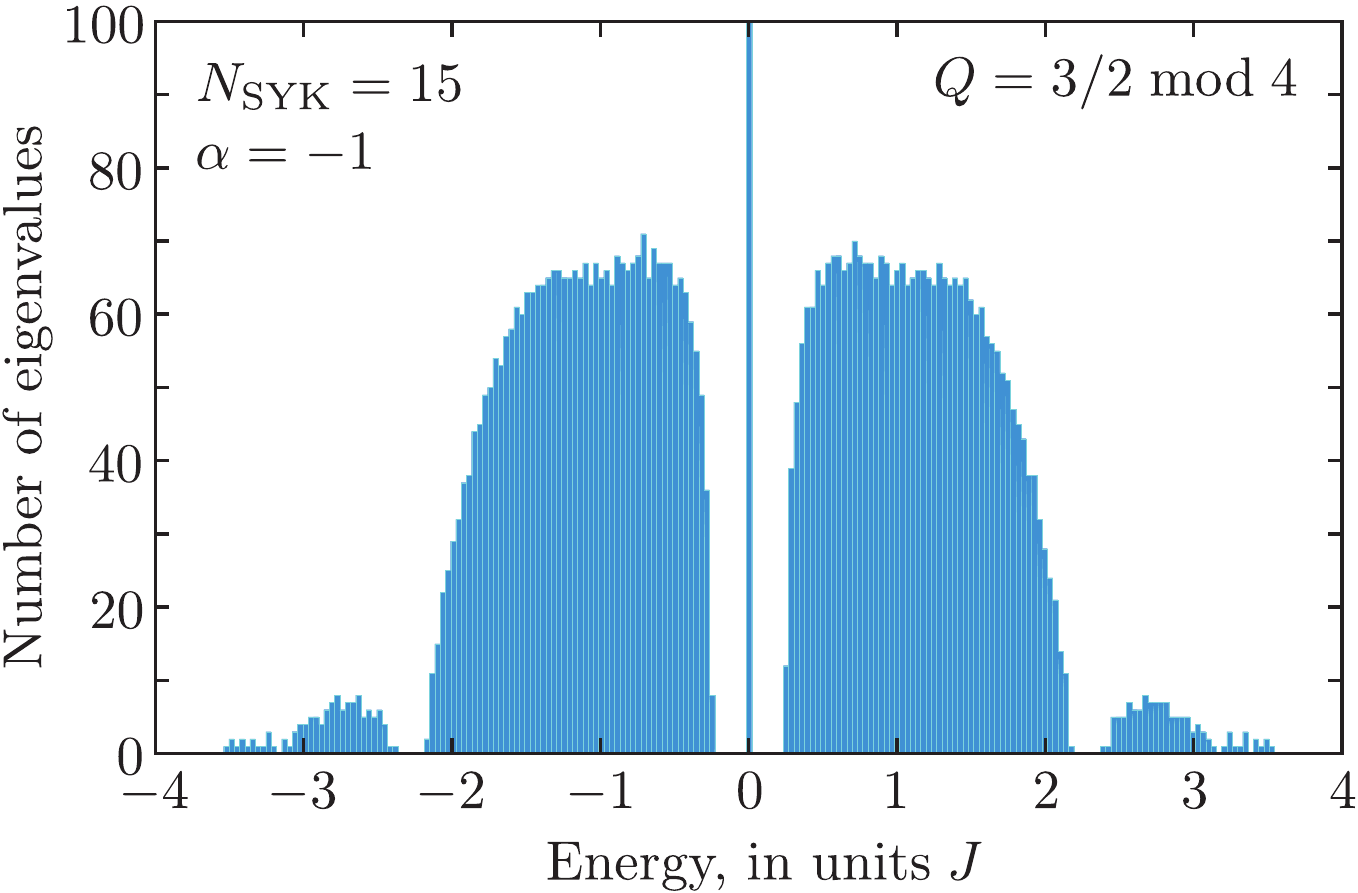}\\
 \vspace{1em}
 \includegraphics [width=0.4\textwidth, angle=0.]{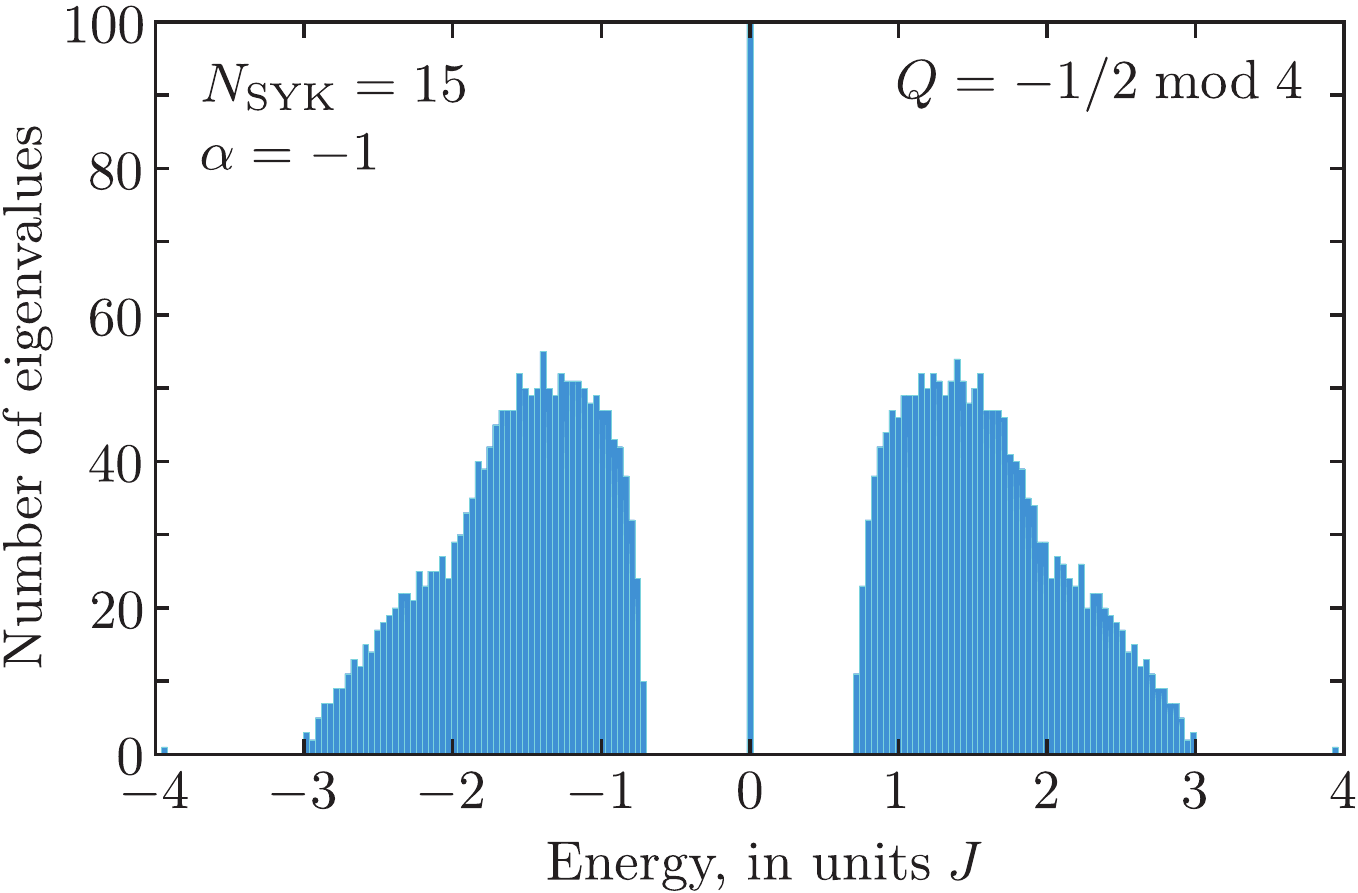}\hspace{3em}
 \includegraphics [width=0.4\textwidth, angle=0.]{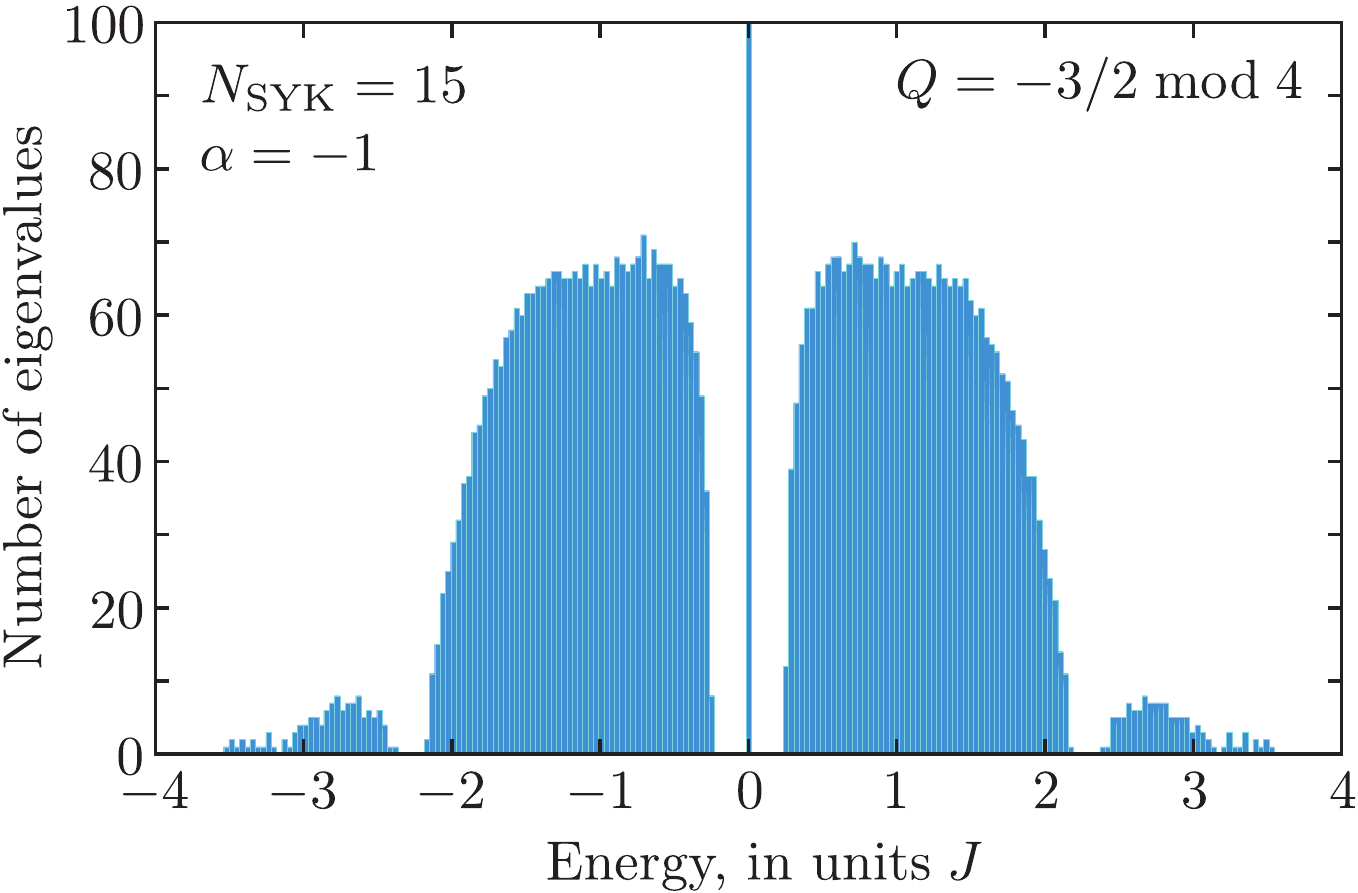}
  \end{center}
  \caption{The spectrum for a single realization of the coupled SYK model with $N_{\rm SYK} = 15$ and $\alpha=-1$ separated into four $Z_4$ symmetry sectors. In each of the sectors the spectrum is symmetric under $E\rightarrow -E$. 
The sectors with $Q=\pm 1/2$ mod $4$ contain the ground state separated by a gap from the rest of the states.  }
  \label{separatespectra15}
\end{figure}

For $N_{\rm SYK}=15$, due to the anomaly in particle-hole symmetry, there are two degenerate ground states, see fig. \ref{espectrum15}.\footnote{If we instead adopt
the  Maldacena-Qi Hamiltonian with a quadratic coupling which breaks the particle-hole symmetry explicitly, there is no such double degeneracy.} In fact, each energy level is doubly degenerate. This is due to the fact that the spectra in the sectors with charges $Q= 1/2$ mod $4$, and with charges $Q=- 1/2$ mod $4$ are identical; similarly, the spectra with
$Q= \pm 3/2$ mod $4$ are identical.
For $-1\leq \alpha < 0$ we observe a gap between the lowest energy level and the next one, as expected. The spectra for $\alpha=-1$ separated into the four sectors are shown in fig. 
 \ref{separatespectra15}.
On the other hand, for $N_{\rm SYK}=16$ there is no exact degeneracy of the ground state, but the first gap is very small, indicating a tendency towards spontaneous symmetry breaking at large $N_{\rm SYK}$.
We show the $N_{\rm SYK}=16$ spectra for $\alpha = -1$ and $\alpha = -0.5$ in fig. \ref{espectrum15}. In both cases, for a typical sampling of the coupling constants $J_{ijkl}$ we observe two closely spaced states followed by a visible gap. For large $N_{\rm SYK}$ the energy gap between the two lowest states is expected to decrease exponentially:
\begin{equation}
-\log \frac{E_1- E_0}{J} \sim N_{\rm SYK} \ .
\end{equation}
For $\alpha \geq 0$ the low-lying spectrum is different -- we observe many closely spaced low-lying states without large gaps, similarly to the standard SYK spectrum.

\begin{figure}[h!]
    \centering
         \includegraphics[width=0.48\textwidth, angle=0.]{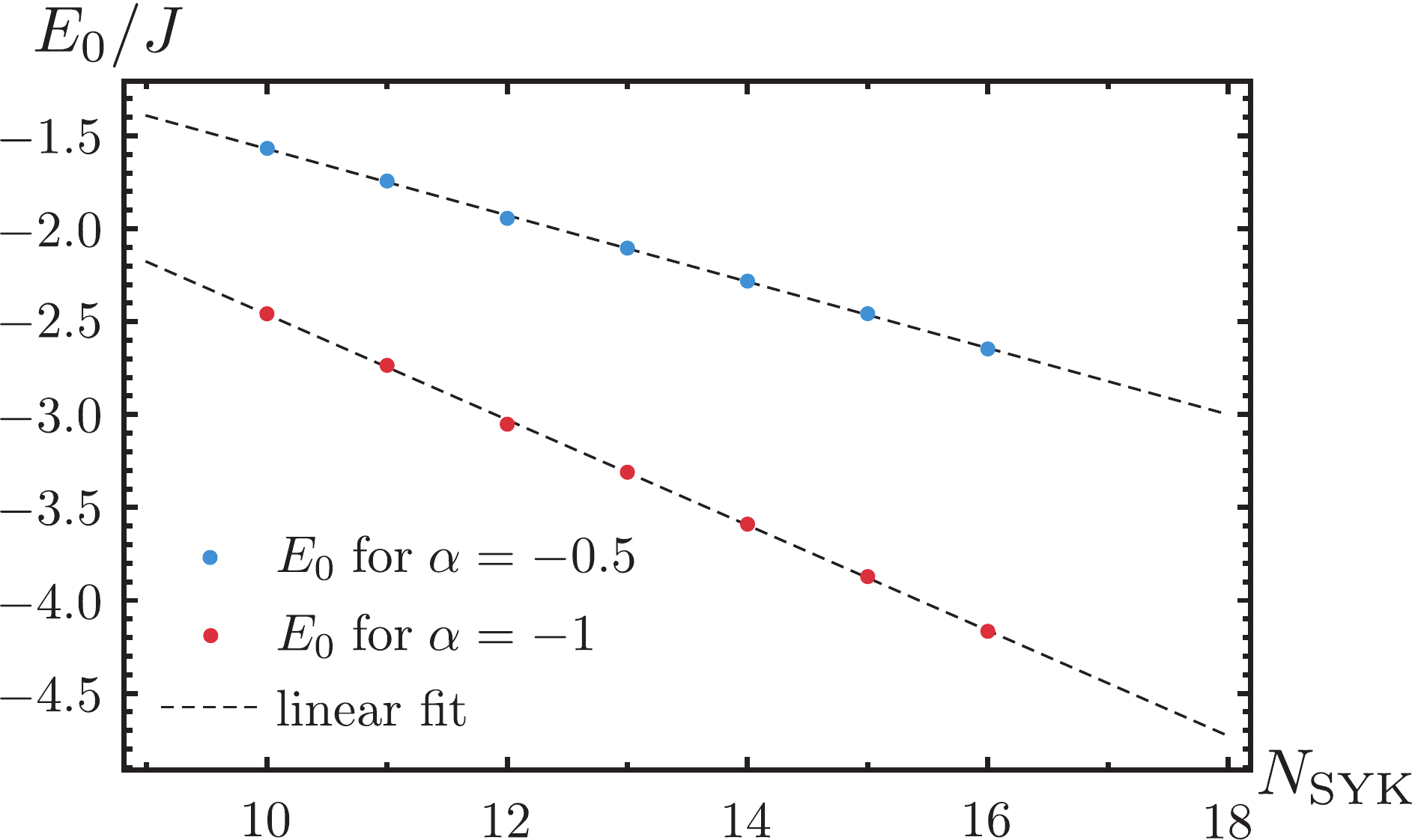}\hspace{2em}
          \includegraphics [width=0.435\textwidth, angle=0.]{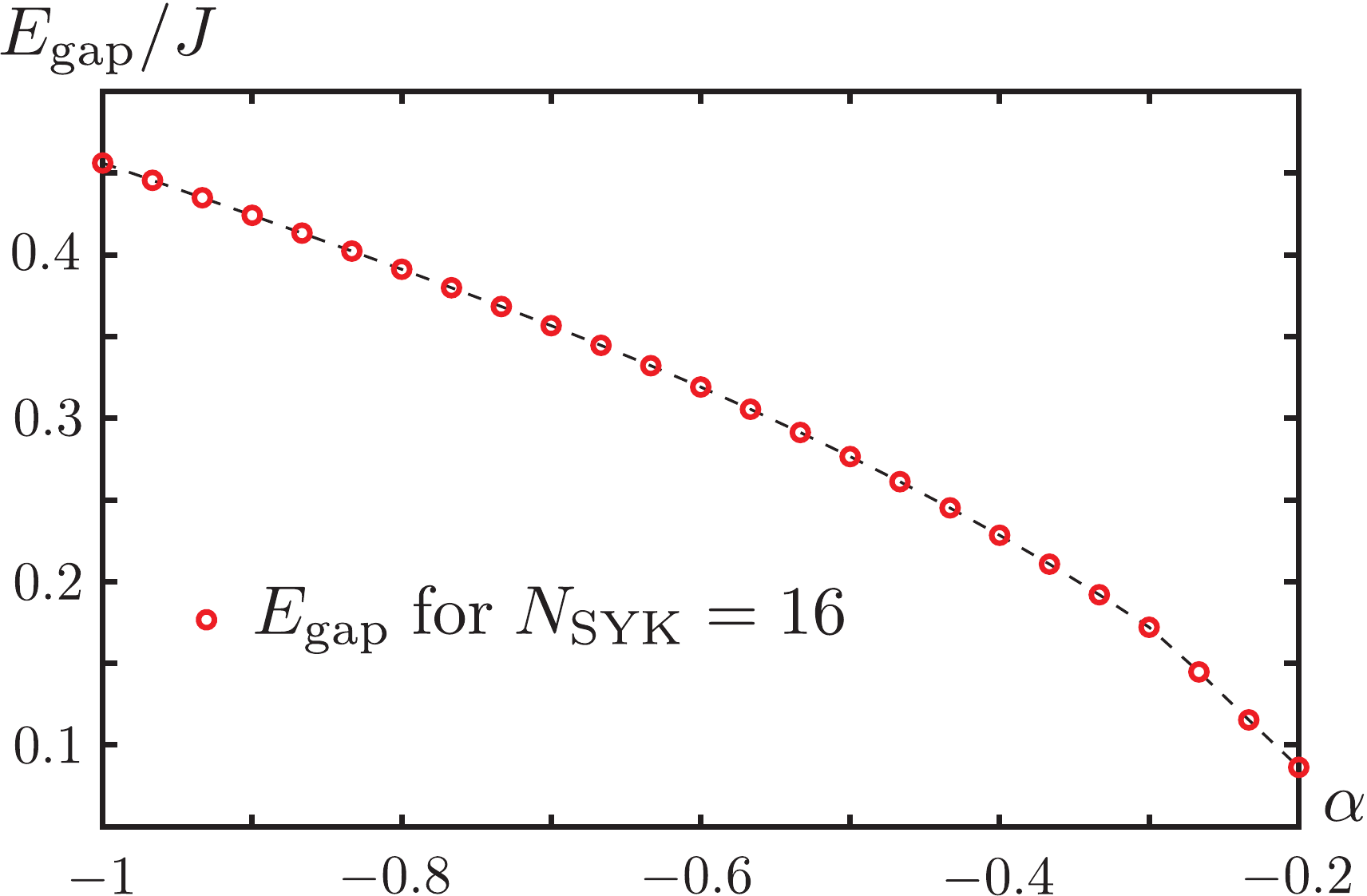}
          \caption{Left: The ground state energy for $\alpha=-0.5,\,-1$ and $N_{\rm SYK} =10,11,\dots,16$ (The number of samples are: $250000, 120000, 50000, 5000, 5000, 2000, 500$). The linear fit is  shown by dashed lines.
          Right: The energy gap between second and third states as a function of $\alpha$ for a single realization of random couplings at $N_{\rm SYK} = 16$.  }
    \label{gstate}
\end{figure}

In fig. \ref{gstate} we plot the ground state energy for $\alpha=-1$ and $\alpha = -0.5$ with $N_{\rm SYK}=10, \ldots, 16$. The plots, where $J$ is set to $1$, are approximately linear, and the fits give
\begin{equation}
\label{gslinear}
E_0^{\alpha=-1}= -0.283  N_{\rm SYK} +0.373\ , \qquad E_0^{\alpha=-0.5}= -0.179  N_{\rm SYK} +0.217\ .
\end{equation}
The limiting values $E_0^{\alpha=-1}/N_{\rm SYK}=-0.283$ and $E_0^{\alpha=-0.5}/N_{\rm SYK}=-0.179$ are in good agreement with the result found from Schwinger-Dyson equations; see fig. \ref{egapvsalpha}.
In figure \ref{gstate} we also exhibit the energy gap between second and third states as a function of $\alpha$. 
As $\alpha$ is increased from $-1$ to $0$, the gap decreases as expected. 

\begin{figure}[h!]
    \centering
    \includegraphics[width=0.45\textwidth, angle=0.]{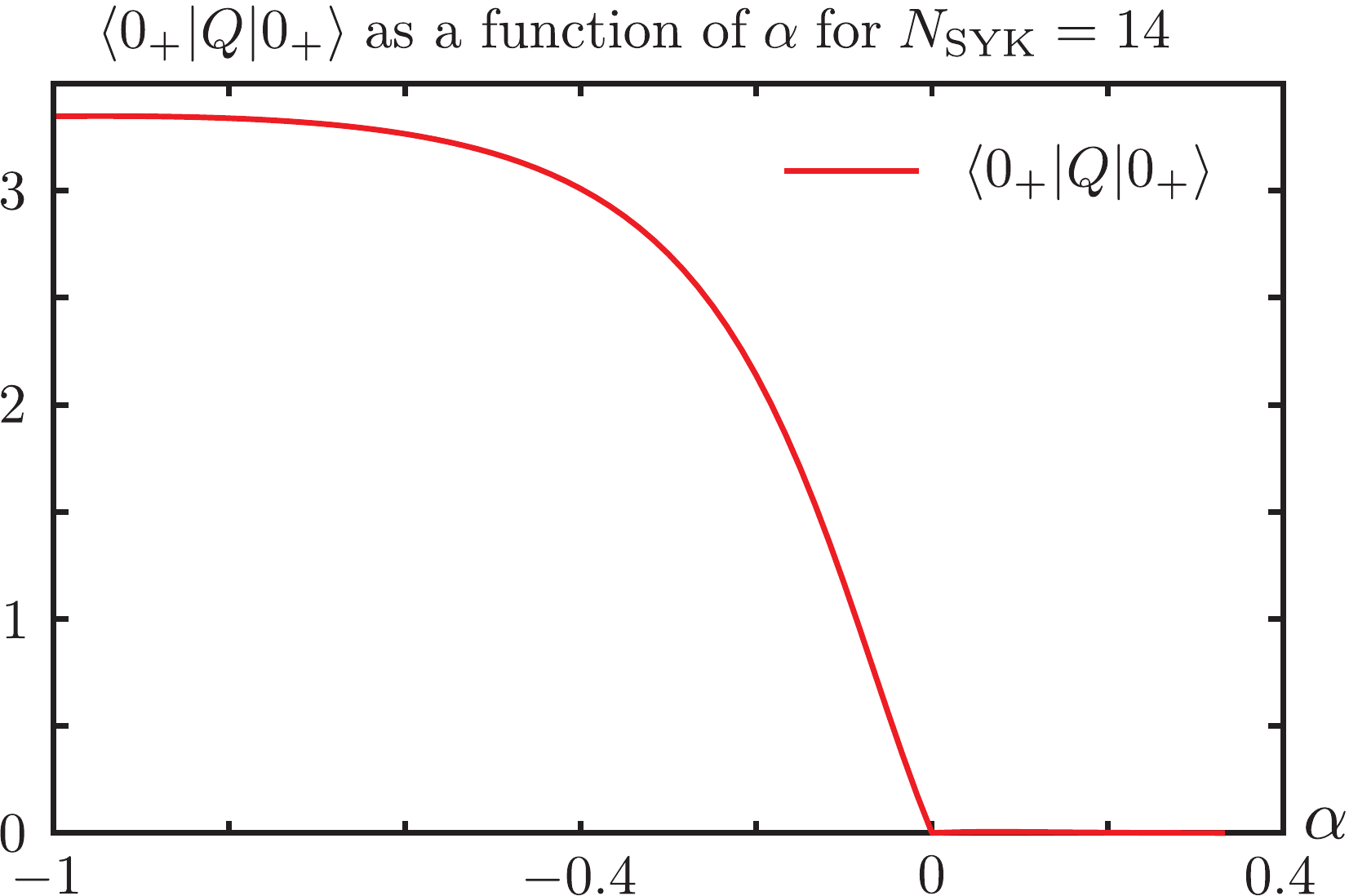}
    \caption{The expectation value $\bra{0_+} Q \ket{0_+}$ as a function of $\alpha$ for a single realization of random couplings at $N_{\rm SYK} = 14$.}
    \label{fig:N14VEV}
\end{figure}

Exact diagonalizations also provide support for the statement that the fermion number $Q$ acquires a vacuum expectation value for $-1 \leq \alpha < 0$. 
For $N_{\rm SYK}$ not divisible by $4$, there are two ground states $\ket{0_\pm}$ which map into each other under the symmetry generator ${\cal P}$.
This can be viewed as anomalous breaking of the 
time-reversal $\mathbb{Z}_2$ symmetry (\ref{phsym}) which occurs for a finite number
of degrees of freedom \cite{2009AIPC.1134...22K,Fidkowski:2009dba,Witten:2015aba,Fu:2016yrv,Cotler:2016fpe}. 
 In figure \ref{fig:N14VEV} the vacuum expectation value as a function of $\alpha$ is plotted for $N_{\rm SYK} = 14$. 
This is the finite $N_{\rm SYK}$ analogue of fig. \ref{condvsalpha}, where the large  $N_{\rm SYK}$ limit of the condensate is plotted.
We also note the qualitative similarity of the plot \ref{fig:N14VEV} and that of the
imaginary part of the scaling dimension of $Q$ in fig. \ref{fgraph}.

\section*{Acknowledgments}

Some of the results presented here are from Jaewon Kim's Princeton University Senior Thesis (May 2018) \cite{Jaewon:2018}.
IRK is grateful to the Kavli Institute for Theoretical Physics at UC, Santa Barbara and the organizers of the program "Chaos and Order: From strongly correlated systems to black holes" for the hospitality and stimulating atmosphere during some of his work on this paper. His research at KITP was supported in part by
the National Science Foundation under Grant No. NSF PH-1748958.
IRK is also grateful to the participants of the program, and especially D. Gross, C.-M. Jian, A. Kitaev, J. Maldacena, D. Stanford, J. Verbaarschot, E. Witten and C. Xu, for very useful discussions.  GT would like to thank D. Jafferis for useful discussions. 
The work of IRK and WZ was supported in part by the US NSF under Grant No.~PHY-1620059. The work of GT was supported by  the MURI grant W911NF-14-1-0003 from ARO, by DOE grant DE-SC0007870 and by DOE Grant No. DE-SC0019030.

\appendix
\section{More on the discrete symmetries }
\label{discretesymmetries}

The model (\ref{SYK2fl}) has the anti-unitary particle-hole $\mathbb{Z}_2$ symmetry generated by (\ref{phsym}).
The operator $K$ is defined to take $z \rightarrow \bar z, \ z \in \mathbb{C}$ but acts as the identity on $\psi$ or $\bar\psi$. 
It may be identified as a kind of time-reversal generator which satisfies $K^2=1$ \cite{2009AIPC.1134...22K,Fidkowski:2009dba,Witten:2015aba}.
It acts by
\begin{equation}
K i K=- i\ , \qquad 
K\chi_1^iK = \chi_1^i\ , \qquad K\chi_2^i K = -\chi_2^i\ ,
\end{equation}
and therefore, satisfies
 \begin{equation}
[K, H]= [K, Q]= 0 \ .
\end{equation}

Note that although $K$ can be anomalous, $K$ is unbroken as it does not change the sign of $Q.$ Another unbroken symmetry is the $\frac{\pi}{2}$ rotation between $\chi^i_1$ and $\chi^i_2.$
\begin{equation}
R=(-1)^{N_{\text{SYK}}/4}2^{-N_{\text{SYK}}/2}\prod_{i}(1-2\chi_1^i\chi_2^i).
\end{equation} 

It satisfies 
\begin{equation}
RR^{\dagger}=1\ , \qquad  R\chi_1^iR^{\dagger}=\chi_2^i\ , \qquad  R\chi_2^iR^{\dagger}=-\chi_1^i\ , \qquad R^4=1. 
\end{equation}

Note $R^2=(-1)^F.$ There are also various reflection $\mathbb{Z}_2$ symmetries that are spontaneously broken by the VEV of $Q.$ In particular, we have the reflection symmetry:
 \begin{equation}
\label{reflectioneven}
P = \begin{cases}(-1)^{N_{\rm SYK}(N_{\rm SYK}-1)/4}2^{N_{\rm SYK}/2} \prod_{i=1}^{N_{\rm SYK}} \chi_1^i & \mbox{if } N_{\rm SYK}=2k, k\in \mathbb{Z}\\ (-1)^{N_{\rm SYK}(N_{\rm SYK}-1)/4}2^{N_{\rm SYK}/2}  \prod_{i=1}^{N_{\rm SYK}} \chi_2^i & \mbox{if } N_{\rm SYK}=2k+1, k\in \mathbb{Z}, \end{cases}
\end{equation}
 such that 
 \begin{equation}
PP^\dagger=1\ , \qquad    P\chi_1^{i} P^\dagger = -\chi_1^{i}\ , \qquad P \chi_2^{i} P^\dagger = \chi_2^{i}\ , \qquad  P^2=1.
\end{equation}
 
 In fact, $R, P, $ and $K$ are enough to generate all discrete symmetries of the model (\ref{SYK2fl}). In particular, all the unitary discrete symmetries form $D_4$, the dihedral group of order 8. To see this, it's enough to check that the group presentation: $R^4=P^2=(RP)^2=1.$ The remaining reflections can be identified with $RP, R^2P$ and $R^3P.$ For a given unitary symmetry we can compose it with $K$ to obtain an anti-unitary one. 
 
In our case, when $N_{\rm SYK}\to\infty$, although multiple $\mathbb{Z}_2$ symmetries are spontaneously broken, we only expect a two-fold ground state degeneracy. In fact, any two broken symmetries that can be related by an unbroken symmetry do not produce any extra ground state degeneracy. To see this, consider for example the reflection symmetry $RP.$ Since $R$ is unbroken, we may assume $R\ket{0}=\ket{0}$ without losing of generality. Then $RP\ket{0}=RPR\ket{0}=P\ket{0}.$
 
At finite $N_{\rm SYK}$, however, certain discrete symmetry can be anomalous and is responsible for an exact two fold degeneracy for certain $N_{\rm SYK}$. For example, the particle-hole symmetry ${\cal P}\sim K P$ acts on the fermions as
\begin{equation} 
{\cal P} \psi^j {\cal P} = \eta \bar \psi^j\ , \qquad 
{\cal P} \bar \psi^j {\cal P} = \eta \psi^j\ , \qquad \eta = (-1)^{(N_{\rm SYK}+2) (N_{\rm SYK}-1)/2}\ .
\end{equation}
The fermion number operator (\ref{fermnum})
is odd under this symmetry: 
\begin{equation}
{\cal P} Q {\cal P} =- {\cal P}^2 Q\ .
\end{equation}
When $N_{\rm SYK}$ is not divisble by $4$, there are two degenerate ground states 
 $\ket{0_\pm}$, and the symmetry generator ${\cal P}$  maps them into each other \cite{2009AIPC.1134...22K,Fidkowski:2009dba,Witten:2015aba,PhysRevB.95.115150,Fu:2016yrv,Cotler:2016fpe}:
\begin{equation}
{\cal P} \ket{0_+}= (-1)^{N_{\rm SYK} (N_{\rm SYK}-1)/4} \ket{0_-}\ , \qquad {\cal P} \ket{0_-}= (-1)^{N_{\rm SYK} (N_{\rm SYK}-1)/4} \ket{0_+}\ .
\label{P0pmEq}
\end{equation}
In this case we can say that the particle-hole symmetry is anomalous.

\section{Zero-energy states in the bipartite model}
\label{zeroenergy}

The bipartite model, which is the $\alpha=-1$ case of the two-flavor tensor or SYK model, has some additional symmetries which make it special. In general the spectrum of the two-flavor SYK is not symmetric under $E\rightarrow -E$ for a given random coupling $J_{ijkl}$. However, for $\alpha=-1$ the spectrum is exactly symmetric for any
choice $J_{ijkl}$ due to the duality symmetry (\ref{duality1}). This symmetry acts by 
\begin{equation} 
\psi^j \rightarrow \frac{1+i}{\sqrt 2}  \bar \psi^j\ , \qquad \bar \psi^j \rightarrow \frac{1-i}{\sqrt 2}  \psi^j\ ,
\end{equation}
and for $\alpha=-1$ this reverses the sign of the Hamiltonian of bipartite model, $H_{\alpha=-1}$, which is given in (\ref{bipartiteHamilt}).

Furthermore, the model with $\alpha=-1$ has a large number of zero-energy states. For the SYK model, the sharp peak at $E=0$ may be seen in fig. \ref{espectrum15}. 
For a generic choice of $J_{ijkl}$ where they are all non-vanishing, the number of $E=0$ states does not depend on their values. In fact, it is not hard to calculate this number separately for each $Z_4$ symmetry sector. The separate sectors may be labeled by $Q=\tilde Q$ mod $4$, where $\tilde Q=0, \pm 1, 2$ when $N$ is even, and
$\tilde Q =\pm 1/2, \pm 3/2$ when $N$ is odd.\footnote{In this Appendix $N$ denotes $N_{\rm SYK}$.}
The general formula for the number of $E=0$ states in sector $\tilde Q$ is
\begin{equation}
\label{numzero}
{\cal N}_{\tilde Q} = \sum_{m=-[(N+2 \tilde Q)/8]}^{[(N-2 \tilde Q)/8]}  (-1)^m { N \choose  \frac{N}{2} + \tilde Q + 4m} \ .
\end{equation}
This formula is applicable to ``generic" bipartite Hamiltonians (\ref{bipartiteHamilt}), where all $J_{ijkl}$ are non-vanishing; in such cases, it does not depend on the specific choice of couplings. 
However, if some couplings $J_{ijkl}$ vanish, then the number of $E=0$ states may be higher than (\ref{numzero}). For example, in the 
$O(N_1)\times O(N_2) \times O(N_3)$ tensor bipartite models, where many quartic couplings vanish \cite{Klebanov:2018fzb}, the number of $E=0$ states is greater than that given by 
(\ref{numzero}) with $N= N_1 N_2 N_3$.

\begin{figure}[h!]
    \centering
    \includegraphics[width=0.5\textwidth, angle=0.]{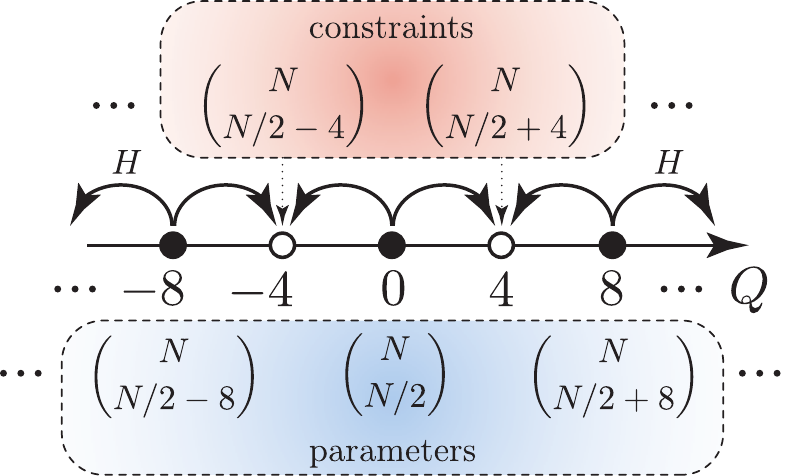}
    \caption{This picture represents the counting of zero-energy states in the $\tilde Q=0$ sector. They are superpositions of states with charges $Q=0\, \textrm{mod}\, 8$.}
    \label{fig:zerstates}
\end{figure}

To explain the origin of the formula (\ref{numzero}), let us consider for example the $\tilde Q=0$ sector of a model with even $N_{SYK}$. In this sector the $E=0$ states may be
obtained from superpositions of states with $Q=0$ mod $8$.\footnote{This may be interpreted as the fact that in the zero-energy sector there is symmetry enhancement from $Z_4$ to $Z_8$.} The dimension of Hilbert space in this sector is  
\begin{equation}
d_{0{\rm mod}8} = \sum_m  { N \choose  \frac{N}{2}+ 8m}
\ .
\end{equation}
When the Hamiltonian of bipartite model acts on such a state, it maps it to a superposition of states with $Q=4$ mod $8$ (see fig. \ref{fig:zerstates}). 
The total number of such states is 
 $\sum_m  { N \choose  \frac{N}{2}+4+ 8m}$, and this is the number of constraints from the requirement that $H_{\alpha=-1}$ annihilates the zero-energy states. 
Subtracting this number of 
constraints from $d_{0{\rm mod}8}$, we arrive at (\ref{numzero}) for the case $\tilde Q=0$. Analogous reasoning provides a derivation of  (\ref{numzero}) for other values of $\tilde Q$.
We have checked numerically that all the $E=0$ wave functions are mixtures of only the states with $Q=\tilde Q$ mod $8$, and that their numbers for any random sampling
of $J_{ijkl}$ are given by  (\ref{numzero}).

For example, for $N=16$ the number of states in the $\tilde Q=0$ sector is
\begin{equation}
{\cal N}_0 = { 16 \choose  8} + 2 { 16 \choose  16} - 2 { 16 \choose  4}= 9232\ .
\end{equation}
The number of states in the $\tilde Q=\pm 1$ sectors is
\begin{equation}
{\cal N}_{1}=
{\cal N}_{-1} = { 16 \choose  7} + { 16 \choose  15} -  { 16 \choose  3}- { 16 \choose  11} = 6528\ .
\end{equation}
The number of states in the $\tilde Q=2$ sector vanishes for any even $N$.

\vspace{10mm}

\bibliographystyle{ssg}
\bibliography{Z2citations}

\end{document}